\documentclass[prb,aps,superscriptaddress,amsmath,twocolumn,amssymb,floatfix,showpacs,superscriptaddress,nofootinbib,longbibliography]{revtex4-2}
\usepackage{braket}
\usepackage[dvipsnames]{xcolor}
\usepackage{float}
\usepackage{subfigure}
\usepackage{tikz}
\usepackage[colorlinks=true,linktoc=page,linkcolor=OliveGreen,citecolor=purple,urlcolor=violet]{hyperref}

\mathchardef\mhyphen="2D 

\newcommand{\ie}{{i.e.,\,\,}}

\newcommand\bea{\begin{eqnarray}}
\newcommand\eea{\end{eqnarray}}
\newcommand\beq{\begin{equation}}  
\newcommand\eeq{\end{equation}}

\usepackage[normalem]{ulem}
\definecolor{lime}{HTML}{A6CE39}
\usepackage{sidecap,tikz}
\DeclareRobustCommand{\orcidicon}{\hspace{-1.0mm}
	\begin{tikzpicture}
		\draw[lime, fill=lime] (0.0,0.0) 
		circle [radius=0.15] 
		node[white] {{\fontfamily{qag}\selectfont \tiny \,ID}};
		\draw[white, fill=white] (-0.0525,0.095) 
		circle [radius=0.007];
	\end{tikzpicture}
	\hspace{-3.0mm}
}
\foreach \x in {A, ..., Z}{\expandafter\xdef\csname orcid\x\endcsname{\noexpand\href{https://orcid.org/\csname orcidauthor\x\endcsname}{\noexpand\orcidicon}}
}

\AtBeginDocument{%
	\newwrite\bibnotes
	\def\bibnotesext{Notes.bib}
	\immediate\openout\bibnotes=\jobname\bibnotesext
	\immediate\write\bibnotes{@CONTROL{REVTEX41Control}}
	\immediate\write\bibnotes{@CONTROL{%
			apsrev41Control,author="08",editor="1",pages="1",title="1",year="1"}}
	\if@filesw
	\immediate\write\@auxout{\string\citation{apsrev41Control}}%
	\fi
}%

\begin{document}

\title{Non-Hermitian  band topology in twisted bilayer graphene aligned with hexagonal boron nitride}  

\author{Kamalesh Bera\orcidA{}}
\email{kamalesh.bera@iopb.res.in}
\affiliation{Institute of Physics, Sachivalaya Marg, Bhubaneswar-751005, India}
\affiliation{Homi Bhabha National Institute, Training School Complex, Anushakti Nagar, Mumbai 400094, India}

\author{Debasish Mondal\orcidB{}}
\email{debashish.m@iopb.res.in}
\affiliation{Institute of Physics, Sachivalaya Marg, Bhubaneswar-751005, India}
\affiliation{Homi Bhabha National Institute, Training School Complex, Anushakti Nagar, Mumbai 400094, India}

\author{Arijit Saha\orcidC{}}
\email{arijit@iopb.res.in}
\affiliation{Institute of Physics, Sachivalaya Marg, Bhubaneswar-751005, India}
\affiliation{Homi Bhabha National Institute, Training School Complex, Anushakti Nagar, Mumbai 400094, India}

\author{Debashree Chowdhury\orcidD{}}
\email{debashreephys@gmail.com}
\affiliation{Centre for Nanotechnology, IIT Roorkee, Roorkee, Uttarakhand 247667, India}

\begin{abstract}
Utilizing the established Bistritzer-MacDonald model for twisted bilayer graphene (tBLG), we theoretically investigate the non-Hermitian (NH) topological properties of this in the presence of non-reciprocal (NR) hopping on both layers and hexagonal boron nitride (hBN) induced mass term incorporated only on the top layer of the tBLG system. It is well known that the hBN mass term breaks the
 \(C_{2}\) symmetry of tBLG and gaps out the Dirac cones inducing a valley Hall insulating phase. However, when NR hopping is introduced, this system transits into a NH valley Hall insulator (NH-VHI). 
 Our analysis reveals that, in the chiral limit, the bandwidth of the system vanishes under NH effects for a wide range of twist angles. Such range can be visibly expanded as we enhance the degree of 
 non-Hermiticity (\(\beta\)). At the magic angle, we observe that enhancement of \(\beta\) inflates the robustness of the gapless Dirac points, requiring a progressively larger mass term to induce a gap
 in the NH tBLG system. Additionally, for a fixed NH parameter, we identify a range of twist angles where gap formation is significantly obstructed. To explore the topological aspects of the NH tBLG, we analyze the direct band gap in the Moir\'e Brillouin zone (mBZ) and compute the Chern number for the NH system. We find that the corresponding topological phase transitions are associated with corresponding direct band gap closings in the mBZ.  
\end{abstract}

\maketitle

\section{Introduction}
Hermiticity is one of the crucial aspects of quantum mechanics, where it is postulated that each observable should be associated with a Hermitian operator. However, in systems where the exchange of energy and particles with the environment is allowed, a violation of the probability conservation is observed, and the operators are no longer Hermitian~\cite{Bender_review}. Non-Hermiticity is a natural consequence in photonic systems~\cite{NH-photonics}, and it has been known for a long time. However, the concept of non-Hermiticity in condensed matter systems is relatively new and presents a fertile ground for exploring interesting physical phenomena related to topology. The properties that make non-Hermitian (NH) systems unprecedented, include the appearance of unique degeneracy, coined as exceptional degeneracy, where, apart from the merging of eigenenergies, the eigenfunctions also merge. Capturing new topological phases in NH systems and the interplay between non-Hermiticity and topology has become a major frontier of research in condensed matter physics~\cite{NH_topo_review, NH_topo1, NH_topo2, NH_chern1, NH_chern2}. Beyond theoretical interest, NH topology also holds promising practical applications, such as enabling exponentially precise sensors, amplifiers, and light funnels~\cite{topological_sensor, quantum_signal, topological_funneling}. 
While most experimental observations of NH topological phases have so far been explored based on ultracold atoms and optical systems~\cite{cold_atom, optical_system1, optical_system2}, a very 
recent experiment has also identified this phase in a condensed matter system~\cite{nh_topo_cmp}. Topology in condensed matter physics began flourishing after the discovery of the quantum Hall effect~\cite{quantum_Hall_effect}. Over time, the discovery of other topological systems, such as topological insulators~\cite{TI_expt}, topological semimetals~\cite{topo_semimetal_expt}, and Weyl semimetals~\cite{WSM_expt} etc., have followed. The exploration of NH topological phases in these systems is already being demonstrated~\cite{NH_TI, NH_SM, NH_HOWSM, NH_HOTSC}. 
The search for new topological phases in NH systems remains one of the primary focus areas in contemporary research.

On the other hand, as far as two-dimensional (2D) materials are concerned, the electronic structure of a single-layer graphene can be enormously modified when the graphene layers are stacked 
on top of each other (also known as van der Waals heterostructures)~\cite{MLG}. Once the two layers of graphene are twisted by a small angle, a Moir\'e pattern emerges~\cite{VHS_tBLG_2010} 
in the real space geometry. The twisted systems are not only highly tunable but can also be fabricated using different types of lattices~\cite{tSL, kagome, dice} and various stacking configurations~\cite{Mono-Bi-graphene1, Park2021-tTLG1, Liu2020-tDBLG1, TMD-homobilayers1, rTTLG+hBN1,tDBLG_soc}. In Recent years, Moir\'e systems have emerged as a highly compelling field of study due to their capacity to manifest a diverse array of exotic physical phenomena, ranging from topology to strong electron correlation~\cite{Cao2018-corr_insulator, Choi2019, Wu2021-chernIns-expt, Nuckolls2020-tBLG_xpt, Choi2021, Xie2021, magnetism1, magnetism2, nematicity1, nematicity2, Cao2018-unconv_sc, Oh2021-tBLG_xpt, Sinha2022}. Importantly, these crucial physical properties are pronounced near various discrete values of twist angles, known as the magic angles, the first of which is reported to be around $1.05^{o}$ for twisted bilayer graphene (tBLG). Initial investigations on tBLG theoretically predicted the presence of low-energy flat bands~\cite{MacDonald-tBLG, Moon-tBLG, Koshino-tBLG, Santos-Peres-tBLG, Shallcross-tBLG}, which were later observed experimentally~\cite{Cao2018-corr_insulator, Cao2018-unconv_sc}. 

While the appearance of flat bands at different magic angles and their impact have been extensively studieded in different systems~\cite{All_Magic_Angle-tBLG, Origin_of_Magic_Angle-tBLG, CIS-tBLG, Magnetism-tBLG, Heavy_fermion-tBLG, Origin_mott, phonon_SC, MIT_tBLG, Floquet}, the NH tBLG and its topological properties remain a less explored avenue~\cite{NH_tBLG1, NH_tBLG2}. Very recently, Ref.~\cite{NH_tBLG1} analyzed the emergence of exceptional magic angles (EMAs) in tBLG considering only the low energy bands in the presence of non-reciprocal (NR) hopping, which introduces non-Hermiticity into the system. Although the role of non-Hermiticity in modifying the magic angles is discussed, the topological aspects of the system remain unaddressed. Furthermore, it is unclear how the emergent flat bands (FBs) at the magic angle in the Bistritzer-MacDonald (BM) model evolve under NH effects. This motivates us to ask the following intriguing questions: (a) can we identify the EMAs in the BM model? (b) can we engineer NH topology in this system? (c) if so, how do the EMAs affect the topology in both the chiral and non-chiral limits? Additionally, (d) how do varying twist angles and the strength of non-Hermiticity influence the system's topology even in presence of hexagonal boron nitride (hBN) substrate?

In this article, we intend to answear the above mentioned questions and for that we propose a model Hamiltonian to realize a NH topological phase in tBLG by introducing non-reciprocal (NR) tunneling 
in each layer. The FBs in tBLG are gapless at the Dirac points of the Moir\'e Brillouin zone (mBZ) and are protected by $C_2 T$-symmetry (where $C_2$ represents the two-fold rotation and $T$ denotes the time reversal operation). Breaking this symmetry can open up a gap in the FBs, transforming them into a topologically non-trivial phase. One possible approach for breaking $C_2$-symmetry is to 
align one layer of graphene with hBN, which induces a mass gap in the FBs and leads to a topological state~\cite{tBLG_hBN1, tBLG_hBN2}. 
First, we analyze the NH band dispersion for both untwisted bi-layer graphene (BLG) and tBLG at different values of non-Hermiticity and hBN-induced mass term. From the behavior of the band gap over the BZ, we observe the emergence of a nodal line in the presence of non-Hermiticity in both BLG and tBLG. In the latter part of our study, we compute the FB bandwidth and identify the non-Hermiticity induced magic angles as a function of inter-layer coupling and non-Hermiticity strength in tBLG. Finally, we investigate the system's topological properties by computing two key quantities: the direct 
band gap and the Chern number. Through several phase diagrams, we examine the interplay between non-Hermiticity and topology, finding that non-Hermiticity generally suppresses topological behavior. 
The direct band gap features are further supported by Chern number calculations, confirming that the gapped regions correspond to a NH valley Hall insulating phase. Results for both the chiral and 
non-chiral limits of tBLG are discussed throughout our analysis.

The remainder of this paper is organized as follows. In Sec.~\ref{Sec:II}, we introduce our model Hamiltonian in the presence of NH ingredients for both tBLG and BLG. In Sec.~\ref{Sec:III}, we present 
the electronic band dispersion of the NH-tBLG and NH-BLG. In Sec.~\ref{Sec:IV}, we discuss the FB bandwidth and identify NH magic angles for different strengths of non-Hermiticity. Sec.~\ref{Sec:V} 
is devoted to the discussion of topological properties of the NH-tBLG by calculating the direct band gap and Chern number for different parameter values. Finally, we summarize and conclude 
our paper in Sec.~\ref{Sec:VI}.

\section{Model}\label{Sec:II}
In this section, we briefly discuss the construction of the model Hamiltonians for NH-BLG and NH-tBLG.

\subsection{NH-BLG}\label{subsec:IIA}
We consider BLG and stack a hBN substrate on top of it in such a way that the top layer of graphene remains aligned with the hBN layer. To investigate the effects of non-Hermiticity in this system, we introduce NR hopping within each layer of BLG. Considering the tight-binding model~\cite{McCann_2011-BilayerReview}, the final Hamiltonian can be written as:
\begin{eqnarray} 
	H^{\text{NH}}_{\text{BLG-hBN}} &=
	\left( \begin{array}{cccc}
		h_{t}  & h_{int}\\
		h_{int}^{\dagger} & h_{b}
	\end{array}\right)\ ,
	\label{Eq8}
\end{eqnarray}	
where the top and bottom layer of NH-BLG denoted by $h_{t}$, $h_{b}$ and the interlayer coupling between the two layers is denoted as $h_{int}$.

Non-Hermiticity in both layers is introduced via non-reciprocal (NR) nearest-neighbor hopping: the forward hopping amplitude is $t + \gamma$, whereas the reverse hopping is $t - \gamma$, as illustrated in Fig.~\ref{schematic}(a). Consequently, the Hamiltonian for the top layer, $h_{t}$, can be written as:

\begin{eqnarray} 
	h_{t} &=
	\left( \begin{array}{cc}
		M_0  & -(t + \gamma) f_{p} \\
		-(t - \gamma) f_{p}^{*} & -M_0 + \Delta
	\end{array}\right)\ .
\end{eqnarray}	
The Hamiltonian for the bottom layer, $h_{b}$, is expressed as:

\begin{eqnarray} 
	h_{b} &=
	\left( \begin{array}{cc}
		\Delta  & -(t + \gamma) f_{p} \\
		-(t - \gamma) f_{p}^{*} & 0
	\end{array}\right)\ .
\end{eqnarray}	
Finally, the tunneling Hamiltonian between the layers ($h_{int}$) is written as, 
\begin{eqnarray} 
	h_{int} &=
	\left( \begin{array}{cc}
		t_{4} f_{p}& -t_{3} f_{p}^{*}\\
		t_{1} & t_{4} f_{p}
	\end{array}\right) \ ,
	\label{EqT}	
\end{eqnarray}	
where, $f_{p} = 1 + 2  \text{exp}(-i 3 p_{y}/a) + \cos(\sqrt{3} p_{x} a /2)$ and $p_{x}$ and $p_{y}$ are the two crystal momentum along $x$ and $y$ directions respectively. In the above, $M_{0}$ represents the hBN induced mass, $t_3$ and $t_4$ correspond to the trigonal warping and particle-hole asymmetry terms in BLG, respectively. The parameter $\Delta$ accounts for the distinction between dimer and non-dimer sites in BLG.

\subsection{NH-tBLG}\label{subsec:IIB}
\begin{figure}
	\subfigure{\includegraphics[width=0.5\textwidth]{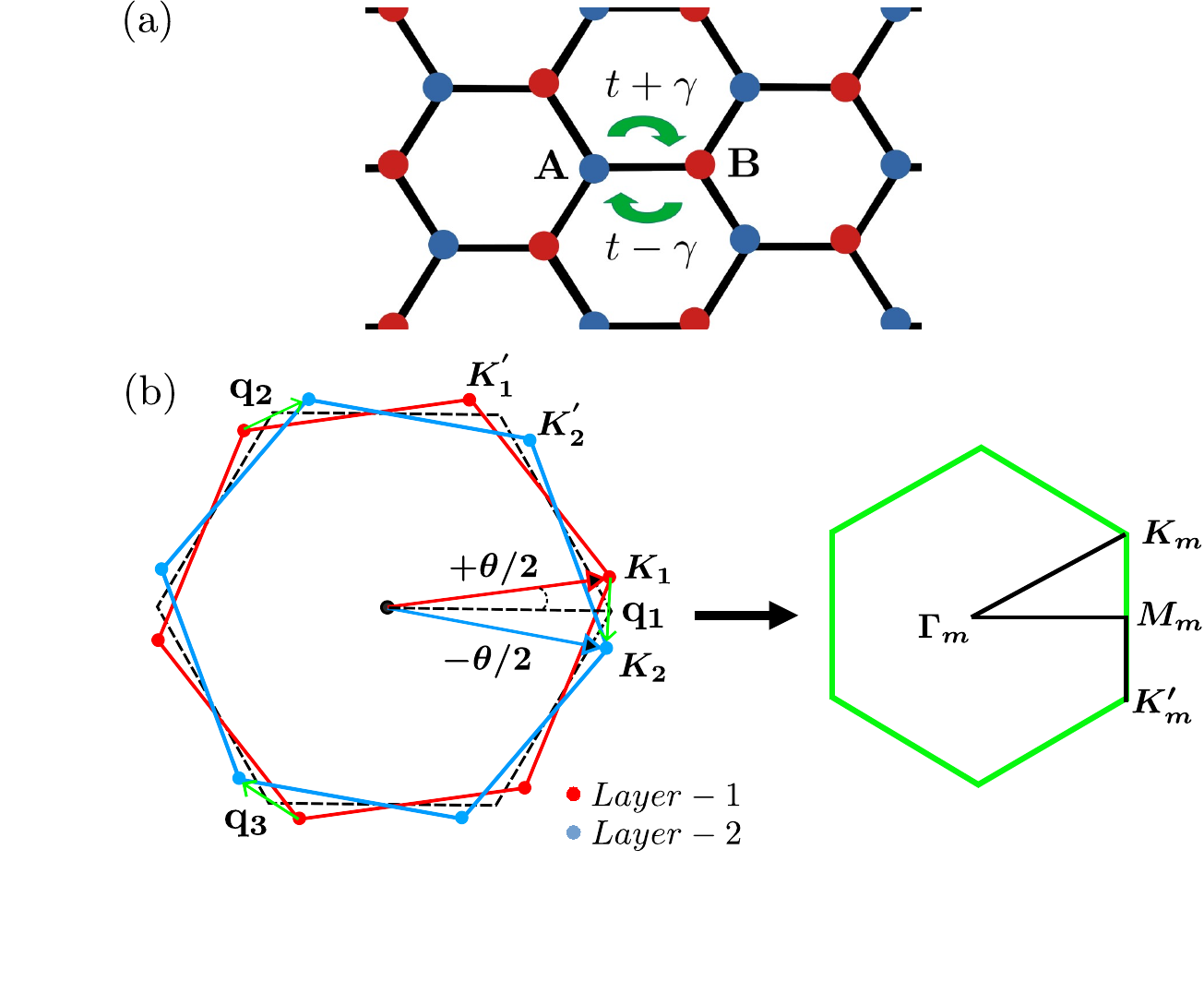}}
	\caption{ (a) Schematic diagram of the single layer graphene with NR hopping [($t + \gamma$) and ($t - \gamma$)] between the nearest neighbour lattice sites (b) Two hexagons (red and cyan) 
	indicate layer-1 and layer-2 of the single layer graphene rotated by $+\theta/2$ and $-\theta/2$ respectively. This rotational misalignment gives rise to three distinct momenta ($\mathbf{q_1}, 
	\mathbf{q_2}, \mathbf{q_3}$) at three equivalent Dirac points of the single layer graphene via which the two layers couple to each other. These three momenta form the mini Brillouin zone
	(highlighted by the green hexagon) as directed by an arrow. The black line over the mBZ manifests the high symmetry path we follow to establish the band dispersions of the twisted system.}
	\label{schematic}
\end{figure}

When two layers of graphene are stacked and slightly rotated by an angle $\theta$, they form tBLG. Further, we stack a hBN substrate on top of tBLG in a way that the top layer of graphene remains aligned with the hBN layer (\ie the angle between the top graphene layer and hBN layer is approximately zero). To investigate the effects of non-Hermiticity in this twisted system, we introduce 
NR hopping within each layer of tBLG. The final Hamiltonian can be written as,
\begin{align}
	H^{\text{NH}}_{\text{tBLG-hBN}} = H_{\text{tBLG}} + H_{\text{hBN}} + H_{\text{NH}}\ ,
	\label{Eq1}
\end{align}	
where, $H_{\text{tBLG}}$, $H_{\text{hBN}}$ and $H_{\text{NH}}$ represent the model Hamiltonians for tBLG, effect of hBN induced mass, and the effects due to non-Hermiticity, respectively. Below, we discuss each of these terms in more detail.

To explore the low-energy physics at small twist angles, we employ the continuum model developed by Bistritzer and MacDonald (BM)~\cite{MacDonald-tBLG,NH_tBLG1}. The corresponding Hamiltonian is given by,
\begin{eqnarray}
	\begin{aligned}
		H_{\text{tBLG}} &= \sum_{\bf k} \sum_{\alpha=t,b}  \psi_{\alpha}^{\dagger}(\mathbf{k}) h_{0}^{\xi}(\theta_{\alpha}) \psi_{\alpha}(\mathbf{k})\\ 
		+ &\sum_{j=1,2,3} \psi_{t}^{\dagger}(\mathbf{k+ q_j}) T^{\xi}_j \psi_{b}(\mathbf{k}) + H.c\ ,
	\end{aligned}
	\label{Eq2}	
\end{eqnarray} 
where, the low-energy Dirac Hamiltonian for each graphene layer can be expressed as,
\begin{eqnarray} 
	h_{0}^{\xi}(\theta_{\alpha}) &=
	\left( \begin{array}{cc}
		0  & -\hbar v_{H} k^{\xi}_-(\theta_{\alpha}) \\
		-\hbar v_{H} k^{\xi}_+(\theta_{\alpha}) & 0
	\end{array}\right)\ .
	\label{Eq3}
\end{eqnarray}	

Here, $k^{\xi}_{\pm} = \xi k_x \pm i k_y$ with $\xi = \pm 1$ denotes valley-$K$ and $K^{'}$ of the single layer graphene. Also, $v_{H}$ corresponds to the Fermi velocity of the low energy Dirac quasiparticles in the graphene layers, the subscript `$H$' stands for Hermiticity. Note that, $\psi_{\alpha}$ is a two component spinor in the basis of $A$- and $B$-sublattices of each layer of graphene
and $\alpha = t, b$ represents the top and bottom layer of the tBLG, respectively. Also, $\mathbf{k}(\theta) = R(\theta)\mathbf{k}$, with $R(\theta)$ being the rotation matrix (by an angle $\theta$) in 2D and $\theta_{\alpha} = \pm \theta/2$ denote the twist angles respectively for top and bottom layers of tBLG. The inter-layer tunnelling matrices can be written as,

\begin{eqnarray}
	T^{\xi}_{1} &=
	\left( \begin{array}{cc}
		u  & u^{'} \\
		u^{'} & u
	\end{array}\right) ~~,~~
	T^{\xi}_{2} 
	&=
	\left( \begin{array}{cc}
		u  & u^{'} \omega^{-\xi} \\
		u^{'}\omega^{\xi} & u
	\end{array}\right) ~~,~~ \nonumber\\
	&T^{\xi}_{3} 
	=&\!\!\!\!\!\!\!\!\!\!\!\!\!\!\!\!\!\!\!\!
	\left( \begin{array}{cc}
		u  & u^{'} \omega^{\xi} \\
		u^{'}\omega^{-\xi} & u
	\end{array}\right).
	\label{Eq4}
\end{eqnarray}

These interlayer terms (\ie $T^{\xi}_1$, $T^{\xi}_2$, $T^{\xi}_3$) couple the two layers of tBLG through the Moir{\'e} basis vectors (\ie $\mathbf{q_1}$, $\mathbf{q_2}$, $\mathbf{q_3}$), which arise due to the misalignment of the BZs of the individual graphene layers (as shown in Fig.~\ref{schematic}(b)). Note that, in our numerical computation, we consider upto the seventh nearest neighbor k-points in the momentum space lattice of the mBZ as the momentum cutoff~\cite{tBLG_cutoff1, tBLG_cutoff2, tDBLG_soc}. This is in contrast to the previous work where only the first nearest neighbor is implemented~\cite{NH_tBLG1}. The k-points inside a circle with the radius equal to the momentum cutoff, centered at the Dirac points, form the basis vectors. Diagonalizing the Hamiltonian in this basis results in the flat bands. Here, $u$ and $u^{'}$ denote the tunneling amplitudes between AA/BB and AB/BA sublattices, respectively and $\omega = e^{2 \pi i/3}$, with $2 \pi/3$ being the angle between three basis vectors of the Moir{\'e} BZ respectively. In the limiting case where the interlayer tunneling amplitude between identical sublattices (\ie AA/BB) vanishes (\ie $u = 0$), the model acquires chiral symmetry. In this idealized limit, the two quasi-flat bands become degenerate and perfectly flat at specific twist angles~\cite{Origin_of_Magic_Angle-tBLG, chiral_model}.

The mass term induced by the coupling between the top graphene layer and the hBN substrate (while neglecting any coupling with the bottom graphene layer) can be expressed as:
\begin{eqnarray}
	\begin{aligned}
		H_{\text{hBN}} &= M_0 \sum_{\bf k}  \psi_{t}^{\dagger}(\mathbf{k}) \sigma_{z} \psi_{t}(\mathbf{k})\ ,
	\end{aligned}
	\label{Eq5}	
\end{eqnarray} 
where, $\sigma_{z}$ denotes the Pauli matrix in the sublattice space. 

To incorporate the effects due to non-Hermiticity, we introduce a NR nearest-neighbor hopping term in each layer of graphene. In the tight-binding model of a single-layer graphene, the hopping amplitudes between the A- and B-sublattices are considered to be asymmetric: the forward hopping is given by ($t + \gamma$) while the reverse hopping is ($t - \gamma$), as illustrated in 
Fig.~\ref{schematic}(a). Thus the low-energy model near the Dirac point, associated with the NH contribution, can be obtained as follows:
\begin{eqnarray} 
	h_{\text{NH}}^{\xi} &=
	\left( \begin{array}{cc}
		0  & -\hbar v_{\text{NH}} k^{\xi}_- \\
		\hbar v_{\text{NH}} k^{\xi}_+ & 0
	\end{array}\right) \ ,
	\label{Eq6}
\end{eqnarray}	
where, $v_{\text{NH}} = 3\gamma / 2a$ and $v_{H}$ in Eq.~(\ref{Eq3}) is $3t/2a$.  Hence, we define a dimensionless parameter $\beta = v_{\text{NH}} /v_{\text{H}} $ to quantify the effect of 
non-Hermiticity.

Finally, we express the effect of the non-Hermitian Hamiltonian term as follows:

\begin{eqnarray}
	\begin{aligned}
		H_{\text{NH}} &=  \sum_{\bf k} \sum_{\alpha=t,b} \psi_{\alpha}^{\dagger}(\mathbf{k}) h_{\text{NH}}^{\xi}(\theta_{\alpha}) \psi_{\alpha}(\mathbf{k})\ .
	\end{aligned}
	\label{Eq7}	
\end{eqnarray} 

In our calculations, we consider $\hbar v_{H}/a =2135.4$  meV \cite{McCann_2011-BilayerReview, Koshino-tBLG}, with $a = 0.246$ nm being the lattice constant of graphene. Throughout this paper, unless otherwise specified, we use $u = 79.7$ meV and $u^{'} = 97.5$ meV~\cite{Moon-tBLG,Koshino-tBLG,corrugation-DFT-uu,Dai2016-uu}. 


\section{Electronic band spectrum}\label{Sec:III}
In this section, we discuss the electronic band spectrum of both untwisted BLG and tBLG in the presence of NR hopping and hBN-induced mass term.
\begin{figure*}[t]
	\centering
	\subfigure{\includegraphics[width=1.0\textwidth,height=0.6\textwidth]{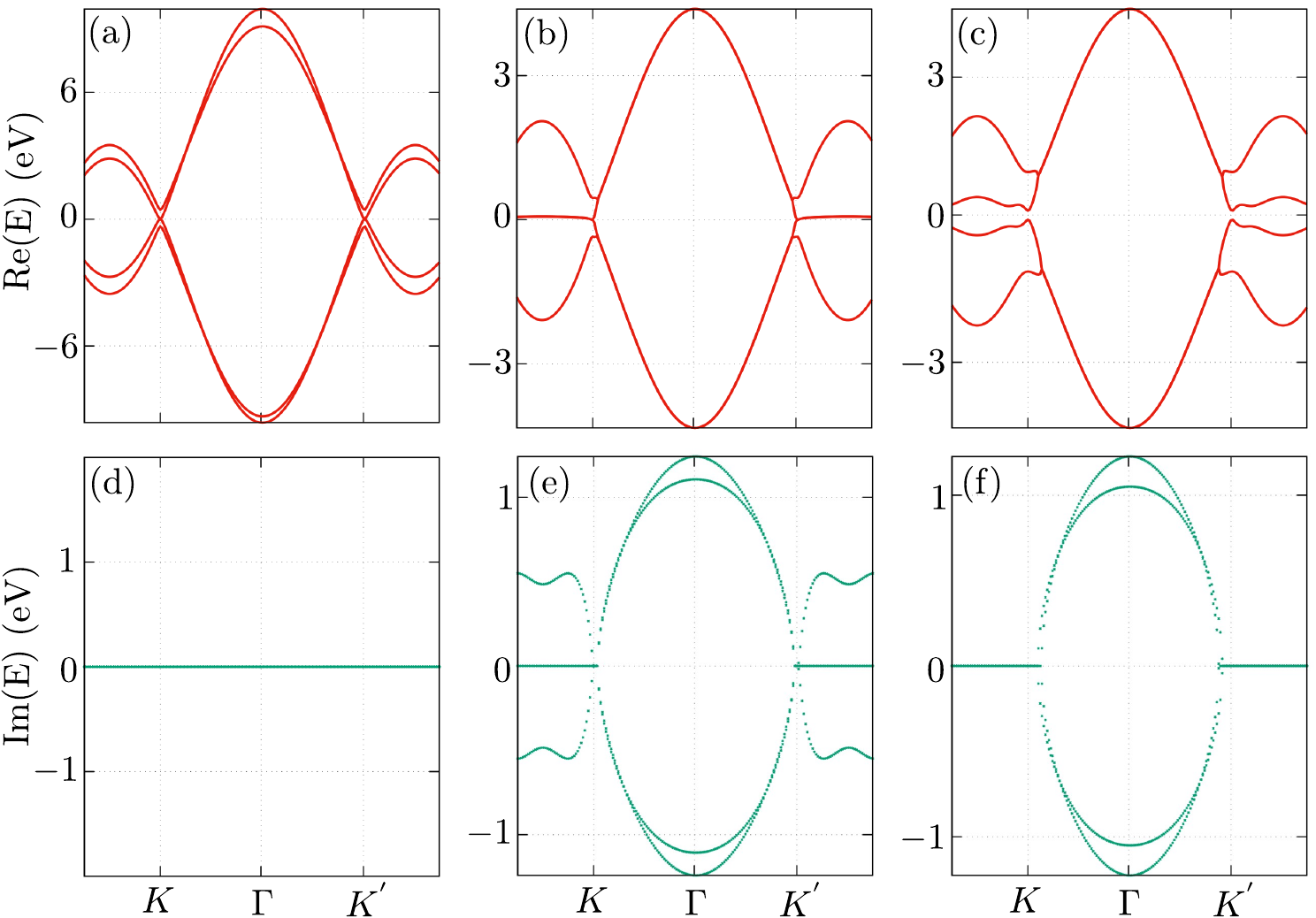}}
	\caption{Energy spectra of untwisted NH-BLG is presented, incorporating non-reciprocity in the nearest neighbour hopping. The band structures are depicted along the corner ($K$ and $K^{\prime}$) 
	and centre ($\Gamma$) of the hexagonal BZ. The Re(E) (upper panel) and Im(E) (lower panel) spectra have been shown for different values of the non-Hermiticity ($\beta$) and hBN-induced mass 
	($M_0$) terms.  In panel (a), $\beta = 0$ and $M_0 = 0$ \ie this corresponds to the Hermitian limit, in panel (b) $\beta = 0.9$ and $M_0 = 0$, two bands out of four become degenerate before 
	valley-$K$ and after valley-$K^{\prime}$ and the band width also decreases. Finally, in panel (c) $\beta = 0.9$ and $M_0 = 1$ eV and the bands become gapped.
	}
	\label{bands_dispers1}
\end{figure*}
\subsection{NH-BLG}
Before discussing the tBLG, let us first discuss the untwisted case \ie the BLG. In Fig.~\ref{bands_dispers1}, we show the band structure of the BLG to understand the impact of NR hopping present in each layer and hBN mass term induced on the top layer only. We fix the same parameter ($t$, $\Delta$, $t_1$, $t_3$, $t_4$) values, as mentioned in the previous section, for analysing the band structure. We diagonalise the Hamiltonian in Eq.~({\ref{Eq8}}) and depict the band dispersions in Fig.~\ref{bands_dispers1} along the path $K^{\prime}$ - $\Gamma$ - $K$ considered in the original hexagonal 
BZ of graphene. 

\begin{figure}[h]
	\subfigure{\includegraphics[width=0.49\textwidth]{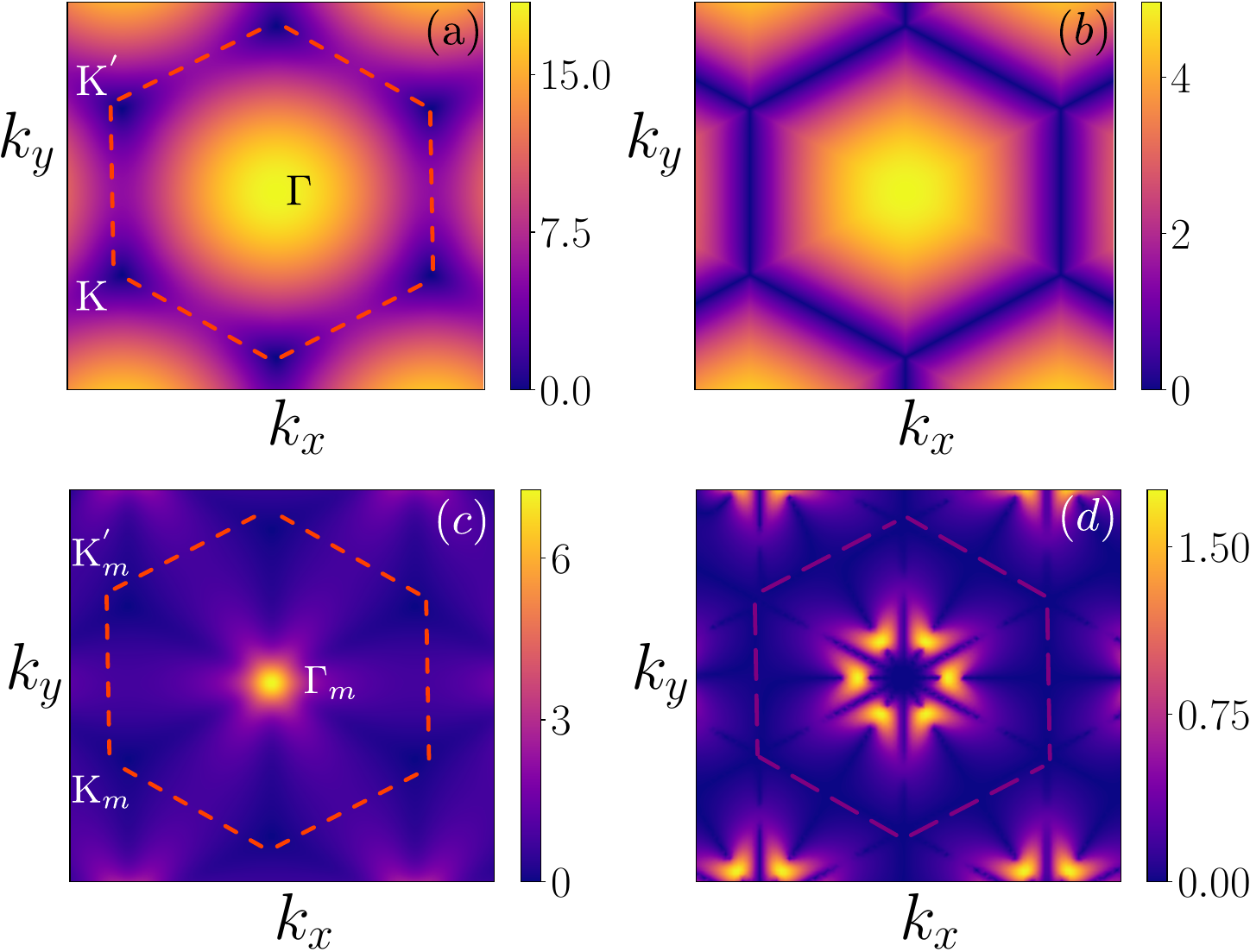}}
	\caption{Density plots for the band gap (in meV), considering the real part of the first valence and conduction bands, are depicted in the $k_x - k_y$ plane within the BZ. In the upper panel (\ie panels (a) and (b)) we show the band gap for NH-BLG choosing 
	different values of the NR hopping strength ($\beta = 0$ and $\beta = 0.9$ respectively) and observe to form a nodal line due to non-Hermiticity. In the lower panel we display the band gap for
	$\beta = 0$ and $\beta = 0.1$ respectively in panels (c) and (d) and observe similar nodal line for NH-tBLG near valley-$K$ when $\beta \neq 0$.
	}
	\label{band_gap}
\end{figure}

At first, we focus on the band structure of BLG in the absence of both the NR hopping and mass term (i.e. $ \beta = M_0 = 0$), which is presented in Fig.~\ref{bands_dispers1}(a). As the strength of the 
NR hopping $\gamma$ increases, the band degeneracy gets augmented. Between the $K$ and $K^{\prime}$ points, the four bands form a pair of doubly degenerate bands. On the other hand, before 
$K$ and after $K^{\prime}$, two out of the four bands move towards the charge neutral point and become degenerate. For example, in Fig.~\ref{bands_dispers1}(b), we showcase the bands (real part of the dispersion) for the non-Hermiticity strength $\beta = 0.9$ with no hBN induced mass (\ie $M_0 = 0$). The band degeneracy near the charge neutral point before $K$ and after $K^{\prime}$ point is reflected as a nodal ring in the hexagonal BZ when we investigate the band gap on the entire BZ. This nodal ring appears gradually as one continuously increases the non-Hermiticity strength $\beta$. Here, the band gap is defined as the difference between the first conduction and the first valence bands, considering the real part of the band dispersion. In Figs.~\ref{band_gap}(a) and (b) we illustrate the band gap on the BZ respectively for $\beta=0$ and $\beta = 0.9$. As the strength of non-Hermiticity ($\beta$) increases,  we observe that the Dirac points gradually smear and eventually form a ring, as shown in Fig.~\ref{band_gap}(b). Finally, in Fig.~\ref{bands_dispers1}(c) we turn on the mass term to open up a band gap while NR hopping is present (\ie $\beta = 0.9$). We observe that at $M_0 = 1$ eV a gap opens and before $K$ and after $K^{\prime}$ the bands become non-degenerate. In Figs.~\ref{bands_dispers1}(d)-(f), we show the corresponding imaginary parts of the band dispersions, while the real parts are displayed in Figs.~\ref{bands_dispers1}(a) - (c). 
 

\subsection{NH-tBLG}
\begin{figure*}[t]
	\centering
	\subfigure{\includegraphics[width=1.0\textwidth,height=0.5\textwidth]{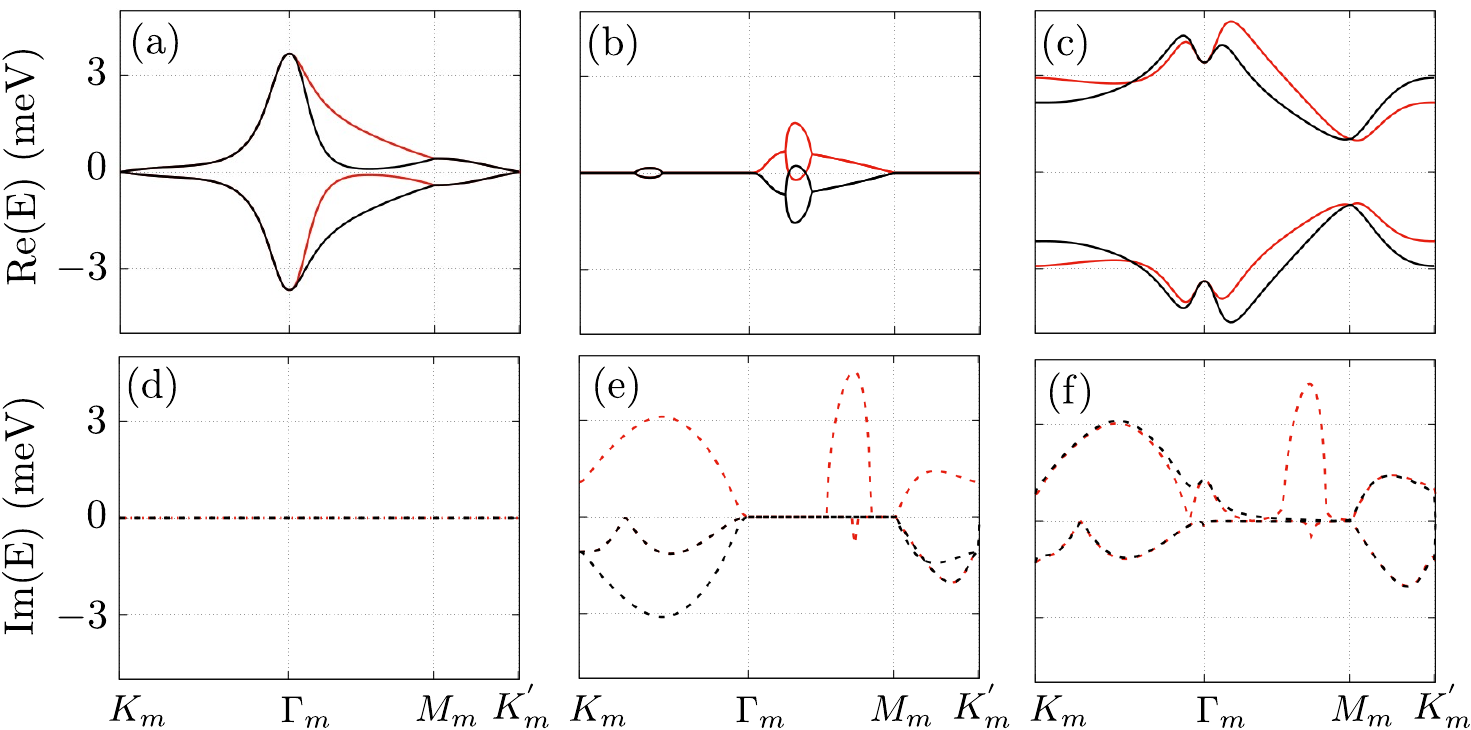}}
	\caption{The energy spectra of NH-tBLG are presented for different values of the non-reciprocal hopping strength ($ \beta $) and the hBN-induced mass ($ M_0 $) at a twist angle of $ \theta = 1.05^\circ $. The band structures are plotted along the high-symmetry path $ K_m - \Gamma_m - M_m - K'_m $ of the mBZ. The red and black lines represent the energy bands corresponding to valleys $ K $ and $ K' $ (i.e., $ \xi = \pm 1 $), respectively. The real and imaginary parts of the spectra are shown as $\text{Re}(E)$ (solid lines, upper panel) and $\text{Im}(E)$ (dashed lines, lower panel).  In panel (a), we set $ \beta = 0 $ and $M_0 = 0$, corresponding to the Hermitian model, where the characteristic tBLG flat bands appear. In panel (b), with $ \beta = 0.1 $ and $M_0 = 0$, the bands from the two valleys become degenerate along the high-symmetry lines, and the bandwidth decreases significantly. Panel (c) corresponds to $ \beta = 0.1 $ and $ M_0 = 20 ,\text{meV} $, where the flat bands are separated and a gap opens. Panels (d)--(f) show the imaginary parts of the band dispersions for the same parameter values as in panels (a)--(c), respectively.	 
	}
	\label{tBLG_band_dispers}
\end{figure*}

Since we are interested in studying the effect of non-Hermiticity on the low-energy flat bands of the tBLG, we confine ourselves to the smaller twist angles and display the band dispersion near the magic angle (\ie $\theta=1.05^{o}$). In Fig.~$\ref{tBLG_band_dispers}$, we show the electronic band dispersions of tBLG around the $K$ and $K'$ valleys, depicted in red and black lines, respectively.  Diagonalizing Eq.~(\ref{Eq1}) we plot the bands along the high symmetry line $K_m$-$\Gamma_m$-$M_m$-$K^{'}_m$ on the Moir{\'e} brillouin zone (mBZ) for different values of the NR hopping strength ($\beta$) and hBN induced mass term ($M_0$).

In Fig.~$\ref{tBLG_band_dispers}$(a), we depict the first conduction and valence bands (\ie the flat bands) of tBLG for $\beta = M_0 = 0$. These flat bands remain gapless at the Dirac points $K_m$ and $K^{'}_m$ and the gapless nature is protected by $C_2 T$-symmetry. When we introduce non-Hermiticity into the system, we observe that the bandwidth of the FBs drastically decreases. It is important to note that, except in two small regions, the flat bands become degenerate along the entire high-symmetry line for both the valleys. This can be seen in Fig.~$\ref{tBLG_band_dispers}$(b), where the real part of band spectrum for $\beta = 0.1$ and $M_0 = 0$ is presented. As a result, the corresponding density of states peaks near the charge-neutral point, resembling a Dirac delta function (see the discussion on the density of states in Appendix.~\ref{AppA} for further insight).  
With the higher values of $\beta,$ the flat bands become even flatter, and other bands move closer to the charge neutral point. Similar to BLG, we consider plotting the band gap for valley-$K$ on the mBZ for tBLG as well (see Fig.~\ref{band_gap}(c) and (d) respectively) for $\beta = 0$ and $\beta = 0.1$ without any induced mass term. A nodal-line like structure, arising from the presence of non-Hermiticity 
(\ie $\beta = 0.1$), is observed in this case as well. Finally, we turn on the hBN induced mass term to lift the band degeneracy and create a band gap. It is observed that any small finite value of the mass does not separate the bands rather it depends on the NR hopping strength ($\beta$), which is discussed more elaborately in Sec.~\ref{Sec:V}. Here, in Fig.~$\ref{tBLG_band_dispers}$(c) we show the bands to completely separated for a induced mass value of $M_{0} = 20$ meV (at $\beta = 0.1$). Figs.~\ref{tBLG_band_dispers}(d) - (f) represent the imaginary parts of the band dispersions (shown with dashed lines) corresponding to the Fig.~\ref{tBLG_band_dispers}(a) - (c), the real parts of the band dispersions.

\section{Exceptional magic angles}\label{Sec:IV}

\begin{figure}[h]
	\subfigure{\includegraphics[width=0.52\textwidth]{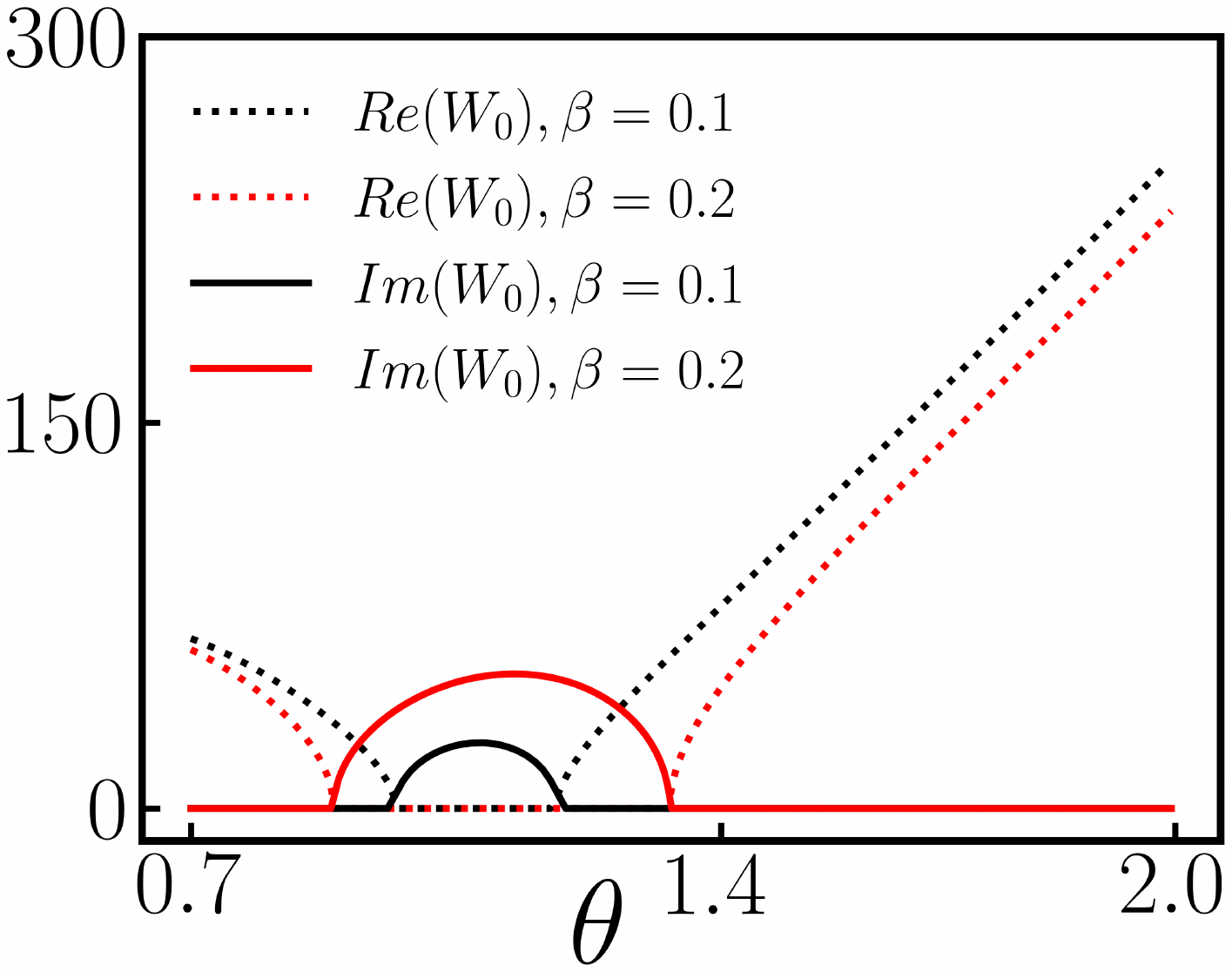}}
	\caption{The Real (represented with dotted lines) and Imaginary (denoted by solid lines) part of the bandwidth (Re($W_{0}$) and Im($W_{0}$)) are depicted with respect to the twist angle ($\theta$) of the NH-tBLG. The black and red color represent two different values of the NR-hopping strength 
		$\beta = 0.1, 0.2$.
	}
	\label{EMA}
\end{figure}
In this section, we briefly discuss the role of non-Hermiticity on magic angles in tBLG. Importantly, one can note a intriguing type of NH magic angles called \textit{exceptional magic angle} (EMA) in case of 
NH-tBLG, as explored in a previous study~\cite{NH_tBLG1}. While the earlier work utilized the eight band BM model (i.e., considering one shell of harmonics) and analytically calculated the effective Fermi velocity for the NH case, here we employ the BM model with higher harmonics and identify different magic angles based on the flat band bandwidth~\cite{Origin_of_Magic_Angle-tBLG, Floquet}. We define the FB bandwidth considering the real and imaginary parts of the band dispersion as following, 
\begin{equation}
	\begin{aligned}
	&Re(W_0) = \text{max}_{\mathbf{k}}\left[\rm{Re}(E_{1}(\mathbf{k}))\right] -
	\text{min}_{\mathbf{k}}\left[\rm{Re}(E_{-1}(\mathbf{k}))\right]\ ,\\
	&Im(W_0) = \text{max}_{\mathbf{k}}\left[\rm{Im}(E_{1}(\mathbf{k}))\right] -
	\text{min}_{\mathbf{k}}\left[\rm{Im}(E_{-1}(\mathbf{k}))\right]\ ,\
	\end{aligned}
\end{equation}
\ie the difference between maxima of first conduction band ($E_{1}(\mathbf{k})$) and minima of first valence band ($E_{-1}(\mathbf{k})$) in the mBZ, considering both the real and imaginary parts of the bands. In Fig.~\ref{EMA}, we present both the real and imaginary parts of the bandwidth, denoted as $\mathrm{Re}(W_{0})$ and $\mathrm{Im}(W_{0})$, respectively, choosing two different values of the NR hopping strength ($\beta = 0.1, 0.2$), represented by black and red lines, respectively, at the chiral limit (i.e., $u = 0$). From Fig.~\ref{EMA}, we can clearly identify the points where both the real and imaginary parts of the bandwidth simultaneously vanish. We refer to these points as the \textit{exceptional magic angles} in our analysis.

Following the expression of the effective group velocity for NH-tBLG in Ref.~\cite{NH_tBLG1}, we introduce an effective bandwidth defined as
\[
W = \sqrt{\mathrm{Re}(W_{0})^{2} - \mathrm{Im}(W_{0})^{2}}\ ,
\]

\begin{figure}[h]
	\subfigure{\includegraphics[width=0.48\textwidth]{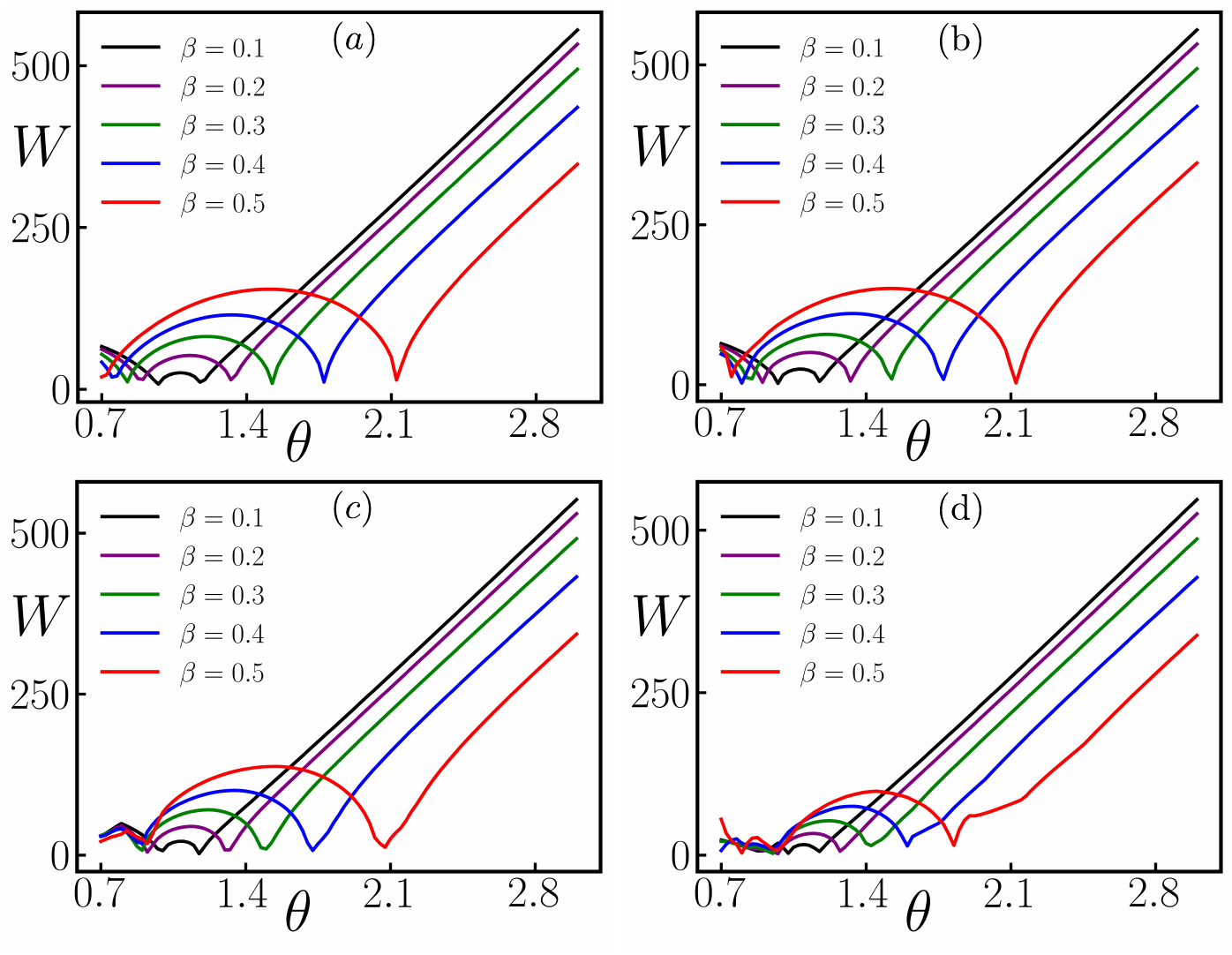}}
	\caption{Effective bandwidth of the flat bands ($W$) for the NH-tBLG (around valley-$K$) are depicted as a function of twist angle ($\theta$) considering different values of the NR hopping strength ($\beta$). Panels (a)-(d) correspond to different values of the interlayer coupling between the sublattices AA/BB, respectively, which are $u = 0, 25, 50, 79.7$ meV.
	}
	\label{bandwidth}
\end{figure}

In Fig.~\ref{bandwidth}, we depict the effective bandwidth of FBs ($W$) as a function of the twist angle ($\theta$) for different values of the NR hopping strength ($\beta$). 
In Fig.~\ref{bandwidth}(a), we first examine the chiral limit (\ie $u = 0$), where we find that for a small NR hopping strength ($\beta = 0.1$, shown as a black solid line), the effective bandwidth of the flat bands ($W$) decreases with reducing twist angle ($\theta$) and eventually hits the zero bandwidth line for two different values of twist angle (say, $\theta_{l}$ and $\theta_{r}$), indicating the two EMAs. We note that as $\beta$ increases--as indicated in Fig.~\ref{bandwidth} by the colors purple, green, blue, and red corresponding to $\beta = 0.2$, $\beta = 0.3$, $\beta = 0.4$, and $\beta = 0.5$, respectively- the width of the region between the two EMAs (say, $d\theta = \theta_{r} - \theta_{l}$) gradually increases. On the other hand, Figs.~\ref{bandwidth}(b)-(d) showcase the results for the non-chiral limit, corresponding to $u = 25, 50, 79.7$ meV, respectively. While each individual plot exhibits a similar trend with increasing $\beta$, we observe that as we jump from the chiral to the 
non-chiral limit, the the width of region between two EMAs (i.e., $d\theta$) progressively decreases. 

For completeness, we also explore the following two cases: (a) considering NR hopping only in one layer of tBLG, and (b) considering NR hopping in the interlayer coupling. In the first case, we observe similar behavior to that discussed in the previous paragraph. However, in the latter case, no EMA appears; instead, the existing magic angle merely shifts. We discuss this scenario in more detail 
in Appendix~\ref{AppB}.

\section{Topological properties}\label{Sec:V}

To characterize the topologcial band properties of the system, we focus on two key quantities: the direct band gap and a topological invariant. Below, we define these quantities and present our findings. Unlike tBLG, in untwisted BLG, opening a mass gap does not lead to a topological phase. In tBLG, the low-energy effective model lacks inter-valley scattering, resulting in valley degeneracy. When the Hamiltonian is formulated around a specific valley, time-reversal symmetry is effectively broken due to the valley-selective description. As a result, when a band gap opens, a finite valley Chern number emerge, indicating a nontrivial topological phase. In contrast, for BLG (see the full tight-binding model in Eq.~(\ref{Eq8}), inter-valley scattering is present. Therefore, even when a band gap is introduced, the system does not develop topological characteristics and instead remains a trivial band insulator. Thus, the NH-BLG with a mass term induced by hBN also remains a trivial band insulator, rather than becoming a topological insulator.

\subsection{Direct Band Gap}
Before computing any topological invariant first we investigate the direct band gap closings between the first valence band and the first conduction band of tBLG. As it is well known that presence of a direct band-gap closing transition indicates a probable topological phase transition. Here we define the direct band gap as following, 
\begin{equation}
	\delta^{\xi}_{\text{dir}} = \text{min}_{\mathbf{k}}\left[\rm{Re}(E_{1,\xi}(\mathbf{k})) - \rm{Re}(E_{-1,\xi}(\mathbf{k}))\right] \in \text{mBZ}\ .
\end{equation}
where, $\rm{Re}(E_{-1,\xi}(\mathbf{k}))$, $\rm{Re}(E_{1,\xi}(\mathbf{k}))$ correspond to the real part of first valence and first conduction band respectively at momentum $\mathbf{k}$ around valley $\xi$.

\begin{figure}[h]
\subfigure{\includegraphics[width=0.5\textwidth]{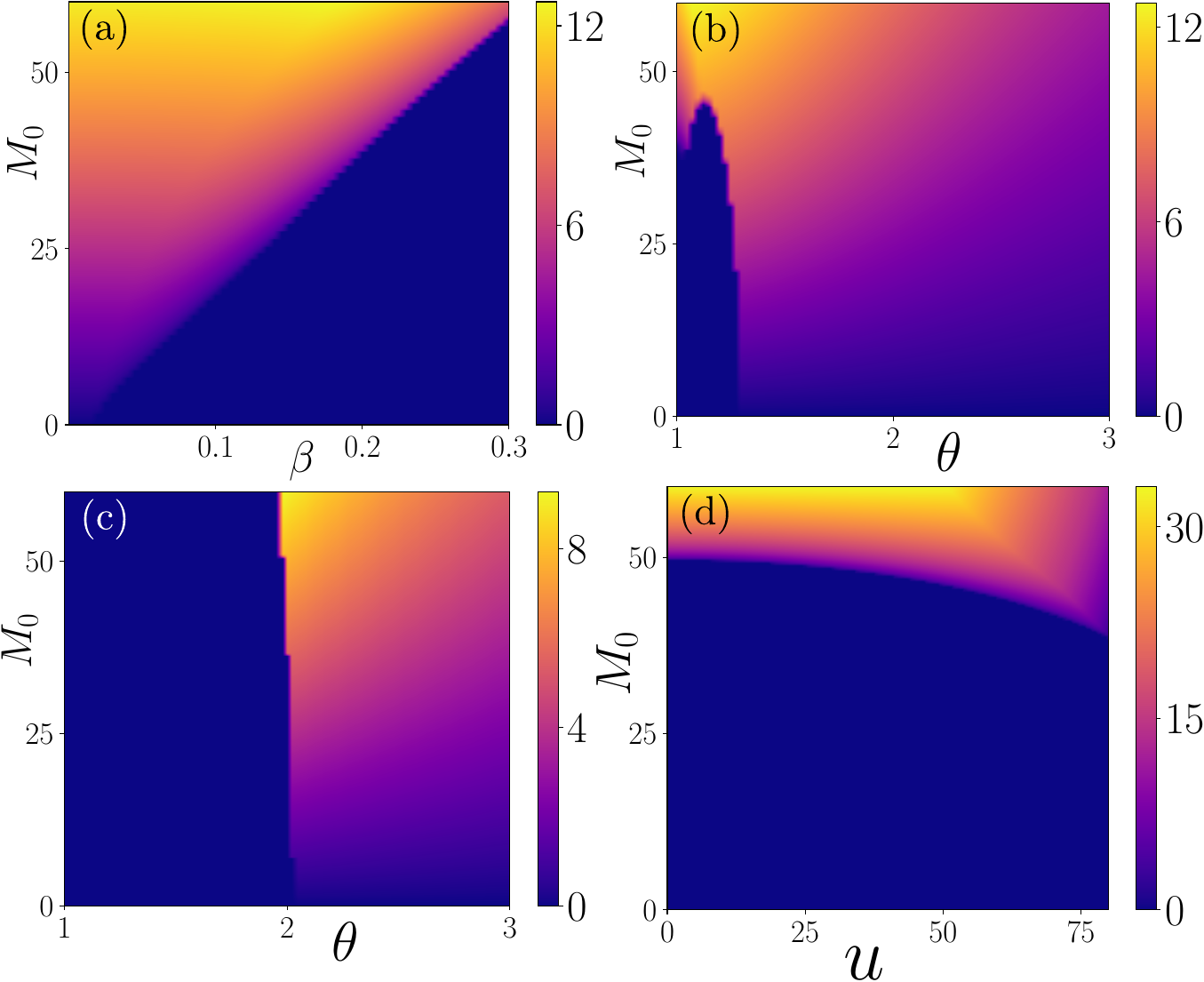}}
	\caption{Density plots of direct band gaps (considering the real part of the band dispersion) for NH-tBLG around valley-$K$ are illustrated in different parameter space. In panel (a), we depict the direct band gap in the plane of hBN induced mass ($M_0$ in meV) 
	and non-reciproal hopping strength ($\beta$) at twist angle $\theta=1.05^{o}$ (\ie at the magic angle). In panels (b) and (c), the direct band gap is shown in $M_0 - \theta$ plane respectively 
	choosing $\beta = 0.2$ and $\beta = 0.5$ respectively. Finally, in panel (d), we show the direct band gap in the $M_0 - u$ plane to emphasize the physics near chiral limit. 
	} 
	\label{dir_BG}
\end{figure}

In Fig.~\ref{dir_BG} we present the results for the direct band gap in NH-tBLG around valley-$K$ in order to understand the effects of NR hopping strength ($\beta$), twist angle ($\theta$), interlayer coupling strength ($u$) and hBN induced mass ($M_0$). At first we consider the phenomena near the magic angle (\ie $\theta = 1.05^{\circ}$) and calculate the direct band gap in the $M_0 - \beta$ plane. We observe that with the increasing NR hopping strength the required hBN induced mass needed to create a band gap increases in NH-tBLG and the nature is also linear as can be seen from Fig.~\ref{dir_BG}(a). In Figs.~\ref{dir_BG}(b) and (c), we depict the direct band gap in the $M_0 - \theta$ plane to analyze the gap profile for two different NR hopping strengths, $\beta = 0.2$ and $0.5$, respectively. It is evident from the plots that as $\beta$ increases, the gapless regions expand in that plane. In other words, there exists a range of twist angles for which obtaining a band gap becomes gradually difficult, even with large values of the hBN-induced mass term. Moreover, the width of this angular range grows with increasing NR hopping strength. However, note that, for small values of $M_{0}$ and $\beta$, $\theta$ 
can drive a topological phase transition in the band (see Fig.~\ref{dir_BG}(b)). At this point, we also examine the band structure in presence of non-Hermiticity and find that while the induced mass act almost linearly at Dirac points, it doesn't act in the similar way in other high symmetry points, leading to such gapless regions. On the other hand, there are regions in the direct band gap profile where the effect of non-Hermiticity is suppressed, and even a small hBN-induced mass is sufficient to open a gap. In fact, any twist angle beyond the dome-shaped gapless regions (shown in dark blue in Fig.~\ref{dir_BG}(b) and (c)) exhibits this characteristic behavior. As an example, in Fig.~\ref{Phase_diag3}(a), we show the direct band gap on the $M_{0} - \beta$ plane for $\theta = 2.0^{\circ}$, and interestingly, we observe that increasing $\beta$ no longer affects the band gap. Finally, we attempt to understand the effect of the interlayer coupling strength between the AA/BB sub-lattices 
(\ie $u$). We plot the direct band gap in the $M_0 - u$ plane for $\beta = 0.2$ (shown in Fig.~{\ref{dir_BG}}(d)). Interestingly, one notes that the height of such gapless region decreases as we translate from the chiral ($u=0$) to non-chiral ($u\neq 0$) limit, \ie the non-chiral limit is more accessible to induce a mass gap as compared to the chiral limit.  

\subsection{Chern Number}
To appropriately characterize the topological phases, we calculate a topological invariant, in this context the NH generalization of Chern number~\cite{NH_chern1,NH_chern2}. The NH version of the 
Chern number can be defined over the mBZ for an energy band. For a separable band of energy $E_n$, the Chern number is defined as,

\begin{eqnarray}
	C_{n,\xi}^{\mu \nu} = \frac{1}{2\pi} \int_{mBZ} d^{2}\mathbf{k}\hspace{2pt} \epsilon_{ij}   i \langle \partial_{i}\psi^{\mu}_{n,\xi}(\mathbf{k}) |  \partial_{j}\psi^{\nu}_{n,\xi}(\mathbf{k}) \rangle\ \ ,
	\label{Eq:Fukui-chern}
\end{eqnarray}
where, $\epsilon_{ij} = - \epsilon_{ji}$, $\xi$ is the valley index and $|\psi^{\nu}_{n,\xi}(\mathbf{k}) \rangle$'s are the eigenvectors with the normalization condition, $\langle \psi^{\mu}_{n,\xi}(\mathbf{k}) | \psi^{\nu}_{n,\xi}(\mathbf{k})\rangle = 1$. Here, $\mu,\nu$ denote the left or right eigenvectors. Additionally, we denote the Chern numbers as $C_{K}$ and $C_{K'}$ for two valleys \ie~$\xi = \pm 1$, respectively. To compute the Chern number, we employ the numerical method proposed by Fukui et al.~\cite{Fukui-chern_no} and evaluate it 
for until the filled valence band.

\begin{figure}[t]
	\subfigure{\includegraphics[width=0.48\textwidth]{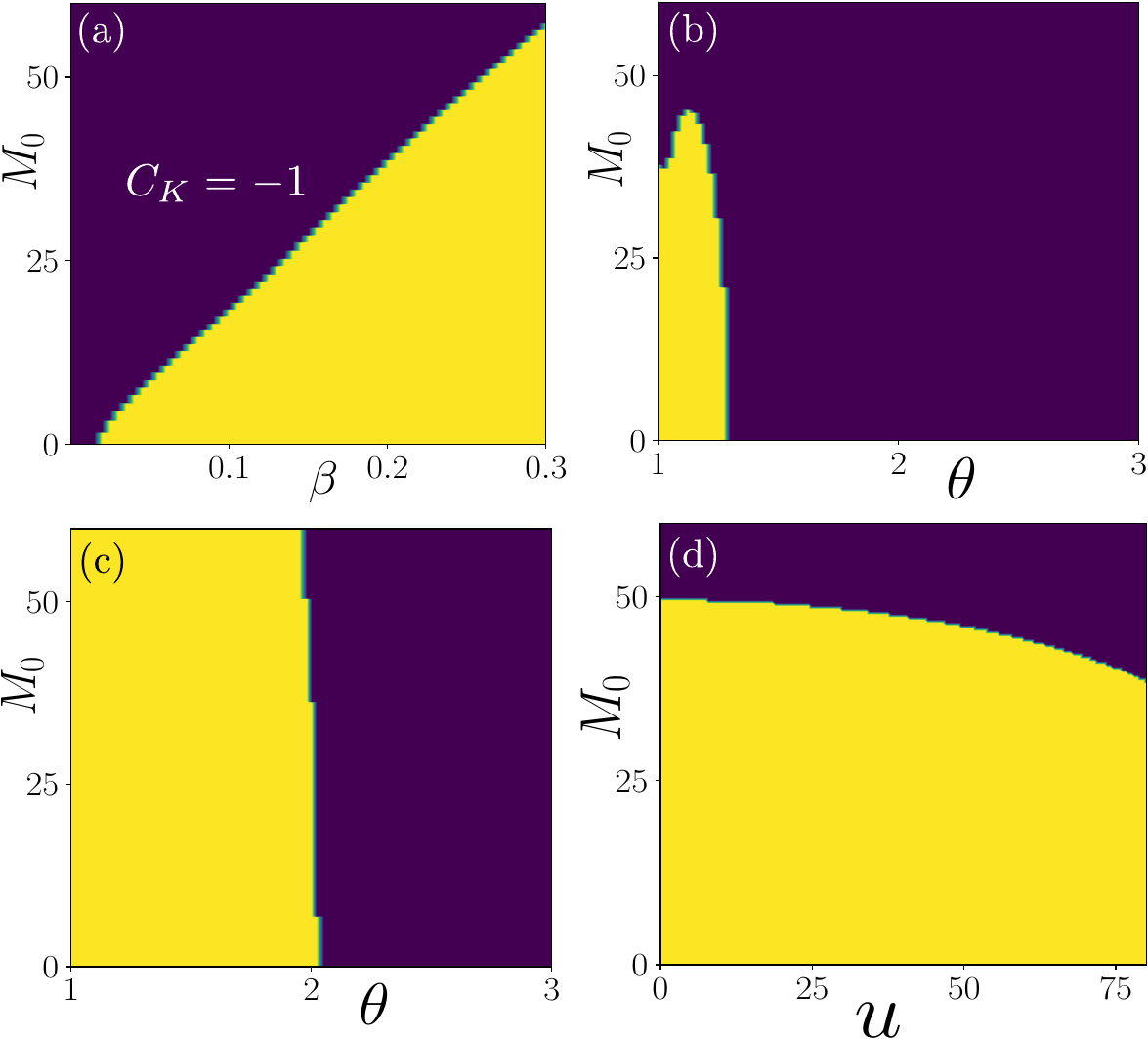}}
	\caption{NH Chern number is demonstrated for NH-tBLG choosing different parameter spaces. In panel (a), we depict the Chern number in the plane of hBN induced mass ($M_0$ in meV) 
	and non-reciproal hopping strength ($\beta$) at twist angle $\theta=1.05^{o}$ (\ie the magic angle). In panels (b) and (c), the Chern number is shown in $M_0 - \theta$ plane for $\beta = 0.2$ 
	and $\beta = 0.5$ respectively. Finally in panel (d), we display the Chern number in the $M_0 - u$ plane to illuminate the physics near chiral limit. Here, the purple color represents a topological region with Chern number $C_{K} = -1$, while the yellow color denotes the gapless region where
	Chern number is not valid.}
	\label{chern_number}
\end{figure}

\begin{figure}[b]
\subfigure{\includegraphics[width=0.49\textwidth]{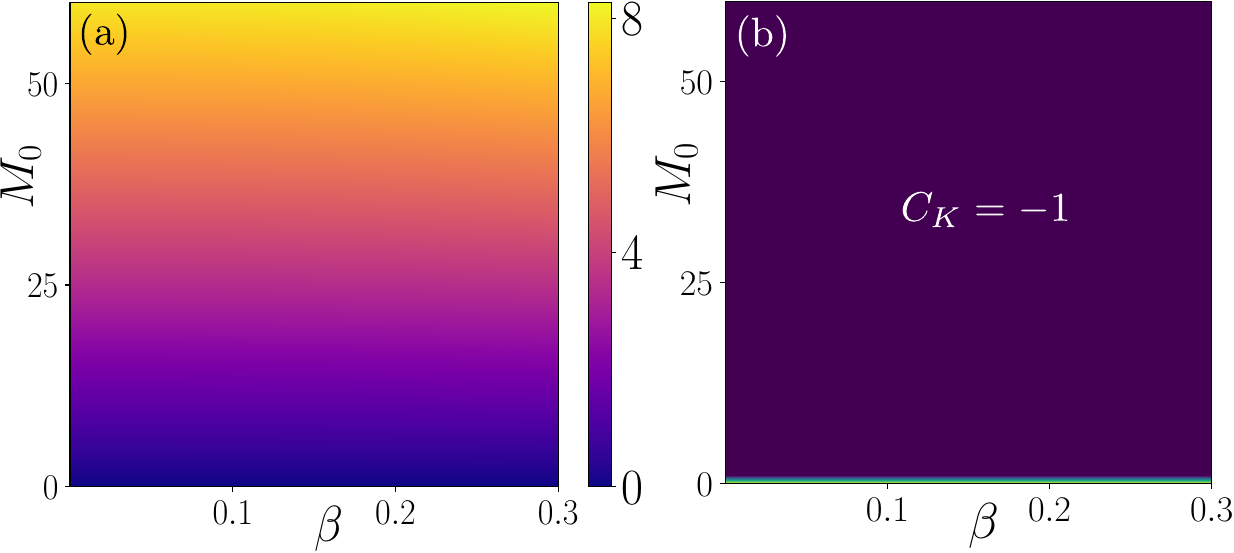}}
	\caption{Density plots for the direct band gap (considering the real part of the band dispersion, see panel (a)) and  Chern number (panel (b)) are depicted in the plane of hBN induced mass ($M_0$ in meV) and NR hopping strength ($\beta$) 
	choosing twist angle $\theta = 2.0^{o}$ and around valley-$K$}.
	
	\label{Phase_diag3}
\end{figure}

Similar to the analysis of the direct band gap, we compute the Chern number (around valley-$K$) in different planes to establish the topological nature of the phases. In Fig.~\ref{chern_number}(a), we present the Chern numbers in the $M_0-\beta$ plane for the twist angle $\theta = 1.05^{\circ}$. In this case, we observe a topological phase characterized by a Chern number $C_{K} = -1$ (indicated by the purple region). The yellow part in Fig.~\ref{chern_number} corresponds to the gapless region, and therefore the Chern number is not well-defined in this area. Interestingly, at the magic angle, the accessible NH topological phase in tBLG diminishes with increasing non-reciprocity in the nearest-neighbor hopping, as a stronger hBN-induced mass is required to sustain it. The results, shown in Fig.~\ref{chern_number}(b) and (c), are consistent with the direct band gap analysis. In these density plots, we compute the Chern numbers on the $M_0-\theta$ plane for $\beta = 0.2$ (see Fig.~\ref{chern_number}(b)) and $\beta = 0.5$ (see Fig.~\ref{chern_number}(c)).
Enhancement in $\beta$ adversely affects the system’s topology by suppressing the topological phases at the magic angle. Finally, in Fig.~\ref{chern_number}(d), we present the Chern number on the $M_0-u$ plane for $\beta = 0.2$. Achieving a topological phase in the chiral limit ($u=0$ line) is more challenging than in the non-chiral limit ($u\neq0$). However, importantly, there exists other twist angles away from the magic angle spectra where an infinitesimally small value of hBN induced mass turns the system into topological. Such a scenario is presented in 
Fig.~\ref{Phase_diag3} for twist angle $\theta = 2^{o}$. The corresponding band gap and Chern number are displayed in Fig.~\ref{Phase_diag3}(a) and Fig.~\ref{Phase_diag3}(b) respectively 
choosing $M_0 - \beta$ plane. Note that, the direct band remains finite for infinitesimal $\beta$ and $M_{0}$ value. The following NH phase remains topological with $C_{K} = -1$ over the same regime. 
In our analysis, the low-energy model Hamiltonian of tBLG, from which the Chern number is computed, is formulated around the valley-$K$. Consequently, for the other valley 
(\ie valley-$K^{\prime}$), which is the time-reversal partner of $K$, the Chern number is opposite (\ie $C_{K^{'}} = +1$). Therefore, we define the valley Chern number~\cite{Ezawa_2012-Diff_chern} as $C_{\text{valley}} = C_{K} - C_{K^{\prime}}$, which takes finite values (e.g., $C_{\text{valley}} = -2$). Hence, the topological phase that we identify is a non-Hermitian valley-Hall insulating phase (NH-VHI).

\section{Summary and Conclusions}\label{Sec:VI}

To summarize, in this article, we propose a model Hamiltonian to realize NH topology in tBLG. We introduce non-Hermiticity via the NR nearest-neighbor hopping within each layer of tBLG. On top of that, the top layer is assumed to be aligned with hBN, which breaks the $C_2$ symmetry and induces a mass term. We begin by discussing the electronic band structure and analyzing the effects of NR hopping strength and the hBN-induced mass on the physical properties of both untwisted BLG and tBLG. We compute the band gap over the mBZ, which exhibits a nodal ring-like structure in the presence of non-Hermiticity for both BLG and tBLG. In the latter part of our study, we compute the 
flat band bandwidth as a function of the twist angle and identify a pair of EMAs for each value of the NR hopping strength ($\beta$). We observe that the EMA plateau widens with increasing $\beta$, while it narrows with higher interlayer coupling strength ($u$). Our findings regarding the EMAs qualitatively match with those reported by Esparza \textit{et al.}~\cite{NH_tBLG1}. However, in 
Ref.~\cite{NH_tBLG1}, the authors employ an eight-band model for tBLG and identify EMAs through the renormalized Fermi velocity. In contrast, our results not only corroborate these earlier observations but also provide more realistic values for the EMAs by utilizing the BM model with higher harmonics rather than the reduced eight-band approximation. Finally, we investigate the topological properties of the NH-tBLG by calculating the direct band gap across various parameter regimes. Our results are illustrated through a series of density plots in the $M_0-\beta$, $M_0-\theta$, and $M_0-u$ planes. Additionally, we compute the NH Chern number over these parameter spaces. We observe good agreement between the Chern numbers and the corresponding gapped regions of the direct band gap, indicating non-trivial topology (NH-VHI phase). However, the influence of non-Hermiticity 
on the topological characteristics is found to be predominantly detrimental around the magic angle.

Finally, we discuss the possible experimental feasibility of our proposed model. Inspired by proposals to emulate tBLG in analogous phononic systems~\cite{twisted_pillared_plates, twist_accoustic}, photonic analogues of bilayer devices have also been introduced~\cite{tBL_photonic1, tBL_photonic2}. In a recent experimental study, an atomic Bose-Einstein condensate was realized using a twisted bilayer optical lattice~\cite{BEC_tBLG}. NR hopping has been experimentally realized in both metamaterials and cold-atomic systems~\cite{RMP_topo_NH, NH_AIP, NH_topo_EP_geometries}. Notably, the exceptional ring structure characteristic of Weyl systems has been demonstrated in photonic experiments~\cite{Weyl_NH1, Weyl_NH2}. Very recenetly, a NH topological phase is observed by exploiting the non-reciprocity of quantum Hall edge states, rather than relying on gain and loss mechanisms in the system~\cite{nh_topo_cmp}. This suggests that, by either designing an optical lattice for tBLG or employing the concept of a synthetic dimension in tBLG~\cite{tBLG_optical_lattices, synthetic_tBLG}, and inducing NR hopping in such systems, it may be possible to simulate the model we have proposed and experimentally verify our results. Based on this discussion, we remain optimistic that our proposed predictions may be possible
to realize in future experiments.

Note that, in a very recent study~\cite{NH_tBLG2}, we come accross topological aspects of NH tBLG. The author introduces non-Hermiticity through gain and loss in the chemical potential.
Also, the reported topological phase is different from ours and is approached via a different route as well.


\subsection*{Acknowledgments}
K.B. and A.S. acknowledge SAMKHYA: High-Performance Computing Facility provided by Institute of Physics, Bhubaneswar, for numerical computations. D.C. acknowledges financial support from DST (project number DST/WISE-PDF/PM-40/2023).


\subsection*{Data Availability Statement}
The datasets generated and analyzed during the current study are available from the corresponding author upon reasonable request.


\appendix
\section{Density of States with non-Hermiticity} \label{AppA}
\begin{figure*}[t]
	\subfigure{\includegraphics[width=0.9\textwidth]{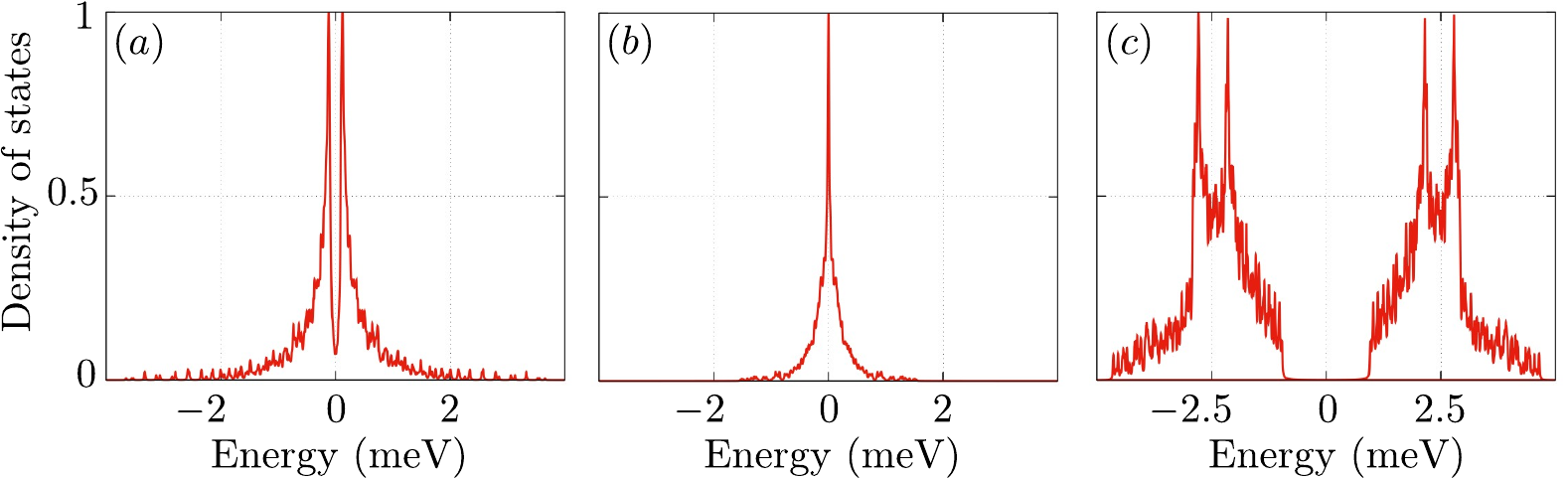}}
	\caption{The total DOS of tBLG is displayed as a function of energy choosing different values of the non-Hermiticity strength ($\beta$) and the hBN-induced mass term ($M_{0}$). 
	Specifically, panel (a) corresponds to $\beta = M_{0} = 0$, whereas panels (b) and (c) refer to $\beta = 0.1, M_{0} = 0$, and $\beta = 0.1, M_{0} = 10$ meV, respectively.}
	\label{DOS}
\end{figure*}

Here, we discuss the features of total density of states (DOS) of tBLG in absence as well as presence of non-Hermiticity $\beta$ and the hBN-induced mass term $M_{0}$. In Figs.~\ref{DOS}(a)-(c), 
we present the DOS for three different parameter sets: (a) $\beta = M_{0} = 0$, (b) $\beta = 0.1, M_{0} = 0$, and (c) $\beta = 0.1, M_{0} = 10$ meV, at a fixed twist angle $\theta = 1.05^{o}$ \ie magic angle. The corresponding band dispersions have already been depicted in Figs.~\ref{tBLG_band_dispers}(a)-(c) of the main text. In absence of both non-Hermiticity and the hBN-induced mass term 
(see Fig.~\ref{DOS}(a)), the two flat bands exhibit distinct van Hove singularities (VHS), one for each band. When non-Hermiticity $\beta$ is introduced, these two VHS merge into a single prominent 
peak at zero energy (as shown in Fig.\ref{DOS}(b)), reflecting the emergence of an extremely flat band structure. Upon further inclusion of the hBN-induced mass term $M_{0}$, the DOS peaks split 
again, indicating a lifting of the degeneracy introduced by the mass term (see Fig.~\ref{DOS}(c)).

In tBLG, an intriguing type of VHS, known as high-order VHS, has been reported~\cite{HO_VHS}. Unlike conventional VHS, where the DOS exhibits a logarithmic divergence, high-order VHS lead to a power-law divergence in the DOS. The presence of such singularities is indicative of enhanced electron correlation effects in the system. While it would be interesting to perform a similar analysis 
for the non Hermiticity induced DOS peak observed in Fig.~\ref{DOS}(b), such an investigation lies beyond the scope of the present study and will be presented elsewhere.

\begin{figure}[h]
	\subfigure{\includegraphics[width=0.49\textwidth]{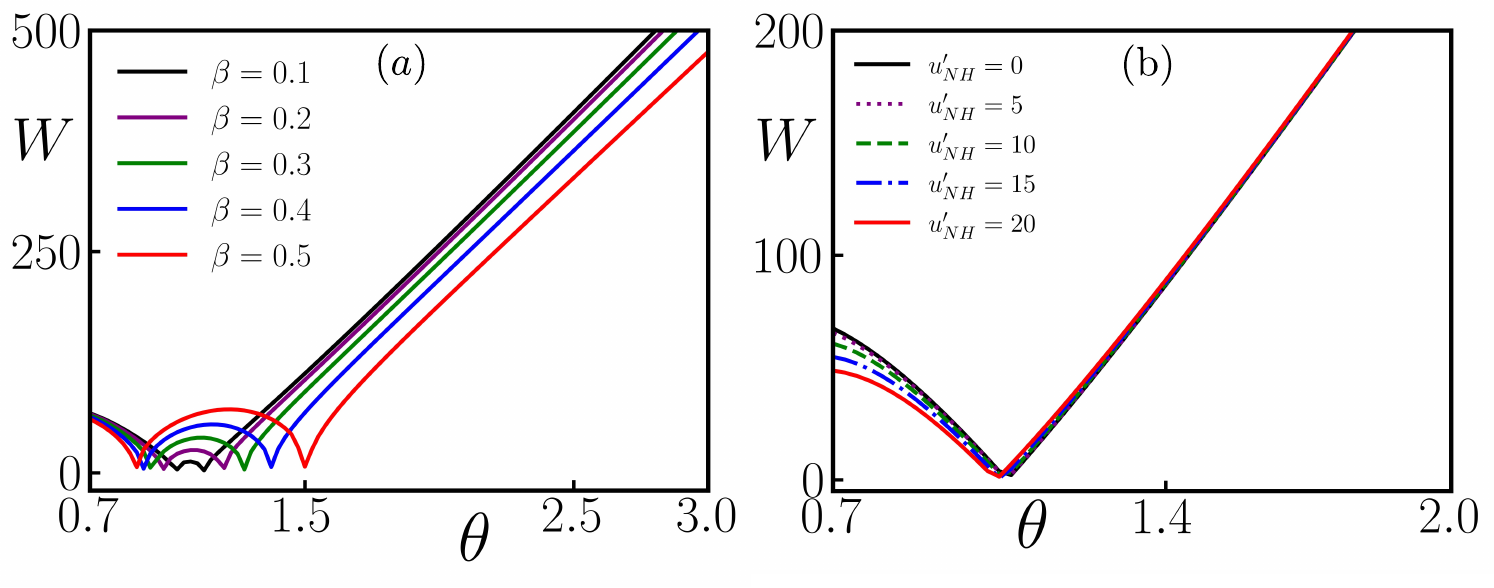}}
	\caption{The real part of the bandwidth of the flat bands ($W$) in NH-tBLG (around valley-$K$) is displayed as a function of the twist angle ($\theta$). In panel (a), the same is presented choosing different values of the NR hopping strength ($\beta$)                                                                                                                                                                                                                                                                                                                                                                                                                        incorporated in one layer of the tBLG. In panel (b), this behavior is predicted considering various values of the NR interlayer coupling, $u_{\text{NH}}' = 0, 5, 10, 15, 20$~meV.}
	\label{BW_App}
\end{figure}

\section{non-Hermiticity introduced in inter layer coupling and one of two layers of tBLG} \label{AppB}
As outlined in the main text, the EMAs (defined by the start and end points (\ie $\theta_{l}$ and $\theta_{r}$) of the bandwidth minimum and shown in Fig.~\ref{bandwidth}(a)-(b) \ie W vs $\theta$ plot)
vary with different values of the NR hopping strength $\beta$. Here, we extend our analysis to two additional scenarios. In the first scenario, we consider NR hopping in only one layer of the tBLG, 
while the other layer has $\beta = 0$. The results in this case remain qualitatively similar to those presented in the main text, where both layers have finite $\beta$. Specifically, the region between the two twist angle increases with increasing non-reciprocal hopping strength, as shown in Fig.~\ref{BW_App}(a) at the chiral limit. However, upon closer inspection, we find that the length 
of the zero bandwidth plateau for a given $\beta$ is smaller than that of observed when both layers have NR hopping. This can be atrributed to the less degree of non-Hermiticity in the present scenario. Consequently, the EMAs also exhibit different values.


In the second scenario, we consider the case where $\beta = 0$ in both layers of tBLG, but a non-reciprocal tunneling is present in the interlayer coupling of the tBLG. The forward and backward 
tunneling amplitudes are considered as $(u' + u'_{\text{NH}})$ and $(u' - u'_{\text{NH}})$, respectively. Unlike the previous cases, increasing the non-Hermiticity does not lead to the formation of a zero bandwidth plateau, even in the chiral limit. Instead, the bandwidth minimum occurs at a single point, as shown in Fig.~\ref{BW_App}(b). Consequently, EMAs are not expected to appear in this case. 
However, the value of the Hermitian magic angle shifts slightly as the non-Hermiticity in the interlayer coupling increases, as illustrated in Fig.~\ref{BW_App}(b). The black, purple, green, blue, and red curves correspond to $u'_{\text{NH}} = 0, 5, 10, 15, 20$, respectively,  and it is evident that the black and red curves reach zero bandwidth at different points.

\bibliographystyle{apsrev4-2mod}
\bibliography{bibfile}

\begin{thebibliography}{85}%
\makeatletter
\providecommand \@ifxundefined [1]{%
 \@ifx{#1\undefined}
}%
\providecommand \@ifnum [1]{%
 \ifnum #1\expandafter \@firstoftwo
 \else \expandafter \@secondoftwo
 \fi
}%
\providecommand \@ifx [1]{%
 \ifx #1\expandafter \@firstoftwo
 \else \expandafter \@secondoftwo
 \fi
}%
\providecommand \natexlab [1]{#1}%
\providecommand \enquote  [1]{``#1''}%
\providecommand \bibnamefont  [1]{#1}%
\providecommand \bibfnamefont [1]{#1}%
\providecommand \citenamefont [1]{#1}%
\providecommand \href@noop [0]{\@secondoftwo}%
\providecommand \href [0]{\begingroup \@sanitize@url \@href}%
\providecommand \@href[1]{\@@startlink{#1}\@@href}%
\providecommand \@@href[1]{\endgroup#1\@@endlink}%
\providecommand \@sanitize@url [0]{\catcode `\\12\catcode `\$12\catcode
  `\&12\catcode `\#12\catcode `\^12\catcode `\_12\catcode `\%12\relax}%
\providecommand \@@startlink[1]{}%
\providecommand \@@endlink[0]{}%
\providecommand \url  [0]{\begingroup\@sanitize@url \@url }%
\providecommand \@url [1]{\endgroup\@href {#1}{\urlprefix }}%
\providecommand \urlprefix  [0]{URL }%
\providecommand \Eprint [0]{\href }%
\providecommand \doibase [0]{http://dx.doi.org/}%
\providecommand \selectlanguage [0]{\@gobble}%
\providecommand \bibinfo  [0]{\@secondoftwo}%
\providecommand \bibfield  [0]{\@secondoftwo}%
\providecommand \translation [1]{[#1]}%
\providecommand \BibitemOpen [0]{}%
\providecommand \bibitemStop [0]{}%
\providecommand \bibitemNoStop [0]{.\EOS\space}%
\providecommand \EOS [0]{\spacefactor3000\relax}%
\providecommand \BibitemShut  [1]{\csname bibitem#1\endcsname}%
\let\auto@bib@innerbib\@empty
\bibitem [{\citenamefont {Bender}(2007)}]{Bender_review}%
  \BibitemOpen
  \bibfield  {author} {\bibinfo {author} {\bibfnamefont {C.~M.}\ \bibnamefont
  {Bender}},\ }\bibfield  {title} {\emph {\enquote {\bibinfo {title} {Making
  sense of non-Hermitian Hamiltonians},}\ }}\href {\doibase
  10.1088/0034-4885/70/6/R03} {\bibfield  {journal} {\bibinfo  {journal}
  {Reports on Progress in Physics}\ }\textbf {\bibinfo {volume} {70}},\
  \bibinfo {pages} {947} (\bibinfo {year} {2007})}\BibitemShut {NoStop}%
\bibitem [{\citenamefont {Feng}\ \emph {et~al.}(2017)\citenamefont {Feng},
  \citenamefont {El-Ganainy},\ and\ \citenamefont {Ge}}]{NH-photonics}%
  \BibitemOpen
  \bibfield  {author} {\bibinfo {author} {\bibfnamefont {L.}~\bibnamefont
  {Feng}}, \bibinfo {author} {\bibfnamefont {R.}~\bibnamefont {El-Ganainy}}, \
  and\ \bibinfo {author} {\bibfnamefont {L.}~\bibnamefont {Ge}},\ }\bibfield
  {title} {\emph {\enquote {\bibinfo {title} {Non-Hermitian photonics based on
  parity--time symmetry},}\ }}\href {\doibase 10.1038/s41566-017-0031-1}
  {\bibfield  {journal} {\bibinfo  {journal} {Nature Photonics}\ }\textbf
  {\bibinfo {volume} {11}},\ \bibinfo {pages} {752} (\bibinfo {year}
  {2017})}\BibitemShut {NoStop}%
\bibitem [{\citenamefont {Okuma}\ and\ \citenamefont
  {Sato}(2023)}]{NH_topo_review}%
  \BibitemOpen
  \bibfield  {author} {\bibinfo {author} {\bibfnamefont {N.}~\bibnamefont
  {Okuma}}\ and\ \bibinfo {author} {\bibfnamefont {M.}~\bibnamefont {Sato}},\
  }\bibfield  {title} {\emph {\enquote {\bibinfo {title} {Non-Hermitian
  Topological Phenomena: A Review},}\ }}\href {\doibase
  https://doi.org/10.1146/annurev-conmatphys-040521-033133} {\bibfield
  {journal} {\bibinfo  {journal} {Annual Review of Condensed Matter Physics}\
  }\textbf {\bibinfo {volume} {14}},\ \bibinfo {pages} {83} (\bibinfo {year}
  {2023})}\BibitemShut {NoStop}%
\bibitem [{\citenamefont {Kawabata}\ \emph
  {et~al.}(2019{\natexlab{a}})\citenamefont {Kawabata}, \citenamefont
  {Shiozaki}, \citenamefont {Ueda},\ and\ \citenamefont {Sato}}]{NH_topo1}%
  \BibitemOpen
  \bibfield  {author} {\bibinfo {author} {\bibfnamefont {K.}~\bibnamefont
  {Kawabata}}, \bibinfo {author} {\bibfnamefont {K.}~\bibnamefont {Shiozaki}},
  \bibinfo {author} {\bibfnamefont {M.}~\bibnamefont {Ueda}}, \ and\ \bibinfo
  {author} {\bibfnamefont {M.}~\bibnamefont {Sato}},\ }\bibfield  {title}
  {\emph {\enquote {\bibinfo {title} {Symmetry and Topology in Non-Hermitian
  Physics},}\ }}\href {\doibase 10.1103/PhysRevX.9.041015} {\bibfield
  {journal} {\bibinfo  {journal} {Phys. Rev. X}\ }\textbf {\bibinfo {volume}
  {9}},\ \bibinfo {pages} {041015} (\bibinfo {year}
  {2019}{\natexlab{a}})}\BibitemShut {NoStop}%
\bibitem [{\citenamefont {Malzard}\ \emph {et~al.}(2015)\citenamefont
  {Malzard}, \citenamefont {Poli},\ and\ \citenamefont {Schomerus}}]{NH_topo2}%
  \BibitemOpen
  \bibfield  {author} {\bibinfo {author} {\bibfnamefont {S.}~\bibnamefont
  {Malzard}}, \bibinfo {author} {\bibfnamefont {C.}~\bibnamefont {Poli}}, \
  and\ \bibinfo {author} {\bibfnamefont {H.}~\bibnamefont {Schomerus}},\
  }\bibfield  {title} {\emph {\enquote {\bibinfo {title} {Topologically
  Protected Defect States in Open Photonic Systems with Non-Hermitian
  Charge-Conjugation and Parity-Time Symmetry},}\ }}\href {\doibase
  10.1103/PhysRevLett.115.200402} {\bibfield  {journal} {\bibinfo  {journal}
  {Phys. Rev. Lett.}\ }\textbf {\bibinfo {volume} {115}},\ \bibinfo {pages}
  {200402} (\bibinfo {year} {2015})}\BibitemShut {NoStop}%
\bibitem [{\citenamefont {Shen}\ \emph {et~al.}(2018)\citenamefont {Shen},
  \citenamefont {Zhen},\ and\ \citenamefont {Fu}}]{NH_chern1}%
  \BibitemOpen
  \bibfield  {author} {\bibinfo {author} {\bibfnamefont {H.}~\bibnamefont
  {Shen}}, \bibinfo {author} {\bibfnamefont {B.}~\bibnamefont {Zhen}}, \ and\
  \bibinfo {author} {\bibfnamefont {L.}~\bibnamefont {Fu}},\ }\bibfield
  {title} {\emph {\enquote {\bibinfo {title} {Topological Band Theory for
  Non-Hermitian Hamiltonians},}\ }}\href {\doibase
  10.1103/PhysRevLett.120.146402} {\bibfield  {journal} {\bibinfo  {journal}
  {Phys. Rev. Lett.}\ }\textbf {\bibinfo {volume} {120}},\ \bibinfo {pages}
  {146402} (\bibinfo {year} {2018})}\BibitemShut {NoStop}%
\bibitem [{\citenamefont {Yao}\ \emph {et~al.}(2018)\citenamefont {Yao},
  \citenamefont {Song},\ and\ \citenamefont {Wang}}]{NH_chern2}%
  \BibitemOpen
  \bibfield  {author} {\bibinfo {author} {\bibfnamefont {S.}~\bibnamefont
  {Yao}}, \bibinfo {author} {\bibfnamefont {F.}~\bibnamefont {Song}}, \ and\
  \bibinfo {author} {\bibfnamefont {Z.}~\bibnamefont {Wang}},\ }\bibfield
  {title} {\emph {\enquote {\bibinfo {title} {Non-Hermitian Chern Bands},}\
  }}\href {\doibase 10.1103/PhysRevLett.121.136802} {\bibfield  {journal}
  {\bibinfo  {journal} {Phys. Rev. Lett.}\ }\textbf {\bibinfo {volume} {121}},\
  \bibinfo {pages} {136802} (\bibinfo {year} {2018})}\BibitemShut {NoStop}%
\bibitem [{\citenamefont {Budich}\ and\ \citenamefont
  {Bergholtz}(2020)}]{topological_sensor}%
  \BibitemOpen
  \bibfield  {author} {\bibinfo {author} {\bibfnamefont {J.~C.}\ \bibnamefont
  {Budich}}\ and\ \bibinfo {author} {\bibfnamefont {E.~J.}\ \bibnamefont
  {Bergholtz}},\ }\bibfield  {title} {\emph {\enquote {\bibinfo {title}
  {Non-Hermitian Topological Sensors},}\ }}\href {\doibase
  10.1103/PhysRevLett.125.180403} {\bibfield  {journal} {\bibinfo  {journal}
  {Phys. Rev. Lett.}\ }\textbf {\bibinfo {volume} {125}},\ \bibinfo {pages}
  {180403} (\bibinfo {year} {2020})}\BibitemShut {NoStop}%
\bibitem [{\citenamefont {Wang}\ \emph {et~al.}(2022)\citenamefont {Wang},
  \citenamefont {Zhu}, \citenamefont {Wang}, \citenamefont {Zhang},\ and\
  \citenamefont {Chong}}]{quantum_signal}%
  \BibitemOpen
  \bibfield  {author} {\bibinfo {author} {\bibfnamefont {Q.}~\bibnamefont
  {Wang}}, \bibinfo {author} {\bibfnamefont {C.}~\bibnamefont {Zhu}}, \bibinfo
  {author} {\bibfnamefont {Y.}~\bibnamefont {Wang}}, \bibinfo {author}
  {\bibfnamefont {B.}~\bibnamefont {Zhang}}, \ and\ \bibinfo {author}
  {\bibfnamefont {Y.~D.}\ \bibnamefont {Chong}},\ }\bibfield  {title} {\emph
  {\enquote {\bibinfo {title} {Amplification of quantum signals by the
  non-Hermitian skin effect},}\ }}\href {\doibase 10.1103/PhysRevB.106.024301}
  {\bibfield  {journal} {\bibinfo  {journal} {Phys. Rev. B}\ }\textbf {\bibinfo
  {volume} {106}},\ \bibinfo {pages} {024301} (\bibinfo {year}
  {2022})}\BibitemShut {NoStop}%
\bibitem [{\citenamefont {Weidemann}\ \emph {et~al.}(2020)\citenamefont
  {Weidemann}, \citenamefont {Kremer}, \citenamefont {Helbig}, \citenamefont
  {Hofmann}, \citenamefont {Stegmaier}, \citenamefont {Greiter}, \citenamefont
  {Thomale},\ and\ \citenamefont {Szameit}}]{topological_funneling}%
  \BibitemOpen
  \bibfield  {author} {\bibinfo {author} {\bibfnamefont {S.}~\bibnamefont
  {Weidemann}}, \bibinfo {author} {\bibfnamefont {M.}~\bibnamefont {Kremer}},
  \bibinfo {author} {\bibfnamefont {T.}~\bibnamefont {Helbig}}, \bibinfo
  {author} {\bibfnamefont {T.}~\bibnamefont {Hofmann}}, \bibinfo {author}
  {\bibfnamefont {A.}~\bibnamefont {Stegmaier}}, \bibinfo {author}
  {\bibfnamefont {M.}~\bibnamefont {Greiter}}, \bibinfo {author} {\bibfnamefont
  {R.}~\bibnamefont {Thomale}}, \ and\ \bibinfo {author} {\bibfnamefont
  {A.}~\bibnamefont {Szameit}},\ }\bibfield  {title} {\emph {\enquote {\bibinfo
  {title} {Topological funneling of light},}\ }}\href {\doibase
  10.1126/science.aaz8727} {\bibfield  {journal} {\bibinfo  {journal}
  {Science}\ }\textbf {\bibinfo {volume} {368}},\ \bibinfo {pages} {311}
  (\bibinfo {year} {2020})}\BibitemShut {NoStop}%
\bibitem [{\citenamefont {Liang}\ \emph {et~al.}(2022)\citenamefont {Liang},
  \citenamefont {Xie}, \citenamefont {Dong}, \citenamefont {Li}, \citenamefont
  {Li}, \citenamefont {Gadway}, \citenamefont {Yi},\ and\ \citenamefont
  {Yan}}]{cold_atom}%
  \BibitemOpen
  \bibfield  {author} {\bibinfo {author} {\bibfnamefont {Q.}~\bibnamefont
  {Liang}}, \bibinfo {author} {\bibfnamefont {D.}~\bibnamefont {Xie}}, \bibinfo
  {author} {\bibfnamefont {Z.}~\bibnamefont {Dong}}, \bibinfo {author}
  {\bibfnamefont {H.}~\bibnamefont {Li}}, \bibinfo {author} {\bibfnamefont
  {H.}~\bibnamefont {Li}}, \bibinfo {author} {\bibfnamefont {B.}~\bibnamefont
  {Gadway}}, \bibinfo {author} {\bibfnamefont {W.}~\bibnamefont {Yi}}, \ and\
  \bibinfo {author} {\bibfnamefont {B.}~\bibnamefont {Yan}},\ }\bibfield
  {title} {\emph {\enquote {\bibinfo {title} {Dynamic Signatures of
  Non-Hermitian Skin Effect and Topology in Ultracold Atoms},}\ }}\href
  {\doibase 10.1103/PhysRevLett.129.070401} {\bibfield  {journal} {\bibinfo
  {journal} {Phys. Rev. Lett.}\ }\textbf {\bibinfo {volume} {129}},\ \bibinfo
  {pages} {070401} (\bibinfo {year} {2022})}\BibitemShut {NoStop}%
\bibitem [{\citenamefont {Diehl}\ \emph {et~al.}(2011)\citenamefont {Diehl},
  \citenamefont {Rico}, \citenamefont {Baranov},\ and\ \citenamefont
  {Zoller}}]{optical_system1}%
  \BibitemOpen
  \bibfield  {author} {\bibinfo {author} {\bibfnamefont {S.}~\bibnamefont
  {Diehl}}, \bibinfo {author} {\bibfnamefont {E.}~\bibnamefont {Rico}},
  \bibinfo {author} {\bibfnamefont {M.~A.}\ \bibnamefont {Baranov}}, \ and\
  \bibinfo {author} {\bibfnamefont {P.}~\bibnamefont {Zoller}},\ }\bibfield
  {title} {\emph {\enquote {\bibinfo {title} {Topology by dissipation in atomic
  quantum wires},}\ }}\href {\doibase 10.1038/nphys2106} {\bibfield  {journal}
  {\bibinfo  {journal} {Nature Physics}\ }\textbf {\bibinfo {volume} {7}},\
  \bibinfo {pages} {971} (\bibinfo {year} {2011})}\BibitemShut {NoStop}%
\bibitem [{\citenamefont {Xiao}\ \emph {et~al.}(2020)\citenamefont {Xiao},
  \citenamefont {Deng}, \citenamefont {Wang}, \citenamefont {Zhu},
  \citenamefont {Wang}, \citenamefont {Yi},\ and\ \citenamefont
  {Xue}}]{optical_system2}%
  \BibitemOpen
  \bibfield  {author} {\bibinfo {author} {\bibfnamefont {L.}~\bibnamefont
  {Xiao}}, \bibinfo {author} {\bibfnamefont {T.}~\bibnamefont {Deng}}, \bibinfo
  {author} {\bibfnamefont {K.}~\bibnamefont {Wang}}, \bibinfo {author}
  {\bibfnamefont {G.}~\bibnamefont {Zhu}}, \bibinfo {author} {\bibfnamefont
  {Z.}~\bibnamefont {Wang}}, \bibinfo {author} {\bibfnamefont {W.}~\bibnamefont
  {Yi}}, \ and\ \bibinfo {author} {\bibfnamefont {P.}~\bibnamefont {Xue}},\
  }\bibfield  {title} {\emph {\enquote {\bibinfo {title} {Non-Hermitian
  bulk--boundary correspondence in quantum dynamics},}\ }}\href {\doibase
  10.1038/s41567-020-0836-6} {\bibfield  {journal} {\bibinfo  {journal} {Nature
  Physics}\ }\textbf {\bibinfo {volume} {16}},\ \bibinfo {pages} {761}
  (\bibinfo {year} {2020})}\BibitemShut {NoStop}%
\bibitem [{\citenamefont {Ochkan}\ \emph {et~al.}(2024)\citenamefont {Ochkan},
  \citenamefont {Chaturvedi}, \citenamefont {K{\"o}nye}, \citenamefont
  {Veyrat}, \citenamefont {Giraud}, \citenamefont {Mailly}, \citenamefont
  {Cavanna}, \citenamefont {Gennser}, \citenamefont {Hankiewicz}, \citenamefont
  {B{\"u}chner}, \citenamefont {van~den Brink}, \citenamefont {Dufouleur},\
  and\ \citenamefont {Fulga}}]{nh_topo_cmp}%
  \BibitemOpen
  \bibfield  {author} {\bibinfo {author} {\bibfnamefont {K.}~\bibnamefont
  {Ochkan}}, \bibinfo {author} {\bibfnamefont {R.}~\bibnamefont {Chaturvedi}},
  \bibinfo {author} {\bibfnamefont {V.}~\bibnamefont {K{\"o}nye}}, \bibinfo
  {author} {\bibfnamefont {L.}~\bibnamefont {Veyrat}}, \bibinfo {author}
  {\bibfnamefont {R.}~\bibnamefont {Giraud}}, \bibinfo {author} {\bibfnamefont
  {D.}~\bibnamefont {Mailly}}, \bibinfo {author} {\bibfnamefont
  {A.}~\bibnamefont {Cavanna}}, \bibinfo {author} {\bibfnamefont
  {U.}~\bibnamefont {Gennser}}, \bibinfo {author} {\bibfnamefont {E.~M.}\
  \bibnamefont {Hankiewicz}}, \bibinfo {author} {\bibfnamefont
  {B.}~\bibnamefont {B{\"u}chner}}, \bibinfo {author} {\bibfnamefont
  {J.}~\bibnamefont {van~den Brink}}, \bibinfo {author} {\bibfnamefont
  {J.}~\bibnamefont {Dufouleur}}, \ and\ \bibinfo {author} {\bibfnamefont
  {I.~C.}\ \bibnamefont {Fulga}},\ }\bibfield  {title} {\emph {\enquote
  {\bibinfo {title} {Non-Hermitian topology in a multi-terminal quantum Hall
  device},}\ }}\href {\doibase 10.1038/s41567-023-02337-4} {\bibfield
  {journal} {\bibinfo  {journal} {Nature Physics}\ }\textbf {\bibinfo {volume}
  {20}},\ \bibinfo {pages} {395} (\bibinfo {year} {2024})}\BibitemShut
  {NoStop}%
\bibitem [{\citenamefont {von Klitzing}(1986)}]{quantum_Hall_effect}%
  \BibitemOpen
  \bibfield  {author} {\bibinfo {author} {\bibfnamefont {K.}~\bibnamefont {von
  Klitzing}},\ }\bibfield  {title} {\emph {\enquote {\bibinfo {title} {The
  quantized Hall effect},}\ }}\href {\doibase 10.1103/RevModPhys.58.519}
  {\bibfield  {journal} {\bibinfo  {journal} {Rev. Mod. Phys.}\ }\textbf
  {\bibinfo {volume} {58}},\ \bibinfo {pages} {519} (\bibinfo {year}
  {1986})}\BibitemShut {NoStop}%
\bibitem [{\citenamefont {Chen}\ \emph {et~al.}(2009)\citenamefont {Chen},
  \citenamefont {Analytis}, \citenamefont {Chu}, \citenamefont {Liu},
  \citenamefont {Mo}, \citenamefont {Qi}, \citenamefont {Zhang}, \citenamefont
  {Lu}, \citenamefont {Dai}, \citenamefont {Fang}, \citenamefont {Zhang},
  \citenamefont {Fisher}, \citenamefont {Hussain},\ and\ \citenamefont
  {Shen}}]{TI_expt}%
  \BibitemOpen
  \bibfield  {author} {\bibinfo {author} {\bibfnamefont {Y.~L.}\ \bibnamefont
  {Chen}}, \bibinfo {author} {\bibfnamefont {J.~G.}\ \bibnamefont {Analytis}},
  \bibinfo {author} {\bibfnamefont {J.-H.}\ \bibnamefont {Chu}}, \bibinfo
  {author} {\bibfnamefont {Z.~K.}\ \bibnamefont {Liu}}, \bibinfo {author}
  {\bibfnamefont {S.-K.}\ \bibnamefont {Mo}}, \bibinfo {author} {\bibfnamefont
  {X.~L.}\ \bibnamefont {Qi}}, \bibinfo {author} {\bibfnamefont {H.~J.}\
  \bibnamefont {Zhang}}, \bibinfo {author} {\bibfnamefont {D.~H.}\ \bibnamefont
  {Lu}}, \bibinfo {author} {\bibfnamefont {X.}~\bibnamefont {Dai}}, \bibinfo
  {author} {\bibfnamefont {Z.}~\bibnamefont {Fang}}, \bibinfo {author}
  {\bibfnamefont {S.~C.}\ \bibnamefont {Zhang}}, \bibinfo {author}
  {\bibfnamefont {I.~R.}\ \bibnamefont {Fisher}}, \bibinfo {author}
  {\bibfnamefont {Z.}~\bibnamefont {Hussain}}, \ and\ \bibinfo {author}
  {\bibfnamefont {Z.-X.}\ \bibnamefont {Shen}},\ }\bibfield  {title} {\emph
  {\enquote {\bibinfo {title} {Experimental Realization of a Three-Dimensional
  Topological Insulator, Bi<sub>2</sub>Te<sub>3</sub>},}\ }}\href {\doibase
  10.1126/science.1173034} {\bibfield  {journal} {\bibinfo  {journal}
  {Science}\ }\textbf {\bibinfo {volume} {325}},\ \bibinfo {pages} {178}
  (\bibinfo {year} {2009})}\BibitemShut {NoStop}%
\bibitem [{\citenamefont {Liu}\ \emph {et~al.}(2014)\citenamefont {Liu},
  \citenamefont {Zhou}, \citenamefont {Zhang}, \citenamefont {Wang},
  \citenamefont {Weng}, \citenamefont {Prabhakaran}, \citenamefont {Mo},
  \citenamefont {Shen}, \citenamefont {Fang}, \citenamefont {Dai},
  \citenamefont {Hussain},\ and\ \citenamefont {Chen}}]{topo_semimetal_expt}%
  \BibitemOpen
  \bibfield  {author} {\bibinfo {author} {\bibfnamefont {Z.~K.}\ \bibnamefont
  {Liu}}, \bibinfo {author} {\bibfnamefont {B.}~\bibnamefont {Zhou}}, \bibinfo
  {author} {\bibfnamefont {Y.}~\bibnamefont {Zhang}}, \bibinfo {author}
  {\bibfnamefont {Z.~J.}\ \bibnamefont {Wang}}, \bibinfo {author}
  {\bibfnamefont {H.~M.}\ \bibnamefont {Weng}}, \bibinfo {author}
  {\bibfnamefont {D.}~\bibnamefont {Prabhakaran}}, \bibinfo {author}
  {\bibfnamefont {S.-K.}\ \bibnamefont {Mo}}, \bibinfo {author} {\bibfnamefont
  {Z.~X.}\ \bibnamefont {Shen}}, \bibinfo {author} {\bibfnamefont
  {Z.}~\bibnamefont {Fang}}, \bibinfo {author} {\bibfnamefont {X.}~\bibnamefont
  {Dai}}, \bibinfo {author} {\bibfnamefont {Z.}~\bibnamefont {Hussain}}, \ and\
  \bibinfo {author} {\bibfnamefont {Y.~L.}\ \bibnamefont {Chen}},\ }\bibfield
  {title} {\emph {\enquote {\bibinfo {title} {Discovery of a Three-Dimensional
  Topological Dirac Semimetal, Na<sub>3</sub>Bi},}\ }}\href {\doibase
  10.1126/science.1245085} {\bibfield  {journal} {\bibinfo  {journal}
  {Science}\ }\textbf {\bibinfo {volume} {343}},\ \bibinfo {pages} {864}
  (\bibinfo {year} {2014})}\BibitemShut {NoStop}%
\bibitem [{\citenamefont {Xu}\ \emph {et~al.}(2015)\citenamefont {Xu},
  \citenamefont {Belopolski}, \citenamefont {Alidoust}, \citenamefont
  {Neupane}, \citenamefont {Bian}, \citenamefont {Zhang}, \citenamefont
  {Sankar}, \citenamefont {Chang}, \citenamefont {Yuan}, \citenamefont {Lee},
  \citenamefont {Huang}, \citenamefont {Zheng}, \citenamefont {Ma},
  \citenamefont {Sanchez}, \citenamefont {Wang}, \citenamefont {Bansil},
  \citenamefont {Chou}, \citenamefont {Shibayev}, \citenamefont {Lin},
  \citenamefont {Jia},\ and\ \citenamefont {Hasan}}]{WSM_expt}%
  \BibitemOpen
  \bibfield  {author} {\bibinfo {author} {\bibfnamefont {S.-Y.}\ \bibnamefont
  {Xu}}, \bibinfo {author} {\bibfnamefont {I.}~\bibnamefont {Belopolski}},
  \bibinfo {author} {\bibfnamefont {N.}~\bibnamefont {Alidoust}}, \bibinfo
  {author} {\bibfnamefont {M.}~\bibnamefont {Neupane}}, \bibinfo {author}
  {\bibfnamefont {G.}~\bibnamefont {Bian}}, \bibinfo {author} {\bibfnamefont
  {C.}~\bibnamefont {Zhang}}, \bibinfo {author} {\bibfnamefont
  {R.}~\bibnamefont {Sankar}}, \bibinfo {author} {\bibfnamefont
  {G.}~\bibnamefont {Chang}}, \bibinfo {author} {\bibfnamefont
  {Z.}~\bibnamefont {Yuan}}, \bibinfo {author} {\bibfnamefont {C.-C.}\
  \bibnamefont {Lee}}, \bibinfo {author} {\bibfnamefont {S.-M.}\ \bibnamefont
  {Huang}}, \bibinfo {author} {\bibfnamefont {H.}~\bibnamefont {Zheng}},
  \bibinfo {author} {\bibfnamefont {J.}~\bibnamefont {Ma}}, \bibinfo {author}
  {\bibfnamefont {D.~S.}\ \bibnamefont {Sanchez}}, \bibinfo {author}
  {\bibfnamefont {B.}~\bibnamefont {Wang}}, \bibinfo {author} {\bibfnamefont
  {A.}~\bibnamefont {Bansil}}, \bibinfo {author} {\bibfnamefont
  {F.}~\bibnamefont {Chou}}, \bibinfo {author} {\bibfnamefont {P.~P.}\
  \bibnamefont {Shibayev}}, \bibinfo {author} {\bibfnamefont {H.}~\bibnamefont
  {Lin}}, \bibinfo {author} {\bibfnamefont {S.}~\bibnamefont {Jia}}, \ and\
  \bibinfo {author} {\bibfnamefont {M.~Z.}\ \bibnamefont {Hasan}},\ }\bibfield
  {title} {\emph {\enquote {\bibinfo {title} {Discovery of a Weyl fermion
  semimetal and topological Fermi arcs},}\ }}\href {\doibase
  10.1126/science.aaa9297} {\bibfield  {journal} {\bibinfo  {journal}
  {Science}\ }\textbf {\bibinfo {volume} {349}},\ \bibinfo {pages} {613}
  (\bibinfo {year} {2015})}\BibitemShut {NoStop}%
\bibitem [{\citenamefont {Denner}\ \emph {et~al.}(2021)\citenamefont {Denner},
  \citenamefont {Skurativska}, \citenamefont {Schindler}, \citenamefont
  {Fischer}, \citenamefont {Thomale}, \citenamefont {Bzdu{\v{s}}ek},\ and\
  \citenamefont {Neupert}}]{NH_TI}%
  \BibitemOpen
  \bibfield  {author} {\bibinfo {author} {\bibfnamefont {M.~M.}\ \bibnamefont
  {Denner}}, \bibinfo {author} {\bibfnamefont {A.}~\bibnamefont {Skurativska}},
  \bibinfo {author} {\bibfnamefont {F.}~\bibnamefont {Schindler}}, \bibinfo
  {author} {\bibfnamefont {M.~H.}\ \bibnamefont {Fischer}}, \bibinfo {author}
  {\bibfnamefont {R.}~\bibnamefont {Thomale}}, \bibinfo {author} {\bibfnamefont
  {T.}~\bibnamefont {Bzdu{\v{s}}ek}}, \ and\ \bibinfo {author} {\bibfnamefont
  {T.}~\bibnamefont {Neupert}},\ }\bibfield  {title} {\emph {\enquote {\bibinfo
  {title} {Exceptional topological insulators},}\ }}\href {\doibase
  10.1038/s41467-021-25947-z} {\bibfield  {journal} {\bibinfo  {journal}
  {Nature Communications}\ }\textbf {\bibinfo {volume} {12}},\ \bibinfo {pages}
  {5681} (\bibinfo {year} {2021})}\BibitemShut {NoStop}%
\bibitem [{\citenamefont {Kawabata}\ \emph
  {et~al.}(2019{\natexlab{b}})\citenamefont {Kawabata}, \citenamefont
  {Bessho},\ and\ \citenamefont {Sato}}]{NH_SM}%
  \BibitemOpen
  \bibfield  {author} {\bibinfo {author} {\bibfnamefont {K.}~\bibnamefont
  {Kawabata}}, \bibinfo {author} {\bibfnamefont {T.}~\bibnamefont {Bessho}}, \
  and\ \bibinfo {author} {\bibfnamefont {M.}~\bibnamefont {Sato}},\ }\bibfield
  {title} {\emph {\enquote {\bibinfo {title} {Classification of Exceptional
  Points and Non-Hermitian Topological Semimetals},}\ }}\href {\doibase
  10.1103/PhysRevLett.123.066405} {\bibfield  {journal} {\bibinfo  {journal}
  {Phys. Rev. Lett.}\ }\textbf {\bibinfo {volume} {123}},\ \bibinfo {pages}
  {066405} (\bibinfo {year} {2019}{\natexlab{b}})}\BibitemShut {NoStop}%
\bibitem [{\citenamefont {Ghorashi}\ \emph {et~al.}(2021)\citenamefont
  {Ghorashi}, \citenamefont {Li},\ and\ \citenamefont {Sato}}]{NH_HOWSM}%
  \BibitemOpen
  \bibfield  {author} {\bibinfo {author} {\bibfnamefont {S.~A.~A.}\
  \bibnamefont {Ghorashi}}, \bibinfo {author} {\bibfnamefont {T.}~\bibnamefont
  {Li}}, \ and\ \bibinfo {author} {\bibfnamefont {M.}~\bibnamefont {Sato}},\
  }\bibfield  {title} {\emph {\enquote {\bibinfo {title} {Non-Hermitian
  higher-order Weyl semimetals},}\ }}\href {\doibase
  10.1103/PhysRevB.104.L161117} {\bibfield  {journal} {\bibinfo  {journal}
  {Phys. Rev. B}\ }\textbf {\bibinfo {volume} {104}},\ \bibinfo {pages}
  {L161117} (\bibinfo {year} {2021})}\BibitemShut {NoStop}%
\bibitem [{\citenamefont {Ghosh}\ and\ \citenamefont {Nag}(2022)}]{NH_HOTSC}%
  \BibitemOpen
  \bibfield  {author} {\bibinfo {author} {\bibfnamefont {A.~K.}\ \bibnamefont
  {Ghosh}}\ and\ \bibinfo {author} {\bibfnamefont {T.}~\bibnamefont {Nag}},\
  }\bibfield  {title} {\emph {\enquote {\bibinfo {title} {Non-Hermitian
  higher-order topological superconductors in two dimensions: Statics and
  dynamics},}\ }}\href {\doibase 10.1103/PhysRevB.106.L140303} {\bibfield
  {journal} {\bibinfo  {journal} {Phys. Rev. B}\ }\textbf {\bibinfo {volume}
  {106}},\ \bibinfo {pages} {L140303} (\bibinfo {year} {2022})}\BibitemShut
  {NoStop}%
\bibitem [{\citenamefont {Min}\ and\ \citenamefont {MacDonald}(2008)}]{MLG}%
  \BibitemOpen
  \bibfield  {author} {\bibinfo {author} {\bibfnamefont {H.}~\bibnamefont
  {Min}}\ and\ \bibinfo {author} {\bibfnamefont {A.~H.}\ \bibnamefont
  {MacDonald}},\ }\bibfield  {title} {\emph {\enquote {\bibinfo {title}
  {Electronic Structure of Multilayer Graphene},}\ }}\href {\doibase
  10.1143/PTPS.176.227} {\bibfield  {journal} {\bibinfo  {journal} {Progress of
  Theoretical Physics Supplement}\ }\textbf {\bibinfo {volume} {176}},\
  \bibinfo {pages} {227} (\bibinfo {year} {2008})}\BibitemShut {NoStop}%
\bibitem [{\citenamefont {Li}\ \emph {et~al.}(2010)\citenamefont {Li},
  \citenamefont {Luican}, \citenamefont {Lopes~dos Santos}, \citenamefont
  {Castro~Neto}, \citenamefont {Reina}, \citenamefont {Kong},\ and\
  \citenamefont {Andrei}}]{VHS_tBLG_2010}%
  \BibitemOpen
  \bibfield  {author} {\bibinfo {author} {\bibfnamefont {G.}~\bibnamefont
  {Li}}, \bibinfo {author} {\bibfnamefont {A.}~\bibnamefont {Luican}}, \bibinfo
  {author} {\bibfnamefont {J.~M.~B.}\ \bibnamefont {Lopes~dos Santos}},
  \bibinfo {author} {\bibfnamefont {A.~H.}\ \bibnamefont {Castro~Neto}},
  \bibinfo {author} {\bibfnamefont {A.}~\bibnamefont {Reina}}, \bibinfo
  {author} {\bibfnamefont {J.}~\bibnamefont {Kong}}, \ and\ \bibinfo {author}
  {\bibfnamefont {E.~Y.}\ \bibnamefont {Andrei}},\ }\bibfield  {title} {\emph
  {\enquote {\bibinfo {title} {Observation of Van Hove singularities in twisted
  graphene layers},}\ }}\href {\doibase 10.1038/nphys1463} {\bibfield
  {journal} {\bibinfo  {journal} {Nature Physics}\ }\textbf {\bibinfo {volume}
  {6}},\ \bibinfo {pages} {109} (\bibinfo {year} {2010})}\BibitemShut {NoStop}%
\bibitem [{\citenamefont {Can}\ \emph {et~al.}(2021)\citenamefont {Can},
  \citenamefont {Tummuru}, \citenamefont {Day}, \citenamefont {Elfimov},
  \citenamefont {Damascelli},\ and\ \citenamefont {Franz}}]{tSL}%
  \BibitemOpen
  \bibfield  {author} {\bibinfo {author} {\bibfnamefont {O.}~\bibnamefont
  {Can}}, \bibinfo {author} {\bibfnamefont {T.}~\bibnamefont {Tummuru}},
  \bibinfo {author} {\bibfnamefont {R.~P.}\ \bibnamefont {Day}}, \bibinfo
  {author} {\bibfnamefont {I.}~\bibnamefont {Elfimov}}, \bibinfo {author}
  {\bibfnamefont {A.}~\bibnamefont {Damascelli}}, \ and\ \bibinfo {author}
  {\bibfnamefont {M.}~\bibnamefont {Franz}},\ }\bibfield  {title} {\emph
  {\enquote {\bibinfo {title} {High-temperature topological superconductivity
  in twisted double-layer copper oxides},}\ }}\href {\doibase
  10.1038/s41567-020-01142-7} {\bibfield  {journal} {\bibinfo  {journal}
  {Nature Physics}\ }\textbf {\bibinfo {volume} {17}},\ \bibinfo {pages} {519}
  (\bibinfo {year} {2021})}\BibitemShut {NoStop}%
\bibitem [{\citenamefont {Danawe}\ \emph {et~al.}(2021)\citenamefont {Danawe},
  \citenamefont {Li}, \citenamefont {Ba'ba'a},\ and\ \citenamefont
  {Tol}}]{kagome}%
  \BibitemOpen
  \bibfield  {author} {\bibinfo {author} {\bibfnamefont {H.}~\bibnamefont
  {Danawe}}, \bibinfo {author} {\bibfnamefont {H.}~\bibnamefont {Li}}, \bibinfo
  {author} {\bibfnamefont {H.~A.}\ \bibnamefont {Ba'ba'a}}, \ and\ \bibinfo
  {author} {\bibfnamefont {S.}~\bibnamefont {Tol}},\ }\bibfield  {title} {\emph
  {\enquote {\bibinfo {title} {Existence of corner modes in elastic twisted
  kagome lattices},}\ }}\href {\doibase 10.1103/PhysRevB.104.L241107}
  {\bibfield  {journal} {\bibinfo  {journal} {Phys. Rev. B}\ }\textbf {\bibinfo
  {volume} {104}},\ \bibinfo {pages} {L241107} (\bibinfo {year}
  {2021})}\BibitemShut {NoStop}%
\bibitem [{\citenamefont {Ma}\ \emph {et~al.}(2024)\citenamefont {Ma},
  \citenamefont {Chen}, \citenamefont {Yu},\ and\ \citenamefont {Luo}}]{dice}%
  \BibitemOpen
  \bibfield  {author} {\bibinfo {author} {\bibfnamefont {D.}~\bibnamefont
  {Ma}}, \bibinfo {author} {\bibfnamefont {Y.-G.}\ \bibnamefont {Chen}},
  \bibinfo {author} {\bibfnamefont {Y.}~\bibnamefont {Yu}}, \ and\ \bibinfo
  {author} {\bibfnamefont {X.}~\bibnamefont {Luo}},\ }\bibfield  {title} {\emph
  {\enquote {\bibinfo {title} {Moir\'e semiconductors on the twisted bilayer
  dice lattice},}\ }}\href {\doibase 10.1103/PhysRevB.109.155159} {\bibfield
  {journal} {\bibinfo  {journal} {Phys. Rev. B}\ }\textbf {\bibinfo {volume}
  {109}},\ \bibinfo {pages} {155159} (\bibinfo {year} {2024})}\BibitemShut
  {NoStop}%
\bibitem [{\citenamefont {Chen}\ \emph {et~al.}(2021)\citenamefont {Chen},
  \citenamefont {He}, \citenamefont {Zhang}, \citenamefont {Hsieh},
  \citenamefont {Fei}, \citenamefont {Watanabe}, \citenamefont {Taniguchi},
  \citenamefont {Cobden}, \citenamefont {Xu}, \citenamefont {Dean},\ and\
  \citenamefont {Yankowitz}}]{Mono-Bi-graphene1}%
  \BibitemOpen
  \bibfield  {author} {\bibinfo {author} {\bibfnamefont {S.}~\bibnamefont
  {Chen}}, \bibinfo {author} {\bibfnamefont {M.}~\bibnamefont {He}}, \bibinfo
  {author} {\bibfnamefont {Y.-H.}\ \bibnamefont {Zhang}}, \bibinfo {author}
  {\bibfnamefont {V.}~\bibnamefont {Hsieh}}, \bibinfo {author} {\bibfnamefont
  {Z.}~\bibnamefont {Fei}}, \bibinfo {author} {\bibfnamefont {K.}~\bibnamefont
  {Watanabe}}, \bibinfo {author} {\bibfnamefont {T.}~\bibnamefont {Taniguchi}},
  \bibinfo {author} {\bibfnamefont {D.~H.}\ \bibnamefont {Cobden}}, \bibinfo
  {author} {\bibfnamefont {X.}~\bibnamefont {Xu}}, \bibinfo {author}
  {\bibfnamefont {C.~R.}\ \bibnamefont {Dean}}, \ and\ \bibinfo {author}
  {\bibfnamefont {M.}~\bibnamefont {Yankowitz}},\ }\bibfield  {title} {\emph
  {\enquote {\bibinfo {title} {Electrically tunable correlated and topological
  states in twisted monolayer--bilayer graphene},}\ }}\href {\doibase
  10.1038/s41567-020-01062-6} {\bibfield  {journal} {\bibinfo  {journal}
  {Nature Physics}\ }\textbf {\bibinfo {volume} {17}},\ \bibinfo {pages} {374}
  (\bibinfo {year} {2021})}\BibitemShut {NoStop}%
\bibitem [{\citenamefont {Park}\ \emph {et~al.}(2021)\citenamefont {Park},
  \citenamefont {Cao}, \citenamefont {Watanabe}, \citenamefont {Taniguchi},\
  and\ \citenamefont {Jarillo-Herrero}}]{Park2021-tTLG1}%
  \BibitemOpen
  \bibfield  {author} {\bibinfo {author} {\bibfnamefont {J.~M.}\ \bibnamefont
  {Park}}, \bibinfo {author} {\bibfnamefont {Y.}~\bibnamefont {Cao}}, \bibinfo
  {author} {\bibfnamefont {K.}~\bibnamefont {Watanabe}}, \bibinfo {author}
  {\bibfnamefont {T.}~\bibnamefont {Taniguchi}}, \ and\ \bibinfo {author}
  {\bibfnamefont {P.}~\bibnamefont {Jarillo-Herrero}},\ }\bibfield  {title}
  {\emph {\enquote {\bibinfo {title} {Tunable strongly coupled
  superconductivity in magic-angle twisted trilayer graphene},}\ }}\href
  {\doibase 10.1038/s41586-021-03192-0} {\bibfield  {journal} {\bibinfo
  {journal} {Nature}\ }\textbf {\bibinfo {volume} {590}},\ \bibinfo {pages}
  {249} (\bibinfo {year} {2021})}\BibitemShut {NoStop}%
\bibitem [{\citenamefont {Liu}\ \emph {et~al.}(2020)\citenamefont {Liu},
  \citenamefont {Hao}, \citenamefont {Khalaf}, \citenamefont {Lee},
  \citenamefont {Ronen}, \citenamefont {Yoo}, \citenamefont {Haei~Najafabadi},
  \citenamefont {Watanabe}, \citenamefont {Taniguchi}, \citenamefont
  {Vishwanath},\ and\ \citenamefont {Kim}}]{Liu2020-tDBLG1}%
  \BibitemOpen
  \bibfield  {author} {\bibinfo {author} {\bibfnamefont {X.}~\bibnamefont
  {Liu}}, \bibinfo {author} {\bibfnamefont {Z.}~\bibnamefont {Hao}}, \bibinfo
  {author} {\bibfnamefont {E.}~\bibnamefont {Khalaf}}, \bibinfo {author}
  {\bibfnamefont {J.~Y.}\ \bibnamefont {Lee}}, \bibinfo {author} {\bibfnamefont
  {Y.}~\bibnamefont {Ronen}}, \bibinfo {author} {\bibfnamefont
  {H.}~\bibnamefont {Yoo}}, \bibinfo {author} {\bibfnamefont {D.}~\bibnamefont
  {Haei~Najafabadi}}, \bibinfo {author} {\bibfnamefont {K.}~\bibnamefont
  {Watanabe}}, \bibinfo {author} {\bibfnamefont {T.}~\bibnamefont {Taniguchi}},
  \bibinfo {author} {\bibfnamefont {A.}~\bibnamefont {Vishwanath}}, \ and\
  \bibinfo {author} {\bibfnamefont {P.}~\bibnamefont {Kim}},\ }\bibfield
  {title} {\emph {\enquote {\bibinfo {title} {Tunable spin-polarized correlated
  states in twisted double bilayer graphene},}\ }}\href {\doibase
  10.1038/s41586-020-2458-7} {\bibfield  {journal} {\bibinfo  {journal}
  {Nature}\ }\textbf {\bibinfo {volume} {583}},\ \bibinfo {pages} {221}
  (\bibinfo {year} {2020})}\BibitemShut {NoStop}%
\bibitem [{\citenamefont {Wu}\ \emph {et~al.}(2019)\citenamefont {Wu},
  \citenamefont {Lovorn}, \citenamefont {Tutuc}, \citenamefont {Martin},\ and\
  \citenamefont {MacDonald}}]{TMD-homobilayers1}%
  \BibitemOpen
  \bibfield  {author} {\bibinfo {author} {\bibfnamefont {F.}~\bibnamefont
  {Wu}}, \bibinfo {author} {\bibfnamefont {T.}~\bibnamefont {Lovorn}}, \bibinfo
  {author} {\bibfnamefont {E.}~\bibnamefont {Tutuc}}, \bibinfo {author}
  {\bibfnamefont {I.}~\bibnamefont {Martin}}, \ and\ \bibinfo {author}
  {\bibfnamefont {A.~H.}\ \bibnamefont {MacDonald}},\ }\bibfield  {title}
  {\emph {\enquote {\bibinfo {title} {Topological Insulators in Twisted
  Transition Metal Dichalcogenide Homobilayers},}\ }}\href {\doibase
  10.1103/PhysRevLett.122.086402} {\bibfield  {journal} {\bibinfo  {journal}
  {Phys. Rev. Lett.}\ }\textbf {\bibinfo {volume} {122}},\ \bibinfo {pages}
  {086402} (\bibinfo {year} {2019})}\BibitemShut {NoStop}%
\bibitem [{\citenamefont {Chen}\ \emph {et~al.}(2019)\citenamefont {Chen},
  \citenamefont {Jiang}, \citenamefont {Wu}, \citenamefont {Lyu}, \citenamefont
  {Li}, \citenamefont {Chittari}, \citenamefont {Watanabe}, \citenamefont
  {Taniguchi}, \citenamefont {Shi}, \citenamefont {Jung}, \citenamefont
  {Zhang},\ and\ \citenamefont {Wang}}]{rTTLG+hBN1}%
  \BibitemOpen
  \bibfield  {author} {\bibinfo {author} {\bibfnamefont {G.}~\bibnamefont
  {Chen}}, \bibinfo {author} {\bibfnamefont {L.}~\bibnamefont {Jiang}},
  \bibinfo {author} {\bibfnamefont {S.}~\bibnamefont {Wu}}, \bibinfo {author}
  {\bibfnamefont {B.}~\bibnamefont {Lyu}}, \bibinfo {author} {\bibfnamefont
  {H.}~\bibnamefont {Li}}, \bibinfo {author} {\bibfnamefont {B.~L.}\
  \bibnamefont {Chittari}}, \bibinfo {author} {\bibfnamefont {K.}~\bibnamefont
  {Watanabe}}, \bibinfo {author} {\bibfnamefont {T.}~\bibnamefont {Taniguchi}},
  \bibinfo {author} {\bibfnamefont {Z.}~\bibnamefont {Shi}}, \bibinfo {author}
  {\bibfnamefont {J.}~\bibnamefont {Jung}}, \bibinfo {author} {\bibfnamefont
  {Y.}~\bibnamefont {Zhang}}, \ and\ \bibinfo {author} {\bibfnamefont
  {F.}~\bibnamefont {Wang}},\ }\bibfield  {title} {\emph {\enquote {\bibinfo
  {title} {Evidence of a gate-tunable Mott insulator in a trilayer graphene
  moir{\'e} superlattice},}\ }}\href {\doibase 10.1038/s41567-018-0387-2}
  {\bibfield  {journal} {\bibinfo  {journal} {Nature Physics}\ }\textbf
  {\bibinfo {volume} {15}},\ \bibinfo {pages} {237} (\bibinfo {year}
  {2019})}\BibitemShut {NoStop}%
\bibitem [{\citenamefont {Bera}\ \emph {et~al.}(2025)\citenamefont {Bera},
  \citenamefont {Mohan},\ and\ \citenamefont {Saha}}]{tDBLG_soc}%
  \BibitemOpen
  \bibfield  {author} {\bibinfo {author} {\bibfnamefont {K.}~\bibnamefont
  {Bera}}, \bibinfo {author} {\bibfnamefont {P.}~\bibnamefont {Mohan}}, \ and\
  \bibinfo {author} {\bibfnamefont {A.}~\bibnamefont {Saha}},\ }\bibfield
  {title} {\emph {\enquote {\bibinfo {title} {Tailoring topological band
  properties of twisted double bilayer graphene: Effects due to spin-orbit
  coupling},}\ }}\href {\doibase 10.1103/PhysRevB.111.045434} {\bibfield
  {journal} {\bibinfo  {journal} {Phys. Rev. B}\ }\textbf {\bibinfo {volume}
  {111}},\ \bibinfo {pages} {045434} (\bibinfo {year} {2025})}\BibitemShut
  {NoStop}%
\bibitem [{\citenamefont {Cao}\ \emph {et~al.}(2018{\natexlab{a}})\citenamefont
  {Cao}, \citenamefont {Fatemi}, \citenamefont {Demir}, \citenamefont {Fang},
  \citenamefont {Tomarken}, \citenamefont {Luo}, \citenamefont
  {Sanchez-Yamagishi}, \citenamefont {Watanabe}, \citenamefont {Taniguchi},
  \citenamefont {Kaxiras}, \citenamefont {Ashoori},\ and\ \citenamefont
  {Jarillo-Herrero}}]{Cao2018-corr_insulator}%
  \BibitemOpen
  \bibfield  {author} {\bibinfo {author} {\bibfnamefont {Y.}~\bibnamefont
  {Cao}}, \bibinfo {author} {\bibfnamefont {V.}~\bibnamefont {Fatemi}},
  \bibinfo {author} {\bibfnamefont {A.}~\bibnamefont {Demir}}, \bibinfo
  {author} {\bibfnamefont {S.}~\bibnamefont {Fang}}, \bibinfo {author}
  {\bibfnamefont {S.~L.}\ \bibnamefont {Tomarken}}, \bibinfo {author}
  {\bibfnamefont {J.~Y.}\ \bibnamefont {Luo}}, \bibinfo {author} {\bibfnamefont
  {J.~D.}\ \bibnamefont {Sanchez-Yamagishi}}, \bibinfo {author} {\bibfnamefont
  {K.}~\bibnamefont {Watanabe}}, \bibinfo {author} {\bibfnamefont
  {T.}~\bibnamefont {Taniguchi}}, \bibinfo {author} {\bibfnamefont
  {E.}~\bibnamefont {Kaxiras}}, \bibinfo {author} {\bibfnamefont {R.~C.}\
  \bibnamefont {Ashoori}}, \ and\ \bibinfo {author} {\bibfnamefont
  {P.}~\bibnamefont {Jarillo-Herrero}},\ }\bibfield  {title} {\emph {\enquote
  {\bibinfo {title} {Correlated insulator behaviour at half-filling in
  magic-angle graphene superlattices},}\ }}\href {\doibase 10.1038/nature26154}
  {\bibfield  {journal} {\bibinfo  {journal} {Nature}\ }\textbf {\bibinfo
  {volume} {556}},\ \bibinfo {pages} {80} (\bibinfo {year}
  {2018}{\natexlab{a}})}\BibitemShut {NoStop}%
\bibitem [{\citenamefont {Choi}\ \emph {et~al.}(2019)\citenamefont {Choi},
  \citenamefont {Kemmer}, \citenamefont {Peng}, \citenamefont {Thomson},
  \citenamefont {Arora}, \citenamefont {Polski}, \citenamefont {Zhang},
  \citenamefont {Ren}, \citenamefont {Alicea}, \citenamefont {Refael},
  \citenamefont {von Oppen}, \citenamefont {Watanabe}, \citenamefont
  {Taniguchi},\ and\ \citenamefont {Nadj-Perge}}]{Choi2019}%
  \BibitemOpen
  \bibfield  {author} {\bibinfo {author} {\bibfnamefont {Y.}~\bibnamefont
  {Choi}}, \bibinfo {author} {\bibfnamefont {J.}~\bibnamefont {Kemmer}},
  \bibinfo {author} {\bibfnamefont {Y.}~\bibnamefont {Peng}}, \bibinfo {author}
  {\bibfnamefont {A.}~\bibnamefont {Thomson}}, \bibinfo {author} {\bibfnamefont
  {H.}~\bibnamefont {Arora}}, \bibinfo {author} {\bibfnamefont
  {R.}~\bibnamefont {Polski}}, \bibinfo {author} {\bibfnamefont
  {Y.}~\bibnamefont {Zhang}}, \bibinfo {author} {\bibfnamefont
  {H.}~\bibnamefont {Ren}}, \bibinfo {author} {\bibfnamefont {J.}~\bibnamefont
  {Alicea}}, \bibinfo {author} {\bibfnamefont {G.}~\bibnamefont {Refael}},
  \bibinfo {author} {\bibfnamefont {F.}~\bibnamefont {von Oppen}}, \bibinfo
  {author} {\bibfnamefont {K.}~\bibnamefont {Watanabe}}, \bibinfo {author}
  {\bibfnamefont {T.}~\bibnamefont {Taniguchi}}, \ and\ \bibinfo {author}
  {\bibfnamefont {S.}~\bibnamefont {Nadj-Perge}},\ }\bibfield  {title} {\emph
  {\enquote {\bibinfo {title} {Electronic correlations in twisted bilayer
  graphene near the magic angle},}\ }}\href {\doibase
  10.1038/s41567-019-0606-5} {\bibfield  {journal} {\bibinfo  {journal} {Nature
  Physics}\ }\textbf {\bibinfo {volume} {15}},\ \bibinfo {pages} {1174}
  (\bibinfo {year} {2019})}\BibitemShut {NoStop}%
\bibitem [{\citenamefont {Wu}\ \emph {et~al.}(2021)\citenamefont {Wu},
  \citenamefont {Zhang}, \citenamefont {Watanabe}, \citenamefont {Taniguchi},\
  and\ \citenamefont {Andrei}}]{Wu2021-chernIns-expt}%
  \BibitemOpen
  \bibfield  {author} {\bibinfo {author} {\bibfnamefont {S.}~\bibnamefont
  {Wu}}, \bibinfo {author} {\bibfnamefont {Z.}~\bibnamefont {Zhang}}, \bibinfo
  {author} {\bibfnamefont {K.}~\bibnamefont {Watanabe}}, \bibinfo {author}
  {\bibfnamefont {T.}~\bibnamefont {Taniguchi}}, \ and\ \bibinfo {author}
  {\bibfnamefont {E.~Y.}\ \bibnamefont {Andrei}},\ }\bibfield  {title} {\emph
  {\enquote {\bibinfo {title} {Chern insulators, van Hove singularities and
  topological flat bands in magic-angle twisted bilayer graphene},}\ }}\href
  {\doibase 10.1038/s41563-020-00911-2} {\bibfield  {journal} {\bibinfo
  {journal} {Nature Materials}\ }\textbf {\bibinfo {volume} {20}},\ \bibinfo
  {pages} {488} (\bibinfo {year} {2021})}\BibitemShut {NoStop}%
\bibitem [{\citenamefont {Nuckolls}\ \emph {et~al.}(2020)\citenamefont
  {Nuckolls}, \citenamefont {Oh}, \citenamefont {Wong}, \citenamefont {Lian},
  \citenamefont {Watanabe}, \citenamefont {Taniguchi}, \citenamefont
  {Bernevig},\ and\ \citenamefont {Yazdani}}]{Nuckolls2020-tBLG_xpt}%
  \BibitemOpen
  \bibfield  {author} {\bibinfo {author} {\bibfnamefont {K.~P.}\ \bibnamefont
  {Nuckolls}}, \bibinfo {author} {\bibfnamefont {M.}~\bibnamefont {Oh}},
  \bibinfo {author} {\bibfnamefont {D.}~\bibnamefont {Wong}}, \bibinfo {author}
  {\bibfnamefont {B.}~\bibnamefont {Lian}}, \bibinfo {author} {\bibfnamefont
  {K.}~\bibnamefont {Watanabe}}, \bibinfo {author} {\bibfnamefont
  {T.}~\bibnamefont {Taniguchi}}, \bibinfo {author} {\bibfnamefont {B.~A.}\
  \bibnamefont {Bernevig}}, \ and\ \bibinfo {author} {\bibfnamefont
  {A.}~\bibnamefont {Yazdani}},\ }\bibfield  {title} {\emph {\enquote {\bibinfo
  {title} {Strongly correlated Chern insulators in magic-angle twisted bilayer
  graphene},}\ }}\href {\doibase 10.1038/s41586-020-3028-8} {\bibfield
  {journal} {\bibinfo  {journal} {Nature}\ }\textbf {\bibinfo {volume} {588}},\
  \bibinfo {pages} {610} (\bibinfo {year} {2020})}\BibitemShut {NoStop}%
\bibitem [{\citenamefont {Choi}\ \emph {et~al.}(2021)\citenamefont {Choi},
  \citenamefont {Kim}, \citenamefont {Peng}, \citenamefont {Thomson},
  \citenamefont {Lewandowski}, \citenamefont {Polski}, \citenamefont {Zhang},
  \citenamefont {Arora}, \citenamefont {Watanabe}, \citenamefont {Taniguchi},
  \citenamefont {Alicea},\ and\ \citenamefont {Nadj-Perge}}]{Choi2021}%
  \BibitemOpen
  \bibfield  {author} {\bibinfo {author} {\bibfnamefont {Y.}~\bibnamefont
  {Choi}}, \bibinfo {author} {\bibfnamefont {H.}~\bibnamefont {Kim}}, \bibinfo
  {author} {\bibfnamefont {Y.}~\bibnamefont {Peng}}, \bibinfo {author}
  {\bibfnamefont {A.}~\bibnamefont {Thomson}}, \bibinfo {author} {\bibfnamefont
  {C.}~\bibnamefont {Lewandowski}}, \bibinfo {author} {\bibfnamefont
  {R.}~\bibnamefont {Polski}}, \bibinfo {author} {\bibfnamefont
  {Y.}~\bibnamefont {Zhang}}, \bibinfo {author} {\bibfnamefont {H.~S.}\
  \bibnamefont {Arora}}, \bibinfo {author} {\bibfnamefont {K.}~\bibnamefont
  {Watanabe}}, \bibinfo {author} {\bibfnamefont {T.}~\bibnamefont {Taniguchi}},
  \bibinfo {author} {\bibfnamefont {J.}~\bibnamefont {Alicea}}, \ and\ \bibinfo
  {author} {\bibfnamefont {S.}~\bibnamefont {Nadj-Perge}},\ }\bibfield  {title}
  {\emph {\enquote {\bibinfo {title} {Correlation-driven topological phases in
  magic-angle twisted bilayer graphene},}\ }}\href {\doibase
  10.1038/s41586-020-03159-7} {\bibfield  {journal} {\bibinfo  {journal}
  {Nature}\ }\textbf {\bibinfo {volume} {589}},\ \bibinfo {pages} {536}
  (\bibinfo {year} {2021})}\BibitemShut {NoStop}%
\bibitem [{\citenamefont {Xie}\ \emph {et~al.}(2021)\citenamefont {Xie},
  \citenamefont {Pierce}, \citenamefont {Park}, \citenamefont {Parker},
  \citenamefont {Khalaf}, \citenamefont {Ledwith}, \citenamefont {Cao},
  \citenamefont {Lee}, \citenamefont {Chen}, \citenamefont {Forrester},
  \citenamefont {Watanabe}, \citenamefont {Taniguchi}, \citenamefont
  {Vishwanath}, \citenamefont {Jarillo-Herrero},\ and\ \citenamefont
  {Yacoby}}]{Xie2021}%
  \BibitemOpen
  \bibfield  {author} {\bibinfo {author} {\bibfnamefont {Y.}~\bibnamefont
  {Xie}}, \bibinfo {author} {\bibfnamefont {A.~T.}\ \bibnamefont {Pierce}},
  \bibinfo {author} {\bibfnamefont {J.~M.}\ \bibnamefont {Park}}, \bibinfo
  {author} {\bibfnamefont {D.~E.}\ \bibnamefont {Parker}}, \bibinfo {author}
  {\bibfnamefont {E.}~\bibnamefont {Khalaf}}, \bibinfo {author} {\bibfnamefont
  {P.}~\bibnamefont {Ledwith}}, \bibinfo {author} {\bibfnamefont
  {Y.}~\bibnamefont {Cao}}, \bibinfo {author} {\bibfnamefont {S.~H.}\
  \bibnamefont {Lee}}, \bibinfo {author} {\bibfnamefont {S.}~\bibnamefont
  {Chen}}, \bibinfo {author} {\bibfnamefont {P.~R.}\ \bibnamefont {Forrester}},
  \bibinfo {author} {\bibfnamefont {K.}~\bibnamefont {Watanabe}}, \bibinfo
  {author} {\bibfnamefont {T.}~\bibnamefont {Taniguchi}}, \bibinfo {author}
  {\bibfnamefont {A.}~\bibnamefont {Vishwanath}}, \bibinfo {author}
  {\bibfnamefont {P.}~\bibnamefont {Jarillo-Herrero}}, \ and\ \bibinfo {author}
  {\bibfnamefont {A.}~\bibnamefont {Yacoby}},\ }\bibfield  {title} {\emph
  {\enquote {\bibinfo {title} {Fractional Chern insulators in magic-angle
  twisted bilayer graphene},}\ }}\href {\doibase 10.1038/s41586-021-04002-3}
  {\bibfield  {journal} {\bibinfo  {journal} {Nature}\ }\textbf {\bibinfo
  {volume} {600}},\ \bibinfo {pages} {439} (\bibinfo {year}
  {2021})}\BibitemShut {NoStop}%
\bibitem [{\citenamefont {Sharpe}\ \emph {et~al.}(2019)\citenamefont {Sharpe},
  \citenamefont {Fox}, \citenamefont {Barnard}, \citenamefont {Finney},
  \citenamefont {Watanabe}, \citenamefont {Taniguchi}, \citenamefont
  {Kastner},\ and\ \citenamefont {Goldhaber-Gordon}}]{magnetism1}%
  \BibitemOpen
  \bibfield  {author} {\bibinfo {author} {\bibfnamefont {A.~L.}\ \bibnamefont
  {Sharpe}}, \bibinfo {author} {\bibfnamefont {E.~J.}\ \bibnamefont {Fox}},
  \bibinfo {author} {\bibfnamefont {A.~W.}\ \bibnamefont {Barnard}}, \bibinfo
  {author} {\bibfnamefont {J.}~\bibnamefont {Finney}}, \bibinfo {author}
  {\bibfnamefont {K.}~\bibnamefont {Watanabe}}, \bibinfo {author}
  {\bibfnamefont {T.}~\bibnamefont {Taniguchi}}, \bibinfo {author}
  {\bibfnamefont {M.~A.}\ \bibnamefont {Kastner}}, \ and\ \bibinfo {author}
  {\bibfnamefont {D.}~\bibnamefont {Goldhaber-Gordon}},\ }\bibfield  {title}
  {\emph {\enquote {\bibinfo {title} {Emergent ferromagnetism near
  three-quarters filling in twisted bilayer graphene},}\ }}\href {\doibase
  10.1126/science.aaw3780} {\bibfield  {journal} {\bibinfo  {journal}
  {Science}\ }\textbf {\bibinfo {volume} {365}},\ \bibinfo {pages} {605}
  (\bibinfo {year} {2019})}\BibitemShut {NoStop}%
\bibitem [{\citenamefont {Lin}\ \emph {et~al.}(2022)\citenamefont {Lin},
  \citenamefont {Zhang}, \citenamefont {Morissette}, \citenamefont {Wang},
  \citenamefont {Liu}, \citenamefont {Rhodes}, \citenamefont {Watanabe},
  \citenamefont {Taniguchi}, \citenamefont {Hone},\ and\ \citenamefont
  {Li}}]{magnetism2}%
  \BibitemOpen
  \bibfield  {author} {\bibinfo {author} {\bibfnamefont {J.-X.}\ \bibnamefont
  {Lin}}, \bibinfo {author} {\bibfnamefont {Y.-H.}\ \bibnamefont {Zhang}},
  \bibinfo {author} {\bibfnamefont {E.}~\bibnamefont {Morissette}}, \bibinfo
  {author} {\bibfnamefont {Z.}~\bibnamefont {Wang}}, \bibinfo {author}
  {\bibfnamefont {S.}~\bibnamefont {Liu}}, \bibinfo {author} {\bibfnamefont
  {D.}~\bibnamefont {Rhodes}}, \bibinfo {author} {\bibfnamefont
  {K.}~\bibnamefont {Watanabe}}, \bibinfo {author} {\bibfnamefont
  {T.}~\bibnamefont {Taniguchi}}, \bibinfo {author} {\bibfnamefont
  {J.}~\bibnamefont {Hone}}, \ and\ \bibinfo {author} {\bibfnamefont
  {J.~I.~A.}\ \bibnamefont {Li}},\ }\bibfield  {title} {\emph {\enquote
  {\bibinfo {title} {Spin-orbit–driven ferromagnetism at half moiré filling
  in magic-angle twisted bilayer graphene},}\ }}\href {\doibase
  10.1126/science.abh2889} {\bibfield  {journal} {\bibinfo  {journal}
  {Science}\ }\textbf {\bibinfo {volume} {375}},\ \bibinfo {pages} {437}
  (\bibinfo {year} {2022})}\BibitemShut {NoStop}%
\bibitem [{\citenamefont {Cao}\ \emph {et~al.}(2021)\citenamefont {Cao},
  \citenamefont {Rodan-Legrain}, \citenamefont {Park}, \citenamefont {Yuan},
  \citenamefont {Watanabe}, \citenamefont {Taniguchi}, \citenamefont
  {Fernandes}, \citenamefont {Fu},\ and\ \citenamefont
  {Jarillo-Herrero}}]{nematicity1}%
  \BibitemOpen
  \bibfield  {author} {\bibinfo {author} {\bibfnamefont {Y.}~\bibnamefont
  {Cao}}, \bibinfo {author} {\bibfnamefont {D.}~\bibnamefont {Rodan-Legrain}},
  \bibinfo {author} {\bibfnamefont {J.~M.}\ \bibnamefont {Park}}, \bibinfo
  {author} {\bibfnamefont {N.~F.~Q.}\ \bibnamefont {Yuan}}, \bibinfo {author}
  {\bibfnamefont {K.}~\bibnamefont {Watanabe}}, \bibinfo {author}
  {\bibfnamefont {T.}~\bibnamefont {Taniguchi}}, \bibinfo {author}
  {\bibfnamefont {R.~M.}\ \bibnamefont {Fernandes}}, \bibinfo {author}
  {\bibfnamefont {L.}~\bibnamefont {Fu}}, \ and\ \bibinfo {author}
  {\bibfnamefont {P.}~\bibnamefont {Jarillo-Herrero}},\ }\bibfield  {title}
  {\emph {\enquote {\bibinfo {title} {Nematicity and competing orders in
  superconducting magic-angle graphene},}\ }}\href {\doibase
  10.1126/science.abc2836} {\bibfield  {journal} {\bibinfo  {journal}
  {Science}\ }\textbf {\bibinfo {volume} {372}},\ \bibinfo {pages} {264}
  (\bibinfo {year} {2021})}\BibitemShut {NoStop}%
\bibitem [{\citenamefont {Rubio-Verd{\'u}}\ \emph {et~al.}(2022)\citenamefont
  {Rubio-Verd{\'u}}, \citenamefont {Turkel}, \citenamefont {Song},
  \citenamefont {Klebl}, \citenamefont {Samajdar}, \citenamefont {Scheurer},
  \citenamefont {Venderbos}, \citenamefont {Watanabe}, \citenamefont
  {Taniguchi}, \citenamefont {Ochoa}, \citenamefont {Xian}, \citenamefont
  {Kennes}, \citenamefont {Fernandes}, \citenamefont {Rubio},\ and\
  \citenamefont {Pasupathy}}]{nematicity2}%
  \BibitemOpen
  \bibfield  {author} {\bibinfo {author} {\bibfnamefont {C.}~\bibnamefont
  {Rubio-Verd{\'u}}}, \bibinfo {author} {\bibfnamefont {S.}~\bibnamefont
  {Turkel}}, \bibinfo {author} {\bibfnamefont {Y.}~\bibnamefont {Song}},
  \bibinfo {author} {\bibfnamefont {L.}~\bibnamefont {Klebl}}, \bibinfo
  {author} {\bibfnamefont {R.}~\bibnamefont {Samajdar}}, \bibinfo {author}
  {\bibfnamefont {M.~S.}\ \bibnamefont {Scheurer}}, \bibinfo {author}
  {\bibfnamefont {J.~W.~F.}\ \bibnamefont {Venderbos}}, \bibinfo {author}
  {\bibfnamefont {K.}~\bibnamefont {Watanabe}}, \bibinfo {author}
  {\bibfnamefont {T.}~\bibnamefont {Taniguchi}}, \bibinfo {author}
  {\bibfnamefont {H.}~\bibnamefont {Ochoa}}, \bibinfo {author} {\bibfnamefont
  {L.}~\bibnamefont {Xian}}, \bibinfo {author} {\bibfnamefont {D.~M.}\
  \bibnamefont {Kennes}}, \bibinfo {author} {\bibfnamefont {R.~M.}\
  \bibnamefont {Fernandes}}, \bibinfo {author} {\bibfnamefont
  {{\'A}.}~\bibnamefont {Rubio}}, \ and\ \bibinfo {author} {\bibfnamefont
  {A.~N.}\ \bibnamefont {Pasupathy}},\ }\bibfield  {title} {\emph {\enquote
  {\bibinfo {title} {Moir{\'e} nematic phase in twisted double bilayer
  graphene},}\ }}\href {\doibase 10.1038/s41567-021-01438-2} {\bibfield
  {journal} {\bibinfo  {journal} {Nature Physics}\ }\textbf {\bibinfo {volume}
  {18}},\ \bibinfo {pages} {196} (\bibinfo {year} {2022})}\BibitemShut
  {NoStop}%
\bibitem [{\citenamefont {Cao}\ \emph {et~al.}(2018{\natexlab{b}})\citenamefont
  {Cao}, \citenamefont {Fatemi}, \citenamefont {Fang}, \citenamefont
  {Watanabe}, \citenamefont {Taniguchi}, \citenamefont {Kaxiras},\ and\
  \citenamefont {Jarillo-Herrero}}]{Cao2018-unconv_sc}%
  \BibitemOpen
  \bibfield  {author} {\bibinfo {author} {\bibfnamefont {Y.}~\bibnamefont
  {Cao}}, \bibinfo {author} {\bibfnamefont {V.}~\bibnamefont {Fatemi}},
  \bibinfo {author} {\bibfnamefont {S.}~\bibnamefont {Fang}}, \bibinfo {author}
  {\bibfnamefont {K.}~\bibnamefont {Watanabe}}, \bibinfo {author}
  {\bibfnamefont {T.}~\bibnamefont {Taniguchi}}, \bibinfo {author}
  {\bibfnamefont {E.}~\bibnamefont {Kaxiras}}, \ and\ \bibinfo {author}
  {\bibfnamefont {P.}~\bibnamefont {Jarillo-Herrero}},\ }\bibfield  {title}
  {\emph {\enquote {\bibinfo {title} {Unconventional superconductivity in
  magic-angle graphene superlattices},}\ }}\href {\doibase 10.1038/nature26160}
  {\bibfield  {journal} {\bibinfo  {journal} {Nature}\ }\textbf {\bibinfo
  {volume} {556}},\ \bibinfo {pages} {43} (\bibinfo {year}
  {2018}{\natexlab{b}})}\BibitemShut {NoStop}%
\bibitem [{\citenamefont {Oh}\ \emph {et~al.}(2021)\citenamefont {Oh},
  \citenamefont {Nuckolls}, \citenamefont {Wong}, \citenamefont {Lee},
  \citenamefont {Liu}, \citenamefont {Watanabe}, \citenamefont {Taniguchi},\
  and\ \citenamefont {Yazdani}}]{Oh2021-tBLG_xpt}%
  \BibitemOpen
  \bibfield  {author} {\bibinfo {author} {\bibfnamefont {M.}~\bibnamefont
  {Oh}}, \bibinfo {author} {\bibfnamefont {K.~P.}\ \bibnamefont {Nuckolls}},
  \bibinfo {author} {\bibfnamefont {D.}~\bibnamefont {Wong}}, \bibinfo {author}
  {\bibfnamefont {R.~L.}\ \bibnamefont {Lee}}, \bibinfo {author} {\bibfnamefont
  {X.}~\bibnamefont {Liu}}, \bibinfo {author} {\bibfnamefont {K.}~\bibnamefont
  {Watanabe}}, \bibinfo {author} {\bibfnamefont {T.}~\bibnamefont {Taniguchi}},
  \ and\ \bibinfo {author} {\bibfnamefont {A.}~\bibnamefont {Yazdani}},\
  }\bibfield  {title} {\emph {\enquote {\bibinfo {title} {Evidence for
  unconventional superconductivity in twisted bilayer graphene},}\ }}\href
  {\doibase 10.1038/s41586-021-04121-x} {\bibfield  {journal} {\bibinfo
  {journal} {Nature}\ }\textbf {\bibinfo {volume} {600}},\ \bibinfo {pages}
  {240} (\bibinfo {year} {2021})}\BibitemShut {NoStop}%
\bibitem [{\citenamefont {Sinha}\ \emph {et~al.}(2022)\citenamefont {Sinha},
  \citenamefont {Adak}, \citenamefont {Chakraborty}, \citenamefont {Das},
  \citenamefont {Debnath}, \citenamefont {Sangani}, \citenamefont {Watanabe},
  \citenamefont {Taniguchi}, \citenamefont {Waghmare}, \citenamefont
  {Agarwal},\ and\ \citenamefont {Deshmukh}}]{Sinha2022}%
  \BibitemOpen
  \bibfield  {author} {\bibinfo {author} {\bibfnamefont {S.}~\bibnamefont
  {Sinha}}, \bibinfo {author} {\bibfnamefont {P.~C.}\ \bibnamefont {Adak}},
  \bibinfo {author} {\bibfnamefont {A.}~\bibnamefont {Chakraborty}}, \bibinfo
  {author} {\bibfnamefont {K.}~\bibnamefont {Das}}, \bibinfo {author}
  {\bibfnamefont {K.}~\bibnamefont {Debnath}}, \bibinfo {author} {\bibfnamefont
  {L.~D.~V.}\ \bibnamefont {Sangani}}, \bibinfo {author} {\bibfnamefont
  {K.}~\bibnamefont {Watanabe}}, \bibinfo {author} {\bibfnamefont
  {T.}~\bibnamefont {Taniguchi}}, \bibinfo {author} {\bibfnamefont {U.~V.}\
  \bibnamefont {Waghmare}}, \bibinfo {author} {\bibfnamefont {A.}~\bibnamefont
  {Agarwal}}, \ and\ \bibinfo {author} {\bibfnamefont {M.~M.}\ \bibnamefont
  {Deshmukh}},\ }\bibfield  {title} {\emph {\enquote {\bibinfo {title} {Berry
  curvature dipole senses topological transition in a moir{\'e}
  superlattice},}\ }}\href {\doibase 10.1038/s41567-022-01606-y} {\bibfield
  {journal} {\bibinfo  {journal} {Nature Physics}\ }\textbf {\bibinfo {volume}
  {18}},\ \bibinfo {pages} {765} (\bibinfo {year} {2022})}\BibitemShut
  {NoStop}%
\bibitem [{\citenamefont {Bistritzer}\ and\ \citenamefont
  {MacDonald}(2011)}]{MacDonald-tBLG}%
  \BibitemOpen
  \bibfield  {author} {\bibinfo {author} {\bibfnamefont {R.}~\bibnamefont
  {Bistritzer}}\ and\ \bibinfo {author} {\bibfnamefont {A.~H.}\ \bibnamefont
  {MacDonald}},\ }\bibfield  {title} {\emph {\enquote {\bibinfo {title} {Moiré
  bands in twisted double-layer graphene},}\ }}\href {\doibase
  10.1073/pnas.1108174108} {\bibfield  {journal} {\bibinfo  {journal}
  {Proceedings of the National Academy of Sciences}\ }\textbf {\bibinfo
  {volume} {108}},\ \bibinfo {pages} {12233} (\bibinfo {year}
  {2011})}\BibitemShut {NoStop}%
\bibitem [{\citenamefont {Moon}\ \emph {et~al.}(2014)\citenamefont {Moon},
  \citenamefont {Son},\ and\ \citenamefont {Koshino}}]{Moon-tBLG}%
  \BibitemOpen
  \bibfield  {author} {\bibinfo {author} {\bibfnamefont {P.}~\bibnamefont
  {Moon}}, \bibinfo {author} {\bibfnamefont {Y.-W.}\ \bibnamefont {Son}}, \
  and\ \bibinfo {author} {\bibfnamefont {M.}~\bibnamefont {Koshino}},\
  }\bibfield  {title} {\emph {\enquote {\bibinfo {title} {Optical absorption of
  twisted bilayer graphene with interlayer potential asymmetry},}\ }}\href
  {\doibase 10.1103/PhysRevB.90.155427} {\bibfield  {journal} {\bibinfo
  {journal} {Phys. Rev. B}\ }\textbf {\bibinfo {volume} {90}},\ \bibinfo
  {pages} {155427} (\bibinfo {year} {2014})}\BibitemShut {NoStop}%
\bibitem [{\citenamefont {Koshino}\ \emph {et~al.}(2018)\citenamefont
  {Koshino}, \citenamefont {Yuan}, \citenamefont {Koretsune}, \citenamefont
  {Ochi}, \citenamefont {Kuroki},\ and\ \citenamefont {Fu}}]{Koshino-tBLG}%
  \BibitemOpen
  \bibfield  {author} {\bibinfo {author} {\bibfnamefont {M.}~\bibnamefont
  {Koshino}}, \bibinfo {author} {\bibfnamefont {N.~F.~Q.}\ \bibnamefont
  {Yuan}}, \bibinfo {author} {\bibfnamefont {T.}~\bibnamefont {Koretsune}},
  \bibinfo {author} {\bibfnamefont {M.}~\bibnamefont {Ochi}}, \bibinfo {author}
  {\bibfnamefont {K.}~\bibnamefont {Kuroki}}, \ and\ \bibinfo {author}
  {\bibfnamefont {L.}~\bibnamefont {Fu}},\ }\bibfield  {title} {\emph {\enquote
  {\bibinfo {title} {Maximally Localized Wannier Orbitals and the Extended
  Hubbard Model for Twisted Bilayer Graphene},}\ }}\href {\doibase
  10.1103/PhysRevX.8.031087} {\bibfield  {journal} {\bibinfo  {journal} {Phys.
  Rev. X}\ }\textbf {\bibinfo {volume} {8}},\ \bibinfo {pages} {031087}
  (\bibinfo {year} {2018})}\BibitemShut {NoStop}%
\bibitem [{\citenamefont {Lopes~dos Santos}\ \emph {et~al.}(2007)\citenamefont
  {Lopes~dos Santos}, \citenamefont {Peres},\ and\ \citenamefont
  {Castro~Neto}}]{Santos-Peres-tBLG}%
  \BibitemOpen
  \bibfield  {author} {\bibinfo {author} {\bibfnamefont {J.~M.~B.}\
  \bibnamefont {Lopes~dos Santos}}, \bibinfo {author} {\bibfnamefont
  {N.~M.~R.}\ \bibnamefont {Peres}}, \ and\ \bibinfo {author} {\bibfnamefont
  {A.~H.}\ \bibnamefont {Castro~Neto}},\ }\bibfield  {title} {\emph {\enquote
  {\bibinfo {title} {Graphene Bilayer with a Twist: Electronic Structure},}\
  }}\href {\doibase 10.1103/PhysRevLett.99.256802} {\bibfield  {journal}
  {\bibinfo  {journal} {Phys. Rev. Lett.}\ }\textbf {\bibinfo {volume} {99}},\
  \bibinfo {pages} {256802} (\bibinfo {year} {2007})}\BibitemShut {NoStop}%
\bibitem [{\citenamefont {Shallcross}\ \emph {et~al.}(2008)\citenamefont
  {Shallcross}, \citenamefont {Sharma},\ and\ \citenamefont
  {Pankratov}}]{Shallcross-tBLG}%
  \BibitemOpen
  \bibfield  {author} {\bibinfo {author} {\bibfnamefont {S.}~\bibnamefont
  {Shallcross}}, \bibinfo {author} {\bibfnamefont {S.}~\bibnamefont {Sharma}},
  \ and\ \bibinfo {author} {\bibfnamefont {O.~A.}\ \bibnamefont {Pankratov}},\
  }\bibfield  {title} {\emph {\enquote {\bibinfo {title} {Quantum Interference
  at the Twist Boundary in Graphene},}\ }}\href {\doibase
  10.1103/PhysRevLett.101.056803} {\bibfield  {journal} {\bibinfo  {journal}
  {Phys. Rev. Lett.}\ }\textbf {\bibinfo {volume} {101}},\ \bibinfo {pages}
  {056803} (\bibinfo {year} {2008})}\BibitemShut {NoStop}%
\bibitem [{\citenamefont {Song}\ \emph {et~al.}(2019)\citenamefont {Song},
  \citenamefont {Wang}, \citenamefont {Shi}, \citenamefont {Li}, \citenamefont
  {Fang},\ and\ \citenamefont {Bernevig}}]{All_Magic_Angle-tBLG}%
  \BibitemOpen
  \bibfield  {author} {\bibinfo {author} {\bibfnamefont {Z.}~\bibnamefont
  {Song}}, \bibinfo {author} {\bibfnamefont {Z.}~\bibnamefont {Wang}}, \bibinfo
  {author} {\bibfnamefont {W.}~\bibnamefont {Shi}}, \bibinfo {author}
  {\bibfnamefont {G.}~\bibnamefont {Li}}, \bibinfo {author} {\bibfnamefont
  {C.}~\bibnamefont {Fang}}, \ and\ \bibinfo {author} {\bibfnamefont {B.~A.}\
  \bibnamefont {Bernevig}},\ }\bibfield  {title} {\emph {\enquote {\bibinfo
  {title} {All Magic Angles in Twisted Bilayer Graphene are Topological},}\
  }}\href {\doibase 10.1103/PhysRevLett.123.036401} {\bibfield  {journal}
  {\bibinfo  {journal} {Phys. Rev. Lett.}\ }\textbf {\bibinfo {volume} {123}},\
  \bibinfo {pages} {036401} (\bibinfo {year} {2019})}\BibitemShut {NoStop}%
\bibitem [{\citenamefont {Tarnopolsky}\ \emph {et~al.}(2019)\citenamefont
  {Tarnopolsky}, \citenamefont {Kruchkov},\ and\ \citenamefont
  {Vishwanath}}]{Origin_of_Magic_Angle-tBLG}%
  \BibitemOpen
  \bibfield  {author} {\bibinfo {author} {\bibfnamefont {G.}~\bibnamefont
  {Tarnopolsky}}, \bibinfo {author} {\bibfnamefont {A.~J.}\ \bibnamefont
  {Kruchkov}}, \ and\ \bibinfo {author} {\bibfnamefont {A.}~\bibnamefont
  {Vishwanath}},\ }\bibfield  {title} {\emph {\enquote {\bibinfo {title}
  {Origin of Magic Angles in Twisted Bilayer Graphene},}\ }}\href {\doibase
  10.1103/PhysRevLett.122.106405} {\bibfield  {journal} {\bibinfo  {journal}
  {Phys. Rev. Lett.}\ }\textbf {\bibinfo {volume} {122}},\ \bibinfo {pages}
  {106405} (\bibinfo {year} {2019})}\BibitemShut {NoStop}%
\bibitem [{\citenamefont {Xie}\ and\ \citenamefont
  {MacDonald}(2020)}]{CIS-tBLG}%
  \BibitemOpen
  \bibfield  {author} {\bibinfo {author} {\bibfnamefont {M.}~\bibnamefont
  {Xie}}\ and\ \bibinfo {author} {\bibfnamefont {A.~H.}\ \bibnamefont
  {MacDonald}},\ }\bibfield  {title} {\emph {\enquote {\bibinfo {title} {Nature
  of the Correlated Insulator States in Twisted Bilayer Graphene},}\ }}\href
  {\doibase 10.1103/PhysRevLett.124.097601} {\bibfield  {journal} {\bibinfo
  {journal} {Phys. Rev. Lett.}\ }\textbf {\bibinfo {volume} {124}},\ \bibinfo
  {pages} {097601} (\bibinfo {year} {2020})}\BibitemShut {NoStop}%
\bibitem [{\citenamefont {Wolf}\ \emph {et~al.}(2019)\citenamefont {Wolf},
  \citenamefont {Lado}, \citenamefont {Blatter},\ and\ \citenamefont
  {Zilberberg}}]{Magnetism-tBLG}%
  \BibitemOpen
  \bibfield  {author} {\bibinfo {author} {\bibfnamefont {T.~M.~R.}\
  \bibnamefont {Wolf}}, \bibinfo {author} {\bibfnamefont {J.~L.}\ \bibnamefont
  {Lado}}, \bibinfo {author} {\bibfnamefont {G.}~\bibnamefont {Blatter}}, \
  and\ \bibinfo {author} {\bibfnamefont {O.}~\bibnamefont {Zilberberg}},\
  }\bibfield  {title} {\emph {\enquote {\bibinfo {title} {Electrically Tunable
  Flat Bands and Magnetism in Twisted Bilayer Graphene},}\ }}\href {\doibase
  10.1103/PhysRevLett.123.096802} {\bibfield  {journal} {\bibinfo  {journal}
  {Phys. Rev. Lett.}\ }\textbf {\bibinfo {volume} {123}},\ \bibinfo {pages}
  {096802} (\bibinfo {year} {2019})}\BibitemShut {NoStop}%
\bibitem [{\citenamefont {Song}\ and\ \citenamefont
  {Bernevig}(2022)}]{Heavy_fermion-tBLG}%
  \BibitemOpen
  \bibfield  {author} {\bibinfo {author} {\bibfnamefont {Z.-D.}\ \bibnamefont
  {Song}}\ and\ \bibinfo {author} {\bibfnamefont {B.~A.}\ \bibnamefont
  {Bernevig}},\ }\bibfield  {title} {\emph {\enquote {\bibinfo {title}
  {Magic-Angle Twisted Bilayer Graphene as a Topological Heavy Fermion
  Problem},}\ }}\href {\doibase 10.1103/PhysRevLett.129.047601} {\bibfield
  {journal} {\bibinfo  {journal} {Phys. Rev. Lett.}\ }\textbf {\bibinfo
  {volume} {129}},\ \bibinfo {pages} {047601} (\bibinfo {year}
  {2022})}\BibitemShut {NoStop}%
\bibitem [{\citenamefont {Po}\ \emph {et~al.}(2018)\citenamefont {Po},
  \citenamefont {Zou}, \citenamefont {Vishwanath},\ and\ \citenamefont
  {Senthil}}]{Origin_mott}%
  \BibitemOpen
  \bibfield  {author} {\bibinfo {author} {\bibfnamefont {H.~C.}\ \bibnamefont
  {Po}}, \bibinfo {author} {\bibfnamefont {L.}~\bibnamefont {Zou}}, \bibinfo
  {author} {\bibfnamefont {A.}~\bibnamefont {Vishwanath}}, \ and\ \bibinfo
  {author} {\bibfnamefont {T.}~\bibnamefont {Senthil}},\ }\bibfield  {title}
  {\emph {\enquote {\bibinfo {title} {Origin of Mott Insulating Behavior and
  Superconductivity in Twisted Bilayer Graphene},}\ }}\href {\doibase
  10.1103/PhysRevX.8.031089} {\bibfield  {journal} {\bibinfo  {journal} {Phys.
  Rev. X}\ }\textbf {\bibinfo {volume} {8}},\ \bibinfo {pages} {031089}
  (\bibinfo {year} {2018})}\BibitemShut {NoStop}%
\bibitem [{\citenamefont {Lian}\ \emph {et~al.}(2019)\citenamefont {Lian},
  \citenamefont {Wang},\ and\ \citenamefont {Bernevig}}]{phonon_SC}%
  \BibitemOpen
  \bibfield  {author} {\bibinfo {author} {\bibfnamefont {B.}~\bibnamefont
  {Lian}}, \bibinfo {author} {\bibfnamefont {Z.}~\bibnamefont {Wang}}, \ and\
  \bibinfo {author} {\bibfnamefont {B.~A.}\ \bibnamefont {Bernevig}},\
  }\bibfield  {title} {\emph {\enquote {\bibinfo {title} {Twisted Bilayer
  Graphene: A Phonon-Driven Superconductor},}\ }}\href {\doibase
  10.1103/PhysRevLett.122.257002} {\bibfield  {journal} {\bibinfo  {journal}
  {Phys. Rev. Lett.}\ }\textbf {\bibinfo {volume} {122}},\ \bibinfo {pages}
  {257002} (\bibinfo {year} {2019})}\BibitemShut {NoStop}%
\bibitem [{\citenamefont {Yuan}\ and\ \citenamefont {Fu}(2018)}]{MIT_tBLG}%
  \BibitemOpen
  \bibfield  {author} {\bibinfo {author} {\bibfnamefont {N.~F.~Q.}\
  \bibnamefont {Yuan}}\ and\ \bibinfo {author} {\bibfnamefont {L.}~\bibnamefont
  {Fu}},\ }\bibfield  {title} {\emph {\enquote {\bibinfo {title} {Model for the
  metal-insulator transition in graphene superlattices and beyond},}\ }}\href
  {\doibase 10.1103/PhysRevB.98.045103} {\bibfield  {journal} {\bibinfo
  {journal} {Phys. Rev. B}\ }\textbf {\bibinfo {volume} {98}},\ \bibinfo
  {pages} {045103} (\bibinfo {year} {2018})}\BibitemShut {NoStop}%
\bibitem [{\citenamefont {Katz}\ \emph {et~al.}(2020)\citenamefont {Katz},
  \citenamefont {Refael},\ and\ \citenamefont {Lindner}}]{Floquet}%
  \BibitemOpen
  \bibfield  {author} {\bibinfo {author} {\bibfnamefont {O.}~\bibnamefont
  {Katz}}, \bibinfo {author} {\bibfnamefont {G.}~\bibnamefont {Refael}}, \ and\
  \bibinfo {author} {\bibfnamefont {N.~H.}\ \bibnamefont {Lindner}},\
  }\bibfield  {title} {\emph {\enquote {\bibinfo {title} {Optically induced
  flat bands in twisted bilayer graphene},}\ }}\href {\doibase
  10.1103/PhysRevB.102.155123} {\bibfield  {journal} {\bibinfo  {journal}
  {Phys. Rev. B}\ }\textbf {\bibinfo {volume} {102}},\ \bibinfo {pages}
  {155123} (\bibinfo {year} {2020})}\BibitemShut {NoStop}%
\bibitem [{\citenamefont {Esparza}\ and\ \citenamefont {Juricic}()}]{NH_tBLG1}%
  \BibitemOpen
  \bibfield  {author} {\bibinfo {author} {\bibfnamefont {J.~P.}\ \bibnamefont
  {Esparza}}\ and\ \bibinfo {author} {\bibfnamefont {V.}~\bibnamefont
  {Juricic}},\ }\bibfield  {title} {\emph {\enquote {\bibinfo {title}
  {Exceptional magic angles in non-Hermitian twisted bilayer graphene},}\
  }}\href {https://arxiv.org/abs/2408.08804} {\ }\Eprint
  {http://arxiv.org/abs/2408.08804}{arXiv:2408.08804
  [cond-mat.mes-hall]}\BibitemShut {NoStop}%
\bibitem [{\citenamefont {Huang}(2025)}]{NH_tBLG2}%
  \BibitemOpen
  \bibfield  {author} {\bibinfo {author} {\bibfnamefont {Y.}~\bibnamefont
  {Huang}},\ }\bibfield  {title} {\emph {\enquote {\bibinfo {title}
  {Exceptional topology in non-Hermitian twisted bilayer graphene},}\ }}\href
  {\doibase 10.1103/PhysRevB.111.085120} {\bibfield  {journal} {\bibinfo
  {journal} {Phys. Rev. B}\ }\textbf {\bibinfo {volume} {111}},\ \bibinfo
  {pages} {085120} (\bibinfo {year} {2025})}\BibitemShut {NoStop}%
\bibitem [{\citenamefont {Zhang}\ \emph {et~al.}(2019)\citenamefont {Zhang},
  \citenamefont {Mao},\ and\ \citenamefont {Senthil}}]{tBLG_hBN1}%
  \BibitemOpen
  \bibfield  {author} {\bibinfo {author} {\bibfnamefont {Y.-H.}\ \bibnamefont
  {Zhang}}, \bibinfo {author} {\bibfnamefont {D.}~\bibnamefont {Mao}}, \ and\
  \bibinfo {author} {\bibfnamefont {T.}~\bibnamefont {Senthil}},\ }\bibfield
  {title} {\emph {\enquote {\bibinfo {title} {Twisted bilayer graphene aligned
  with hexagonal boron nitride: Anomalous Hall effect and a lattice model},}\
  }}\href {\doibase 10.1103/PhysRevResearch.1.033126} {\bibfield  {journal}
  {\bibinfo  {journal} {Phys. Rev. Res.}\ }\textbf {\bibinfo {volume} {1}},\
  \bibinfo {pages} {033126} (\bibinfo {year} {2019})}\BibitemShut {NoStop}%
\bibitem [{\citenamefont {Cea}\ \emph {et~al.}(2020)\citenamefont {Cea},
  \citenamefont {Pantale\'on},\ and\ \citenamefont {Guinea}}]{tBLG_hBN2}%
  \BibitemOpen
  \bibfield  {author} {\bibinfo {author} {\bibfnamefont {T.}~\bibnamefont
  {Cea}}, \bibinfo {author} {\bibfnamefont {P.~A.}\ \bibnamefont
  {Pantale\'on}}, \ and\ \bibinfo {author} {\bibfnamefont {F.}~\bibnamefont
  {Guinea}},\ }\bibfield  {title} {\emph {\enquote {\bibinfo {title} {Band
  structure of twisted bilayer graphene on hexagonal boron nitride},}\ }}\href
  {\doibase 10.1103/PhysRevB.102.155136} {\bibfield  {journal} {\bibinfo
  {journal} {Phys. Rev. B}\ }\textbf {\bibinfo {volume} {102}},\ \bibinfo
  {pages} {155136} (\bibinfo {year} {2020})}\BibitemShut {NoStop}%
\bibitem [{\citenamefont {McCann}(2011)}]{McCann_2011-BilayerReview}%
  \BibitemOpen
  \bibfield  {author} {\bibinfo {author} {\bibfnamefont {E.}~\bibnamefont
  {McCann}},\ }in\ \href {\doibase 10.1007/978-3-642-22984-8_8} {\emph
  {\bibinfo {booktitle} {Graphene Nanoelectronics}}}\ (\bibinfo  {publisher}
  {Springer Berlin Heidelberg},\ \bibinfo {year} {2011})\ pp.\ \bibinfo {pages}
  {237--275}\BibitemShut {NoStop}%
\bibitem [{\citenamefont {Miao}\ \emph {et~al.}(2023)\citenamefont {Miao},
  \citenamefont {Li}, \citenamefont {Han}, \citenamefont {Pan},\ and\
  \citenamefont {Dai}}]{tBLG_cutoff1}%
  \BibitemOpen
  \bibfield  {author} {\bibinfo {author} {\bibfnamefont {W.}~\bibnamefont
  {Miao}}, \bibinfo {author} {\bibfnamefont {C.}~\bibnamefont {Li}}, \bibinfo
  {author} {\bibfnamefont {X.}~\bibnamefont {Han}}, \bibinfo {author}
  {\bibfnamefont {D.}~\bibnamefont {Pan}}, \ and\ \bibinfo {author}
  {\bibfnamefont {X.}~\bibnamefont {Dai}},\ }\bibfield  {title} {\emph
  {\enquote {\bibinfo {title} {Truncated atomic plane wave method for subband
  structure calculations of moir\'e systems},}\ }}\href {\doibase
  10.1103/PhysRevB.107.125112} {\bibfield  {journal} {\bibinfo  {journal}
  {Phys. Rev. B}\ }\textbf {\bibinfo {volume} {107}},\ \bibinfo {pages}
  {125112} (\bibinfo {year} {2023})}\BibitemShut {NoStop}%
\bibitem [{\citenamefont {Xie}\ and\ \citenamefont {Liu}(2023)}]{tBLG_cutoff2}%
  \BibitemOpen
  \bibfield  {author} {\bibinfo {author} {\bibfnamefont {B.}~\bibnamefont
  {Xie}}\ and\ \bibinfo {author} {\bibfnamefont {J.}~\bibnamefont {Liu}},\
  }\bibfield  {title} {\emph {\enquote {\bibinfo {title} {Lattice distortions,
  moir\'e phonons, and relaxed electronic band structures in magic-angle
  twisted bilayer graphene},}\ }}\href {\doibase 10.1103/PhysRevB.108.094115}
  {\bibfield  {journal} {\bibinfo  {journal} {Phys. Rev. B}\ }\textbf {\bibinfo
  {volume} {108}},\ \bibinfo {pages} {094115} (\bibinfo {year}
  {2023})}\BibitemShut {NoStop}%
\bibitem [{\citenamefont {Wang}\ \emph {et~al.}(2021)\citenamefont {Wang},
  \citenamefont {Zheng}, \citenamefont {Millis},\ and\ \citenamefont
  {Cano}}]{chiral_model}%
  \BibitemOpen
  \bibfield  {author} {\bibinfo {author} {\bibfnamefont {J.}~\bibnamefont
  {Wang}}, \bibinfo {author} {\bibfnamefont {Y.}~\bibnamefont {Zheng}},
  \bibinfo {author} {\bibfnamefont {A.~J.}\ \bibnamefont {Millis}}, \ and\
  \bibinfo {author} {\bibfnamefont {J.}~\bibnamefont {Cano}},\ }\bibfield
  {title} {\emph {\enquote {\bibinfo {title} {Chiral approximation to twisted
  bilayer graphene: Exact intravalley inversion symmetry, nodal structure, and
  implications for higher magic angles},}\ }}\href {\doibase
  10.1103/PhysRevResearch.3.023155} {\bibfield  {journal} {\bibinfo  {journal}
  {Phys. Rev. Res.}\ }\textbf {\bibinfo {volume} {3}},\ \bibinfo {pages}
  {023155} (\bibinfo {year} {2021})}\BibitemShut {NoStop}%
\bibitem [{\citenamefont {Uchida}\ \emph {et~al.}(2014)\citenamefont {Uchida},
  \citenamefont {Furuya}, \citenamefont {Iwata},\ and\ \citenamefont
  {Oshiyama}}]{corrugation-DFT-uu}%
  \BibitemOpen
  \bibfield  {author} {\bibinfo {author} {\bibfnamefont {K.}~\bibnamefont
  {Uchida}}, \bibinfo {author} {\bibfnamefont {S.}~\bibnamefont {Furuya}},
  \bibinfo {author} {\bibfnamefont {J.-I.}\ \bibnamefont {Iwata}}, \ and\
  \bibinfo {author} {\bibfnamefont {A.}~\bibnamefont {Oshiyama}},\ }\bibfield
  {title} {\emph {\enquote {\bibinfo {title} {Atomic corrugation and electron
  localization due to Moir\'e patterns in twisted bilayer graphenes},}\ }}\href
  {\doibase 10.1103/PhysRevB.90.155451} {\bibfield  {journal} {\bibinfo
  {journal} {Phys. Rev. B}\ }\textbf {\bibinfo {volume} {90}},\ \bibinfo
  {pages} {155451} (\bibinfo {year} {2014})}\BibitemShut {NoStop}%
\bibitem [{\citenamefont {Dai}\ \emph {et~al.}(2016)\citenamefont {Dai},
  \citenamefont {Xiang},\ and\ \citenamefont {Srolovitz}}]{Dai2016-uu}%
  \BibitemOpen
  \bibfield  {author} {\bibinfo {author} {\bibfnamefont {S.}~\bibnamefont
  {Dai}}, \bibinfo {author} {\bibfnamefont {Y.}~\bibnamefont {Xiang}}, \ and\
  \bibinfo {author} {\bibfnamefont {D.~J.}\ \bibnamefont {Srolovitz}},\
  }\bibfield  {title} {\emph {\enquote {\bibinfo {title} {Twisted Bilayer
  Graphene: Moir{\'e} with a Twist},}\ }}\href {\doibase
  10.1021/acs.nanolett.6b02870} {\bibfield  {journal} {\bibinfo  {journal}
  {Nano Letters}\ }\textbf {\bibinfo {volume} {16}},\ \bibinfo {pages} {5923}
  (\bibinfo {year} {2016})}\BibitemShut {NoStop}%
\bibitem [{\citenamefont {Fukui}\ \emph {et~al.}(2005)\citenamefont {Fukui},
  \citenamefont {Hatsugai},\ and\ \citenamefont {Suzuki}}]{Fukui-chern_no}%
  \BibitemOpen
  \bibfield  {author} {\bibinfo {author} {\bibfnamefont {T.}~\bibnamefont
  {Fukui}}, \bibinfo {author} {\bibfnamefont {Y.}~\bibnamefont {Hatsugai}}, \
  and\ \bibinfo {author} {\bibfnamefont {H.}~\bibnamefont {Suzuki}},\
  }\bibfield  {title} {\emph {\enquote {\bibinfo {title} {Chern Numbers in
  Discretized Brillouin Zone: Efficient Method of Computing (Spin) Hall
  Conductances},}\ }}\href {\doibase 10.1143/JPSJ.74.1674} {\bibfield
  {journal} {\bibinfo  {journal} {Journal of the Physical Society of Japan}\
  }\textbf {\bibinfo {volume} {74}},\ \bibinfo {pages} {1674} (\bibinfo {year}
  {2005})}\BibitemShut {NoStop}%
\bibitem [{\citenamefont {Ezawa}(2012)}]{Ezawa_2012-Diff_chern}%
  \BibitemOpen
  \bibfield  {author} {\bibinfo {author} {\bibfnamefont {M.}~\bibnamefont
  {Ezawa}},\ }\bibfield  {title} {\emph {\enquote {\bibinfo {title}
  {Quasi-Topological Insulator and Trigonal Warping in Gated Bilayer
  Silicene},}\ }}\href {\doibase 10.1143/jpsj.81.104713} {\bibfield  {journal}
  {\bibinfo  {journal} {Journal of the Physical Society of Japan}\ }\textbf
  {\bibinfo {volume} {81}},\ \bibinfo {pages} {104713} (\bibinfo {year}
  {2012})}\BibitemShut {NoStop}%
\bibitem [{\citenamefont {Oudich}\ \emph {et~al.}(2022)\citenamefont {Oudich},
  \citenamefont {Deng},\ and\ \citenamefont {Jing}}]{twisted_pillared_plates}%
  \BibitemOpen
  \bibfield  {author} {\bibinfo {author} {\bibfnamefont {M.}~\bibnamefont
  {Oudich}}, \bibinfo {author} {\bibfnamefont {Y.}~\bibnamefont {Deng}}, \ and\
  \bibinfo {author} {\bibfnamefont {Y.}~\bibnamefont {Jing}},\ }\bibfield
  {title} {\emph {\enquote {\bibinfo {title} {Twisted pillared phononic crystal
  plates},}\ }}\href {\doibase 10.1063/5.0097082} {\bibfield  {journal}
  {\bibinfo  {journal} {Applied Physics Letters}\ }\textbf {\bibinfo {volume}
  {120}},\ \bibinfo {pages} {232202} (\bibinfo {year} {2022})}\BibitemShut
  {NoStop}%
\bibitem [{\citenamefont {Gardezi}\ \emph {et~al.}(2021)\citenamefont
  {Gardezi}, \citenamefont {Pirie}, \citenamefont {Carr}, \citenamefont
  {Dorrell},\ and\ \citenamefont {Hoffman}}]{twist_accoustic}%
  \BibitemOpen
  \bibfield  {author} {\bibinfo {author} {\bibfnamefont {S.~M.}\ \bibnamefont
  {Gardezi}}, \bibinfo {author} {\bibfnamefont {H.}~\bibnamefont {Pirie}},
  \bibinfo {author} {\bibfnamefont {S.}~\bibnamefont {Carr}}, \bibinfo {author}
  {\bibfnamefont {W.}~\bibnamefont {Dorrell}}, \ and\ \bibinfo {author}
  {\bibfnamefont {J.~E.}\ \bibnamefont {Hoffman}},\ }\bibfield  {title} {\emph
  {\enquote {\bibinfo {title} {Simulating twistronics in acoustic
  metamaterials},}\ }}\href {\doibase 10.1088/2053-1583/abf252} {\bibfield
  {journal} {\bibinfo  {journal} {2D Materials}\ }\textbf {\bibinfo {volume}
  {8}},\ \bibinfo {pages} {031002} (\bibinfo {year} {2021})}\BibitemShut
  {NoStop}%
\bibitem [{\citenamefont {Lou}\ \emph {et~al.}(2022)\citenamefont {Lou},
  \citenamefont {Wang}, \citenamefont {Rodríguez}, \citenamefont {Cappelli},\
  and\ \citenamefont {Fan}}]{tBL_photonic1}%
  \BibitemOpen
  \bibfield  {author} {\bibinfo {author} {\bibfnamefont {B.}~\bibnamefont
  {Lou}}, \bibinfo {author} {\bibfnamefont {B.}~\bibnamefont {Wang}}, \bibinfo
  {author} {\bibfnamefont {J.~A.}\ \bibnamefont {Rodríguez}}, \bibinfo
  {author} {\bibfnamefont {M.}~\bibnamefont {Cappelli}}, \ and\ \bibinfo
  {author} {\bibfnamefont {S.}~\bibnamefont {Fan}},\ }\bibfield  {title} {\emph
  {\enquote {\bibinfo {title} {Tunable guided resonance in twisted bilayer
  photonic crystal},}\ }}\href {\doibase 10.1126/sciadv.add4339} {\bibfield
  {journal} {\bibinfo  {journal} {Science Advances}\ }\textbf {\bibinfo
  {volume} {8}},\ \bibinfo {pages} {eadd4339} (\bibinfo {year}
  {2022})}\BibitemShut {NoStop}%
\bibitem [{\citenamefont {Tang}\ \emph {et~al.}(2023)\citenamefont {Tang},
  \citenamefont {Lou}, \citenamefont {Du}, \citenamefont {Zhang}, \citenamefont
  {Ni}, \citenamefont {Xu}, \citenamefont {Jin}, \citenamefont {Fan},\ and\
  \citenamefont {Mazur}}]{tBL_photonic2}%
  \BibitemOpen
  \bibfield  {author} {\bibinfo {author} {\bibfnamefont {H.}~\bibnamefont
  {Tang}}, \bibinfo {author} {\bibfnamefont {B.}~\bibnamefont {Lou}}, \bibinfo
  {author} {\bibfnamefont {F.}~\bibnamefont {Du}}, \bibinfo {author}
  {\bibfnamefont {M.}~\bibnamefont {Zhang}}, \bibinfo {author} {\bibfnamefont
  {X.}~\bibnamefont {Ni}}, \bibinfo {author} {\bibfnamefont {W.}~\bibnamefont
  {Xu}}, \bibinfo {author} {\bibfnamefont {R.}~\bibnamefont {Jin}}, \bibinfo
  {author} {\bibfnamefont {S.}~\bibnamefont {Fan}}, \ and\ \bibinfo {author}
  {\bibfnamefont {E.}~\bibnamefont {Mazur}},\ }\bibfield  {title} {\emph
  {\enquote {\bibinfo {title} {Experimental probe of twist angle–dependent
  band structure of on-chip optical bilayer photonic crystal},}\ }}\href
  {\doibase 10.1126/sciadv.adh8498} {\bibfield  {journal} {\bibinfo  {journal}
  {Science Advances}\ }\textbf {\bibinfo {volume} {9}},\ \bibinfo {pages}
  {eadh8498} (\bibinfo {year} {2023})}\BibitemShut {NoStop}%
\bibitem [{\citenamefont {Meng}\ \emph
  {et~al.}(2023{\natexlab{a}})\citenamefont {Meng}, \citenamefont {Wang},
  \citenamefont {Han}, \citenamefont {Liu}, \citenamefont {Wen}, \citenamefont
  {Gao}, \citenamefont {Wang}, \citenamefont {Chin},\ and\ \citenamefont
  {Zhang}}]{BEC_tBLG}%
  \BibitemOpen
  \bibfield  {author} {\bibinfo {author} {\bibfnamefont {Z.}~\bibnamefont
  {Meng}}, \bibinfo {author} {\bibfnamefont {L.}~\bibnamefont {Wang}}, \bibinfo
  {author} {\bibfnamefont {W.}~\bibnamefont {Han}}, \bibinfo {author}
  {\bibfnamefont {F.}~\bibnamefont {Liu}}, \bibinfo {author} {\bibfnamefont
  {K.}~\bibnamefont {Wen}}, \bibinfo {author} {\bibfnamefont {C.}~\bibnamefont
  {Gao}}, \bibinfo {author} {\bibfnamefont {P.}~\bibnamefont {Wang}}, \bibinfo
  {author} {\bibfnamefont {C.}~\bibnamefont {Chin}}, \ and\ \bibinfo {author}
  {\bibfnamefont {J.}~\bibnamefont {Zhang}},\ }\bibfield  {title} {\emph
  {\enquote {\bibinfo {title} {Atomic Bose--Einstein condensate in
  twisted-bilayer optical lattices},}\ }}\href {\doibase
  10.1038/s41586-023-05695-4} {\bibfield  {journal} {\bibinfo  {journal}
  {Nature}\ }\textbf {\bibinfo {volume} {615}},\ \bibinfo {pages} {231}
  (\bibinfo {year} {2023}{\natexlab{a}})}\BibitemShut {NoStop}%
\bibitem [{\citenamefont {Bergholtz}\ \emph {et~al.}(2021)\citenamefont
  {Bergholtz}, \citenamefont {Budich},\ and\ \citenamefont
  {Kunst}}]{RMP_topo_NH}%
  \BibitemOpen
  \bibfield  {author} {\bibinfo {author} {\bibfnamefont {E.~J.}\ \bibnamefont
  {Bergholtz}}, \bibinfo {author} {\bibfnamefont {J.~C.}\ \bibnamefont
  {Budich}}, \ and\ \bibinfo {author} {\bibfnamefont {F.~K.}\ \bibnamefont
  {Kunst}},\ }\bibfield  {title} {\emph {\enquote {\bibinfo {title}
  {Exceptional topology of non-Hermitian systems},}\ }}\href {\doibase
  10.1103/RevModPhys.93.015005} {\bibfield  {journal} {\bibinfo  {journal}
  {Rev. Mod. Phys.}\ }\textbf {\bibinfo {volume} {93}},\ \bibinfo {pages}
  {015005} (\bibinfo {year} {2021})}\BibitemShut {NoStop}%
\bibitem [{\citenamefont {Ashida}\ \emph {et~al.}(2020)\citenamefont {Ashida},
  \citenamefont {Gong},\ and\ \citenamefont {and}}]{NH_AIP}%
  \BibitemOpen
  \bibfield  {author} {\bibinfo {author} {\bibfnamefont {Y.}~\bibnamefont
  {Ashida}}, \bibinfo {author} {\bibfnamefont {Z.}~\bibnamefont {Gong}}, \ and\
  \bibinfo {author} {\bibfnamefont {M.~U.}\ \bibnamefont {and}},\ }\bibfield
  {title} {\emph {\enquote {\bibinfo {title} {Non-Hermitian physics},}\ }}\href
  {\doibase 10.1080/00018732.2021.1876991} {\bibfield  {journal} {\bibinfo
  {journal} {Advances in Physics}\ }\textbf {\bibinfo {volume} {69}},\ \bibinfo
  {pages} {249} (\bibinfo {year} {2020})}\BibitemShut {NoStop}%
\bibitem [{\citenamefont {Ding}\ \emph {et~al.}(2022)\citenamefont {Ding},
  \citenamefont {Fang},\ and\ \citenamefont {Ma}}]{NH_topo_EP_geometries}%
  \BibitemOpen
  \bibfield  {author} {\bibinfo {author} {\bibfnamefont {K.}~\bibnamefont
  {Ding}}, \bibinfo {author} {\bibfnamefont {C.}~\bibnamefont {Fang}}, \ and\
  \bibinfo {author} {\bibfnamefont {G.}~\bibnamefont {Ma}},\ }\bibfield
  {title} {\emph {\enquote {\bibinfo {title} {Non-Hermitian topology and
  exceptional-point geometries},}\ }}\href {\doibase
  10.1038/s42254-022-00516-5} {\bibfield  {journal} {\bibinfo  {journal}
  {Nature Reviews Physics}\ }\textbf {\bibinfo {volume} {4}},\ \bibinfo {pages}
  {745} (\bibinfo {year} {2022})}\BibitemShut {NoStop}%
\bibitem [{\citenamefont {Song}\ \emph {et~al.}(2023)\citenamefont {Song},
  \citenamefont {Wu}, \citenamefont {Chen}, \citenamefont {Chen}, \citenamefont
  {Huang}, \citenamefont {Yuan}, \citenamefont {Zhu},\ and\ \citenamefont
  {Li}}]{Weyl_NH1}%
  \BibitemOpen
  \bibfield  {author} {\bibinfo {author} {\bibfnamefont {W.}~\bibnamefont
  {Song}}, \bibinfo {author} {\bibfnamefont {S.}~\bibnamefont {Wu}}, \bibinfo
  {author} {\bibfnamefont {C.}~\bibnamefont {Chen}}, \bibinfo {author}
  {\bibfnamefont {Y.}~\bibnamefont {Chen}}, \bibinfo {author} {\bibfnamefont
  {C.}~\bibnamefont {Huang}}, \bibinfo {author} {\bibfnamefont
  {L.}~\bibnamefont {Yuan}}, \bibinfo {author} {\bibfnamefont {S.}~\bibnamefont
  {Zhu}}, \ and\ \bibinfo {author} {\bibfnamefont {T.}~\bibnamefont {Li}},\
  }\bibfield  {title} {\emph {\enquote {\bibinfo {title} {Observation of Weyl
  Interface States in Non-Hermitian Synthetic Photonic Systems},}\ }}\href
  {\doibase 10.1103/PhysRevLett.130.043803} {\bibfield  {journal} {\bibinfo
  {journal} {Phys. Rev. Lett.}\ }\textbf {\bibinfo {volume} {130}},\ \bibinfo
  {pages} {043803} (\bibinfo {year} {2023})}\BibitemShut {NoStop}%
\bibitem [{\citenamefont {Cerjan}\ \emph {et~al.}(2019)\citenamefont {Cerjan},
  \citenamefont {Huang}, \citenamefont {Wang}, \citenamefont {Chen},
  \citenamefont {Chong},\ and\ \citenamefont {Rechtsman}}]{Weyl_NH2}%
  \BibitemOpen
  \bibfield  {author} {\bibinfo {author} {\bibfnamefont {A.}~\bibnamefont
  {Cerjan}}, \bibinfo {author} {\bibfnamefont {S.}~\bibnamefont {Huang}},
  \bibinfo {author} {\bibfnamefont {M.}~\bibnamefont {Wang}}, \bibinfo {author}
  {\bibfnamefont {K.~P.}\ \bibnamefont {Chen}}, \bibinfo {author}
  {\bibfnamefont {Y.}~\bibnamefont {Chong}}, \ and\ \bibinfo {author}
  {\bibfnamefont {M.~C.}\ \bibnamefont {Rechtsman}},\ }\bibfield  {title}
  {\emph {\enquote {\bibinfo {title} {Experimental realization of a Weyl
  exceptional ring},}\ }}\href {\doibase 10.1038/s41566-019-0453-z} {\bibfield
  {journal} {\bibinfo  {journal} {Nature Photonics}\ }\textbf {\bibinfo
  {volume} {13}},\ \bibinfo {pages} {623} (\bibinfo {year} {2019})}\BibitemShut
  {NoStop}%
\bibitem [{\citenamefont {Meng}\ \emph
  {et~al.}(2023{\natexlab{b}})\citenamefont {Meng}, \citenamefont {Wang},
  \citenamefont {Han}, \citenamefont {Liu}, \citenamefont {Wen}, \citenamefont
  {Gao}, \citenamefont {Wang}, \citenamefont {Chin},\ and\ \citenamefont
  {Zhang}}]{tBLG_optical_lattices}%
  \BibitemOpen
  \bibfield  {author} {\bibinfo {author} {\bibfnamefont {Z.}~\bibnamefont
  {Meng}}, \bibinfo {author} {\bibfnamefont {L.}~\bibnamefont {Wang}}, \bibinfo
  {author} {\bibfnamefont {W.}~\bibnamefont {Han}}, \bibinfo {author}
  {\bibfnamefont {F.}~\bibnamefont {Liu}}, \bibinfo {author} {\bibfnamefont
  {K.}~\bibnamefont {Wen}}, \bibinfo {author} {\bibfnamefont {C.}~\bibnamefont
  {Gao}}, \bibinfo {author} {\bibfnamefont {P.}~\bibnamefont {Wang}}, \bibinfo
  {author} {\bibfnamefont {C.}~\bibnamefont {Chin}}, \ and\ \bibinfo {author}
  {\bibfnamefont {J.}~\bibnamefont {Zhang}},\ }\bibfield  {title} {\emph
  {\enquote {\bibinfo {title} {Atomic Bose--Einstein condensate in
  twisted-bilayer optical lattices},}\ }}\href {\doibase
  10.1038/s41586-023-05695-4} {\bibfield  {journal} {\bibinfo  {journal}
  {Nature}\ }\textbf {\bibinfo {volume} {615}},\ \bibinfo {pages} {231}
  (\bibinfo {year} {2023}{\natexlab{b}})}\BibitemShut {NoStop}%
\bibitem [{\citenamefont {Salamon}\ \emph {et~al.}(2020)\citenamefont
  {Salamon}, \citenamefont {Celi}, \citenamefont {Chhajlany}, \citenamefont
  {Fr\'erot}, \citenamefont {Lewenstein}, \citenamefont {Tarruell},\ and\
  \citenamefont {Rakshit}}]{synthetic_tBLG}%
  \BibitemOpen
  \bibfield  {author} {\bibinfo {author} {\bibfnamefont {T.}~\bibnamefont
  {Salamon}}, \bibinfo {author} {\bibfnamefont {A.}~\bibnamefont {Celi}},
  \bibinfo {author} {\bibfnamefont {R.~W.}\ \bibnamefont {Chhajlany}}, \bibinfo
  {author} {\bibfnamefont {I.}~\bibnamefont {Fr\'erot}}, \bibinfo {author}
  {\bibfnamefont {M.}~\bibnamefont {Lewenstein}}, \bibinfo {author}
  {\bibfnamefont {L.}~\bibnamefont {Tarruell}}, \ and\ \bibinfo {author}
  {\bibfnamefont {D.}~\bibnamefont {Rakshit}},\ }\bibfield  {title} {\emph
  {\enquote {\bibinfo {title} {Simulating Twistronics without a Twist},}\
  }}\href {\doibase 10.1103/PhysRevLett.125.030504} {\bibfield  {journal}
  {\bibinfo  {journal} {Phys. Rev. Lett.}\ }\textbf {\bibinfo {volume} {125}},\
  \bibinfo {pages} {030504} (\bibinfo {year} {2020})}\BibitemShut {NoStop}%
\bibitem [{\citenamefont {Yuan}\ \emph {et~al.}(2019)\citenamefont {Yuan},
  \citenamefont {Isobe},\ and\ \citenamefont {Fu}}]{HO_VHS}%
  \BibitemOpen
  \bibfield  {author} {\bibinfo {author} {\bibfnamefont {N.~F.~Q.}\
  \bibnamefont {Yuan}}, \bibinfo {author} {\bibfnamefont {H.}~\bibnamefont
  {Isobe}}, \ and\ \bibinfo {author} {\bibfnamefont {L.}~\bibnamefont {Fu}},\
  }\bibfield  {title} {\emph {\enquote {\bibinfo {title} {Magic of high-order
  van Hove singularity},}\ }}\href {\doibase 10.1038/s41467-019-13670-9}
  {\bibfield  {journal} {\bibinfo  {journal} {Nature Communications}\ }\textbf
  {\bibinfo {volume} {10}},\ \bibinfo {pages} {5769} (\bibinfo {year}
  {2019})}\BibitemShut {NoStop}%
\end{thebibliography}%


\end{document}